\documentclass[floatfix,twocolumn]{aastex631}

\newcommand{\re}{$r_{\rm e}$}

\defcitealias{Kewley2001}{KE01}
\defcitealias{Kauffmann2003}{KA03}
\defcitealias{stasinska2006}{ST06}
\defcitealias{Sanchez2014}{SA14}
\defcitealias{cidfernandes2011}{CF11}
\defcitealias{Sanchez-Menguiano2016}{SM16}
\defcitealias{Sanchez-Menguiano2018}{SM18}
\defcitealias{lacerda2018}{LA18}

\begin{document}

\title{Analysis of the Internal  Radial Gradient of Chemical  Abundances in Spiral Galaxies from CALIFA}

\correspondingauthor{Andr\'e F. S. Cardoso}
\email{afelipe2992@gmail.com}

\author[0000-0003-1097-3247]{A. F. S. Cardoso}
\affiliation{Instituto de F{\'i}sica e Qu{\'i}mica, Universidade Federal de Itajub{\'a} \\
Av. BPS, 1303, 37500-903 \\
Itajub{\'a}-MG, Brazil}

\author[0000-0002-7103-8036]{O. Cavichia}
\affiliation{Instituto de F{\'i}sica e Qu{\'i}mica, Universidade Federal de Itajub{\'a} \\
Av. BPS, 1303, 37500-903 \\
Itajub{\'a}-MG, Brazil}


\author[0000-0003-0817-581X]{M. Moll{\'a}}
\affiliation{Departamento de Investigaci\'{o}n B\'{a}sica, CIEMAT \\
Avda. Complutense 40, E-28040 \\
Madrid, Spain}
\author[0000-0003-1888-6578]{L. S{\'a}nchez-Menguiano}
\affiliation{Universidad de Granada, Departamento de F{\'i}sica Te{\'o}rica y del Cosmos \\
Campus Fuente Nueva, Edificio Mecenas, 18071 \\
Granada, Spain}
\affiliation{Instituto Carlos I de F{\'i}sica Te{\'o}rica y Computacional \\
Facultad de Ciencias, 18071 \\
Granada, Spain}




\begin{abstract}

The study of chemical evolution is of paramount importance for understanding the galaxies evolution. Models and observations propose an inside-out mechanism in the formation of spiral galaxy disks implying a negative radial gradient of elemental abundances when represented in logarithmic scale. However, observed chemical abundance gradients, in some instances, deviate from a single linear negative straight line, revealing inner drops or outer flattenings, particularly in more massive galaxies. This study analyzes oxygen abundance gradients in spiral galaxies based on observations from the Calar Alto Legacy Integral Field Area (CALIFA) survey. Our focus is specifically on examining oxygen abundance gradient profiles, as obtained with data from H{\sc ii} regions, with a special emphasis on the inner radial gradient. We employ an automated fitting procedure to establish correlations between the physical properties of galaxies and bulges and the presence of these inner drops, seeking for potential explanations for these gradient variations.
We find that the different criteria used in the literature to distinguish H{\sc ii} regions from other ionization sources in the galaxy, such as Active Galactic Nuclei, significantly impact the results, potentially altering abundance gradient profiles and uncovering galaxies with distinct inner drops. Additionally, we analyze the abundance radial gradients to investigate the impact of diffuse ionized gas (DIG) decontamination on oxygen abundances over the featuring inner drops. We observe that DIG, concentrated mainly in the central regions of galaxies, can modify oxygen abundance gradient profiles if left unaddressed.

\end{abstract}

\keywords{Chemical abundances ---  Galaxy abundances --- Galaxy bulges --- Galaxy disks --- Galaxy chemical evolution --- H{\sc ii} regions}


\section{Introduction}

The study of the chemical evolution of spiral galaxies has become extremely important for our understanding of the formation and evolution of galaxies. The chemical abundance gradient is defined as the variation of the abundance of a certain chemical element as a function of the distance to the center of the galaxy. This measure provides information of the chemical enrichment along galactic disks. Observations and analyses of chemical abundance gradients have confirmed the presence of negative gradients across the disks of spiral galaxies, in agreement with the inside-out model proposed by \citet{laceyfall1985, matteucci1989}. According to this model, it is expected a more intense and earlier star formation in the central regions of the galaxies than in the outer ones, indicating that the central regions of the galaxy formed first, or that there was a greater amount of gas available for star formation there.


The abundance gradients in spiral galaxies can be obtained from H{\sc ii} regions, which provide information about the current chemical enrichment in the galaxy \citep[][and many others]{peimbert_1969, searle.1971, pagel_1992, martin1994, vilchez_1996, stasinska2006, izotov_2006, Sanchez2014, ho2015, berg2020, zurita2021, mendez-delgado22}. Through the observation of their nebular emission lines, they provide valuable information about the chemical composition of spiral galaxies, with oxygen serving as a proxy for gas metallicity
\citep{aller_1984, osterbroack2006}. However, the heating of the gas in the galaxy and their corresponding nebular emission lines, could be attributed to various ionization sources, such as young OB-type stars associated with star-forming regions or Active Galactic Nuclei (AGNs) \citep{kewley2006}. Additionally, other sources can also ionize the gas, generating the so-called Diffuse Ionized Gas (DIG), located in regions with different properties,  such as low densities and high electron temperatures, than the H{\sc ii} regions, but which can be erroneously classified as those ones. The sources that ionize the DIG can be hot low-mass evolved stars (HOLMES) \citep[e.g.,][]{stasinska2008, floresfajardo2011, cidfernandes2011}, shocks \citep{collins2001}, or cosmic rays \citep{reynolds2001}. 

In later-type galaxies (Sb or later), there are evidences that the DIG is concentrated in their central regions \citep[][hereafter LA18]{lacerda2018}, located within $R < 1$\,HLR\footnote{HLR corresponds to the half-light radius, defined as the length of the elliptical aperture along the major axis that contains half of the total flux at 5635\,\AA\ \citep{gonzales_delgado_2016}.}, 
where $R$ is the distance to the galaxy center. According to \citetalias{lacerda2018}, the identification of the DIG in the central regions of galaxies is interpreted as a higher proportion of old stellar populations in galactic bulges. Since AGNs can be potentially another source of ionization of the gas at the inner regions of the spiral galaxies, these regions are more likely to be contaminated by other ionization sources than star-formation \citep{Sanchez2012b}. Several models have been proposed in the literature to distinguish between these different ionization sources. In this paper, we will discuss the DIG, addressing the criterion of \citetalias{lacerda2018}, who used the equivalent width of $\mbox{H}\alpha$ ($\mbox{EW(H}\alpha)$) to analyze the DIG in CALIFA galaxies. Additionally, we will address the BPT diagram, proposed by \citet{Baldwin1981}; the WHAN diagram, developed by \citet{cidfernandes2011}; as well as the criterion for selecting H{\sc ii} regions presented by \citet{Sanchez2014}, which considers the luminosity fraction of young stars.

In the past, several authors have studied abundance gradients in the disks of the spiral galaxies and have found that the slope of the abundance gradients is related to galaxy properties, such as (i) the presence of a bar, where non-barred galaxies also show steeper gradient slopes \citep[e.g.,][]{Zaritsky1994, roy1996}; (ii) the morphology, where late-type galaxies exhibit steeper slopes \citep[e.g.,][]{McCall1985, vilacostas1992}; and (iii) the mass, where less massive galaxies show steeper gradient slopes \citep[e.g.,][]{Zaritsky1994, martin1994, garnett1998}. However, the development of integral field spectroscopic (IFS) techniques has enabled a more comprehensive study of the spatial abundances distribution in galaxies. The use of IFS in H{\sc ii} regions has allowed for the investigation of extensive samples of galaxies, providing statistically robust results, as this technique enables spectroscopic analysis in two spatial dimensions. On the other hand, the determination of abundance gradients from IFS studies have thus far relied on strong-line abundance calibrations and, therefore, have systematic uncertainties \citep[e.g.][]{kewley2008, bresolin2009, maiolino19}. 

In this regard, the CALIFA survey \citep{Sanchez2012a} was a collaboration aimed at acquiring spectroscopic information from galaxies using Integral Field Units (IFU), which enabled the collection of a wide range of data, including morphological type, color, and mass of galaxies, generating two-dimensional maps addressing these features. CALIFA data allows the analysis of ionized gas distribution, chemical abundances, excitation mechanisms, as well as information related to stellar populations, such as their ages and metallicities, and kinematic properties of both the stellar component and ionized gas \citep{Sanchez2012a}. Using IFS data, \citet{Sanchez2012b} and \citet{Sanchez2014} identify that spiral galaxies exhibit a characteristic abundance gradient when normalized by the effective radius (\re), regardless of the morphological type of these galaxies, with the gradients showing a very similar slope for all galaxies of approximately $-0.1 \ \mbox{dex}$\,\re$^{-1}$. 

However, the use of a single negative gradient to describe the oxygen abundance radial distribution of spiral galaxies is not always the best fit and there are some observational works reporting deviations from these single negative gradients, especially in more massive spiral galaxies \cite[e.g.,][]{martin.1995, vilchez1996, bresolin.2009, bresolin.2012, Sanchez-Menguiano2016, Sanchez-Menguiano2018}. In the work from \citet[][hereafter SM16]{Sanchez-Menguiano2016}, the authors identify some galaxies that exhibit a radial distribution of oxygen abundances which  become flat or, in some cases, even positive in the radial region beyond 2.0~\re, in agreement with previous studies based on individual H{\sc ii} regions \citep{martin.1995, vilchez1996, bresolin.2009, bresolin.2012}.
Several authors have proposed different mechanisms to explain the origin of these outer flattenings, such as radial motions of both gas and stars \citep{goetz_1992, ferguson_2001, sellwood_2002, minchev_2010, bilitewski_2012, roskar_2012, daniel_2015}; a radial dependence of star formation efficiency at large galactocentric distances \citep{bresolin.2012, esteban_2013}; satellite accretion and minor mergers \citep{quillen_2009, qu_2011, bird_2012}; or a balance between outflows and inflows with the intergalactic medium \citep{oppenheimer_2008, oppenheimer_2010, dave_2011, dave_2012}.
However, the nature of this flattening in the external gradient is still a matter of debate, as it is not clear whether this phenomenon is a common feature in galactic disks. Additionally, \citetalias{Sanchez-Menguiano2016} also observe a break in the abundance gradient associated with the innermost region in the most massive galaxies, at approximately $0.5$\,\re, showing an inner drop that had also been observed in previous studies \cite[e.g.,][]{belley.roy.1992, rosales.ortega.2011, Sanchez2012b}. These deviations
in the gradients in the inner regions do not appear to be related to bar effects, luminosity, or galaxy morphology. Subsequently, \citet[][hereafter SM18]{Sanchez-Menguiano2018} using data of higher spatial resolution from MUSE confirm the existence of this inner drop in the gradient at $0.5$\,\re\, for some spiral galaxies. When the presence of an inner drop is detected, a slightly steepened main gradient is observed, suggesting that radial gas motions may play a significant role in shaping chemical abundance profiles. In both works, \citetalias{Sanchez-Menguiano2016} and \citetalias{Sanchez-Menguiano2018}, no significant differences are find in the gradient slope due to the presence or not of a stellar bar. On the contrary, in a sample of 51 galaxies, both barred and unbarred, \citet{zurita2021} claim that barred galaxies have shallower gradients than unbarred galaxies for low-luminosity galaxies, but the number of low-luminosity barred galaxies in their sample (7) is low to reach a firm conclusion. Thus, given that the origin of the inner drop in the gradient is still unclear, a more thorough investigation of these internal radial gradients is necessary.

The aim of this paper is, therefore, to conduct a comprehensive and detailed analysis of the internal oxygen abundance gradient in a statistically significant sample of spiral galaxies from CALIFA, using information derived from the H{\sc ii} regions within these galaxies. By applying contemporary statistical tools, we identified potential deviations in abundance gradients automatically and without human supervision, requiring only the radial distributions and abundances of the H{\sc ii} regions of each galaxy, as well as setting a few parameters, for the tool to automatically detect these potential deviations. Then, possible connections will be established between the inner drop in the gradient and the galaxy properties, such as their total mass and bulge mass, as well as different criteria of H{\sc ii} regions selection and DIG decontamination. The study sample consisted of the largest possible number of galaxies classified as spirals (S) in the CALIFA sample, as long as they are not involved in merger or interaction processes and exhibit a suitable inclination for analysis. 


The paper is organized as follows. In Section \ref{sec:sample}, the galaxy sample, bulge parameters determination, H{\sc ii} regions classification and the diffuse ionized gas decontamination are presented. In Section \ref{sec:abundances.and.fitting}, we present the abundance determination and the gradient fitting methodology. In Section \ref{sec:results}, the obtained results are presented and in Section \ref{sec:discussion} the discussion of the results. Finally, the conclusions and outline prospects for future studies are given in Section \ref{sec:conclusions}.

\section{The galaxy sample and H{\sc ii} regions selection methods \label{sec:sample}}

The galaxies included in our sample were obtained from CALIFA DR3 \citep{Sanchez2016}. The final sample of the CALIFA survey covered two different configurations for data acquisition, V500 and V1200, encompassing the wavelength range between 3745-7300\,\AA\ and 3400-4750\,\AA, respectively. This selection was applied to galaxies with an absolute magnitude in the range of $-19 > M_{r} > -23.1$ \citep{Walcher2014}. In the analysis and generation of data cubes, only spaxels with S/N greater than 3 were considered, resulting in a final spectral resolution of approximately 6.5\,\AA\ in FWHM and an average spatial resolution of about 1\,kpc for galaxies with a redshift range between $0.005 < z < 0.03$ \citep{Sanchez2012a, Walcher2014, Sanchez2016}. Furthermore, we established specific criteria for the selection of our sample of galaxies: 

\begin{enumerate}
    \item Galaxies must be of morphological types ranging from Sa to Sm, including those with bars; 
    \item These galaxies were not involved in interaction or merging processes -- that is, must be classified as isolated (I) in the CALIFA DR3 catalog; 
    \item The inclination of the galaxies must be smaller than $65^{\circ}$ to avoid uncertainties arising from inclination effect;
    \item A minimum requirement of 10 H{\sc ii} regions per galaxy must exist to ensure a reliable analysis of the abundance gradient.
\end{enumerate}

The choice to adopt an inclination of $i < 65^{\circ}$ represents an intermediate criterion, compared to the studies of \citetalias{Sanchez-Menguiano2016} and \citetalias{Sanchez-Menguiano2018}, which employed inclinations of $i < 60^{\circ}$ and $i < 70^{\circ},$ respectively, also with the aim of avoiding the uncertainties arising from the inclination effect. The effective radii were provided by Dr. S. S{\'a}nchez (private communication) and correspond to results by \citet{Walcher2014}. Based on the conditions established above, we selected a total of 147 spiral galaxies. Detailed information about the main characteristics of these galaxies is available in Table \ref{tab:sample_selection}. 

We used the catalog of multi-component two-dimensional photometric decomposition of galaxies conducted by \citet{mendez.abreu.2017} to determine the bulge parameters. This catalog comprises a total of 404 galaxies from CALIFA and provides information on the disk, bar, central source, and bulge, such as the effective radius and surface brightness at the effective radius of the bulge. We adopted the effective radius of the bulge in the $g$-band, which coincides with the wavelength range observed by CALIFA, to determine the bulge mass, following \citet{beifiori.2012}:

\begin{equation}
M_{\rm bulge} = \frac{\alpha r_{e, {\rm bulge}} \sigma_{e, {\rm bulge}}^2}{\mbox{G}},
\label{eq:bulge_mass}
\end{equation}

\noindent where G represents the gravitational constant, $\sigma_{e, {\rm bulge}}$ is the velocity dispersion within the effective radius $r_{e, {\rm bulge}}$, and $\alpha$ is a parameter related to the S{\'e}rsic index that takes into account the galaxy structure. The parameter $\alpha$ can be obtained through a relationship derived by \citet{prugniel.1997} between $\alpha$ and the S{\'e}rsic index. To determine the velocity dispersion at the effective radius of the bulge $\sigma_{e, {\rm bulge}}$, we followed \citet{sani.2011} through the expression:
\begin{eqnarray}
\log(r_{e, {\rm bulge}}) = 1.55\log(\sigma_{e, {\rm bulge}})
-0.89\log(\langle I_{e, {\rm bulge}} \rangle) \nonumber \\
- 9.89, \, 
\end{eqnarray} 

\noindent where $\langle I_{e, {\rm bulge}} \rangle$ is the surface brightness at the effective radius of the bulge in units of solar luminosity. The conversion from $\langle I_{e, {\rm bulge}} \rangle$ from the surface brightness at the effective radius of the bulge in terms of magnitude ($\mu_{e, {\rm bulge}}$) was done \citep{binney1998} with:
\begin{equation}
\mu_e [\text{mag/\arcsec}^{2}] = 21.572 + \text{M}_{\sun,}{g} - 2.5\log(\langle I_e \rangle [\text{L}{\sun}/\text{pc}^2]),
\end{equation}

\noindent where $\text{M}_{\sun,}{g} = 5.07$  is the absolute magnitude of the Sun in the $g$-band. Thus, using $\mu_{e, {\rm bulge}}$ from \citet{mendez.abreu.2017}, we determined $\sigma_{e, {\rm bulge}}$ for each bulge in each galaxy. Finally, once determined the values of $\alpha$, together with $r_{e,{\rm bulge}}$ and $\sigma_{e,{\rm bulge}}$, we estimate the bulge mass of each galaxy from equation \ref{eq:bulge_mass}, whose results are available in Table \ref{tab:sample_selection}. It is important to note that some parameters of the bulge for 22 galaxies are not available in the catalog of \citet{mendez.abreu.2017}, making impossible to estimate the bulge mass for these galaxies. These cases are identified by the symbol ``-'' in Table \ref{tab:sample_selection}.

\begin{table*}
    \begin{center}
    \setlength\tabcolsep{0.14cm}
    \caption{Fundamental properties of the galaxies in the sample. \label{tab:sample_selection}}
    \begin{tabular}{ccclccccccrrcc}
    \hline
    \hline
       Name & RA & DEC & Morph & $\log{M_{*}}$ & $r_{e,*}$ & $\log{M_{\text{bulge}}}$ & $r_{\text{e,bulge}}$ & $\mu_{\text{bulge}}$ & $z$ & Dist & PA & $b/a$ & $i$ \\
     & (h) & ($\deg$) & type & (M$_{\sun}$) & (kpc) & (M$_{\odot}$) & (kpc) & (mag/\arcsec$^2$) & & (Mpc) & ($\deg$) & & ($\deg$) \\
    (a) & (b) & (c) & (d) & (e) & (f) & (g) & (h) & (i) & (j) & (k) & (l) & (m) & (n) \\
    \hline
	IC 0159 & 01.77 & -08.64 & SBdm & 9.8 & 5.7 & 9.2 & 0.9 & 21.1 & 0.013 & 57 & 16 & 0.78 & 40 \\
	IC 0674 & 11.19 & +43.63 & SBab & 10.9 & 10.9 & 9.7 & 2.1 & 21.4 & 0.026 & 107 & 125 & 0.65 & 50 \\
	IC 0776 & 12.32 & +08.86 & SAdm & 9.3 & 7.0 & 8.7 & 1.4 & 23.1 & 0.010 & 35 & 85 & 0.56 & 56 \\
	IC 1151 & 15.98 & +17.44 & SBcd & 9.8 & 4.7 & 8.0 & 0.3 & 21.2 & 0.009 & 31 & 39 & 0.49 & 61 \\
	IC 1256 & 17.40 & +26.49 & SABb & 10.3 & 6.0 & - & - & - & 0.018 & 66 & 90 & 0.59 & 54 \\
 \hline
    \end{tabular} 
    \end{center}
    \tablecomments{
    The complete Table \ref{tab:sample_selection} is available in Appendix \ref{ap01}. A portion is shown here for guidance regarding its form and content. The columns of the table correspond to the following identifications:  (a) galaxy name; (b) right ascension coordinate in hours; (c) declination coordinate in degrees; (d) morphological type from the Hubble classification, indicating barred galaxies (B), non-barred galaxies (A), and intermediate galaxies that may or may not have a bar (AB); (e) integrated stellar mass of the galaxy in $\log(\mbox{M}_{\sun})$; (f) effective radius of the galaxy in units of kpc; (g) mass of the bulge in $\log(\mbox{M}_{\sun})$; (h) effective radius of the bulge in units of kpc; (i) surface brightness at the effective radius of the bulge in units of mag/\arcsec$^2$; (j) redshift; (k) galaxy distance in Mpc; (l) position angle of the galaxy disk in degrees; (m) ratio between the semi-minor and semi-major axes of the galaxy; (n) inclination of the galaxy with respect to the line of sight in degrees.}
\end{table*}

\begin{figure*}
\centering
\includegraphics[width=0.66\columnwidth]{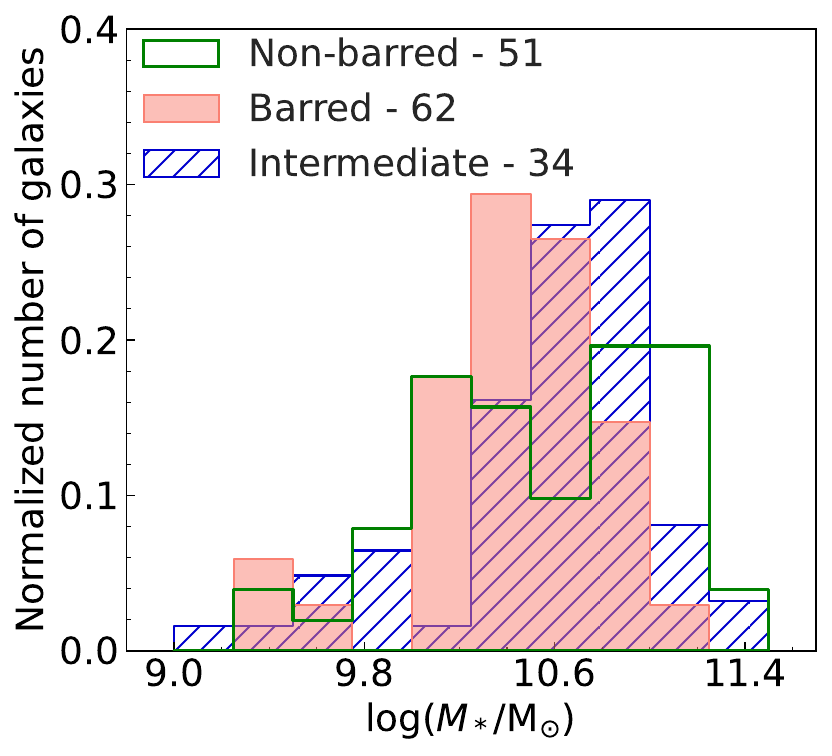}
\includegraphics[width=0.66\columnwidth]{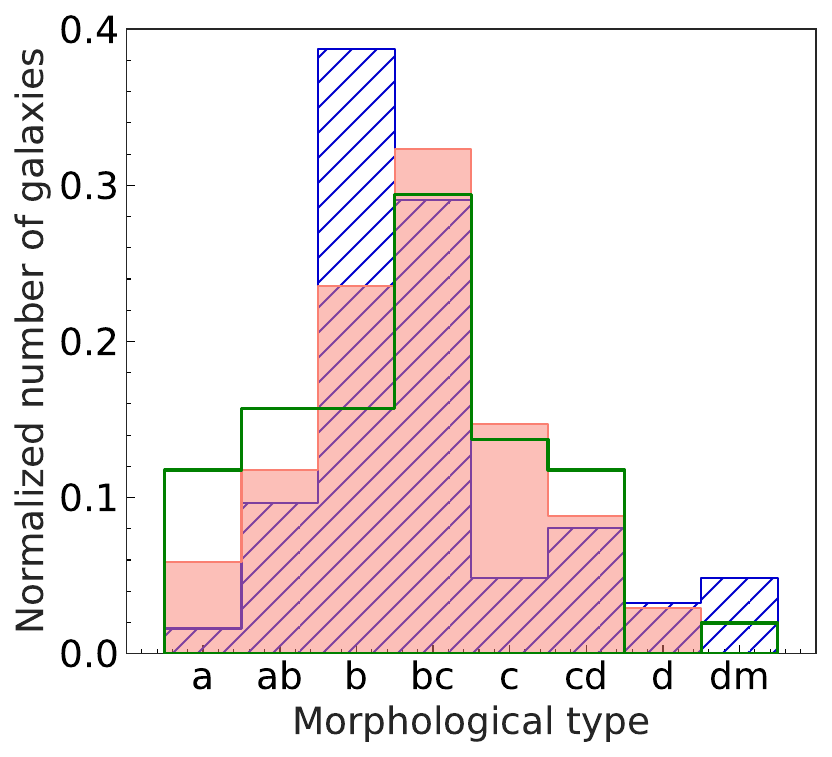}
\includegraphics[width=0.66\columnwidth]{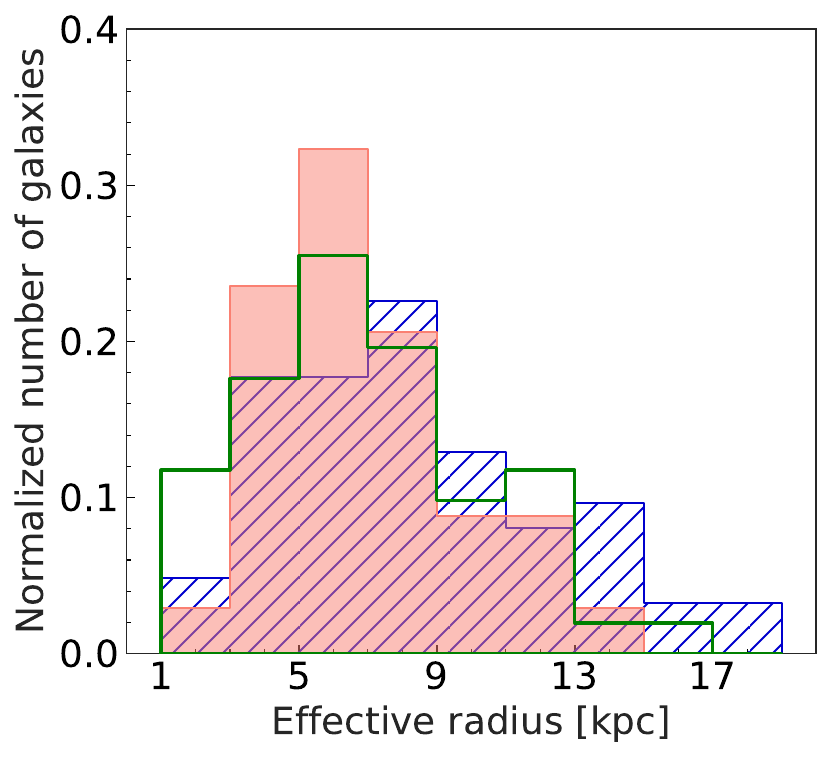}
\\
\includegraphics[width=0.66\columnwidth]{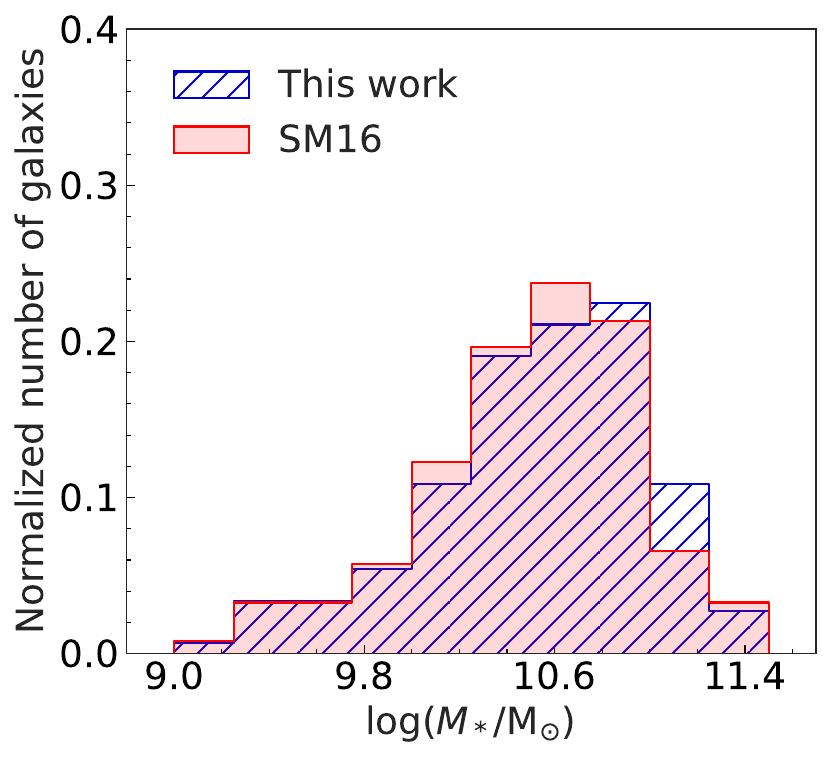}
\includegraphics[width=0.66\columnwidth]{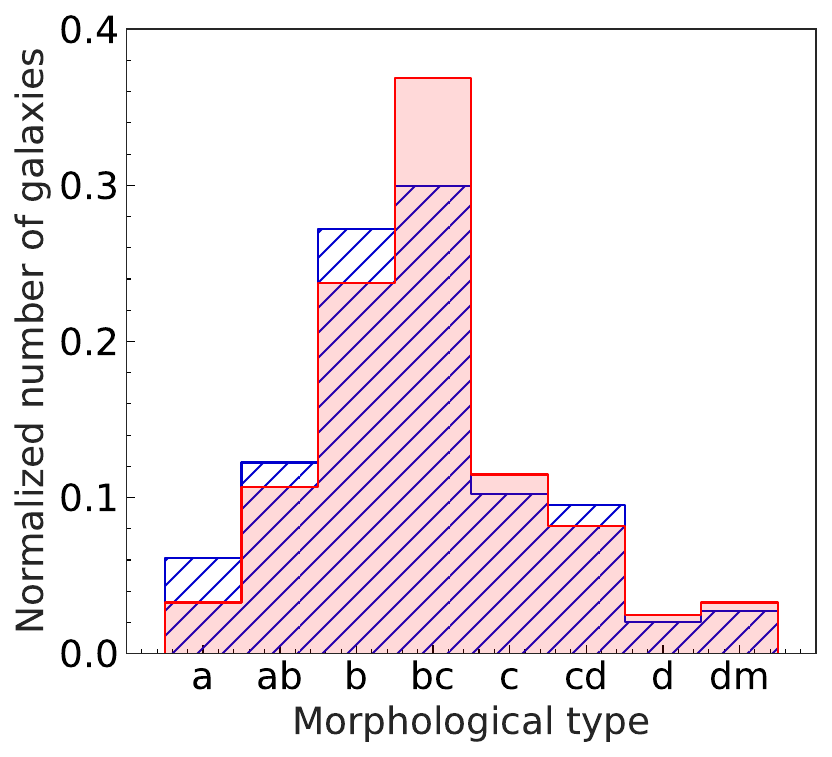}
\includegraphics[width=0.66\columnwidth]{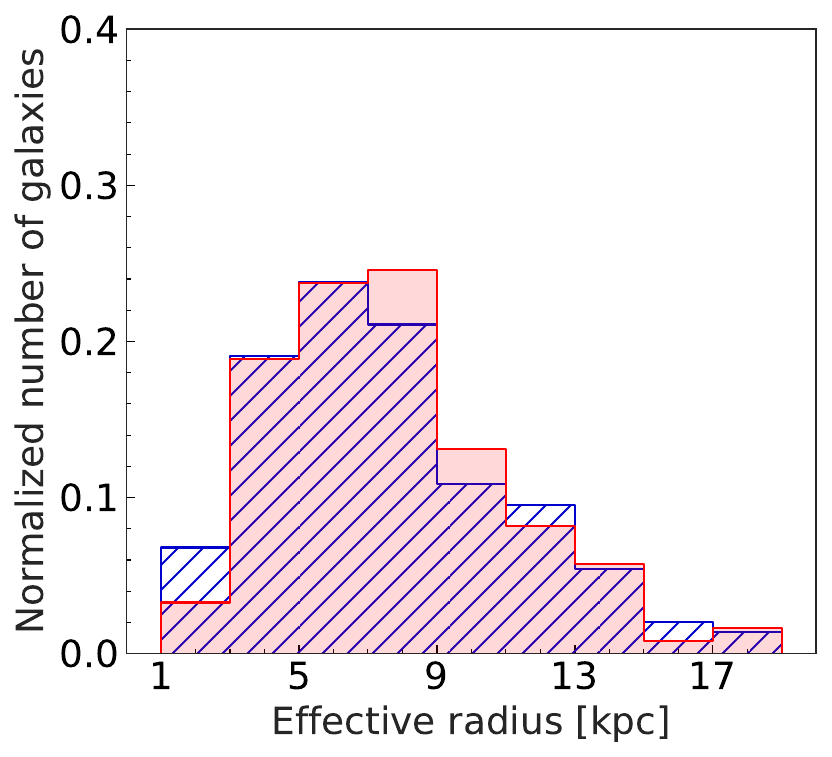}
\caption{
Distribution of properties of galaxies in our sample according to $\log(M_{*}/\mbox{M}_{\sun})$, morphological type, and effective radii $r_e$, shown in the left, central, and right panels, respectively. The morphological classification in the central panels, from ``a" to ``dm", corresponds to spiral galaxies in the Hubble diagram classified as early-type to late-type galaxies, respectively. The top panels show the distribution of the sample galaxies according to the presence or absence of a bar, according to the visual classification described in \citet{Walcher2014}. The bottom panels show a comparison between our sample and the sample from \citetalias{Sanchez-Menguiano2016}, both using CALIFA data.
}
\label{amostra.andre}
\end{figure*}

Considering the criteria described above, our sample is initially composed of galaxies with masses between $10^{9.07}$ and $10^{11.30}\,\mbox{M}_{\sun}$, and effective radii between $2.24 < r_{\rm e} < 17.77$\,kpc. The stellar mass is provided by CALIFA and was derived from the growth curve magnitudes using the methods described in \citet{walcher2008}. We have to stress that this is the largest sample employed in the study of radial gradient using IFS data so far. The top panels of Figure~\ref{amostra.andre} illustrates the distribution of the galaxies from our sample in terms of mass, morphological type, and effective radius, distinguishing between barred (B), non-barred (A), and those classified as intermediate (AB).
The B, A, and AB classifications are provided by the CALIFA survey itself and were obtained through human visual inspection, as described in \citet{Walcher2014}, where five authors classified all the galaxies in the sample as B for barred, A for non-barred, and AB if there was uncertainty.
Since \citetalias{Sanchez-Menguiano2016} also used CALIFA data in the study of oxygen abundance radial gradients in a sample of 122 galaxies, we made a comparison between some properties of the galaxies in our sample and the sample from that study, shown in bottom panels of Figure~\ref{amostra.andre}. From this figure, we can see that the current sample is quite similar to the sample from \citetalias{Sanchez-Menguiano2016}. However, we can notice that the current sample is composed by a slightly higher fraction of galaxies with large stellar masses and also a slightly higher number of early type spiral galaxies. Moreover, the galaxies from our sample tend to have slightly lower effective radii than those from \citetalias{Sanchez-Menguiano2016}.

\subsection{Criteria for Selecting {\rm H}{\sc ii} Regions}
\label{subsec:hii_regions_selection}

The CALIFA data are distributed reduced using the technique developed for IFS data reduction in the R3D pipeline \citep{sanchez2006}. Recently, \citet{Espinosaponce2020} provided a catalog containing the spectroscopic information of clumpy ionized regions detected in 988 galaxies observed with CALIFA\footnote{Available in the CALIFA project web page at 
\url{http://ifs.astroscu.unam.mx/CALIFA/HII_regions/}.}. 
For the segregation of H{\sc ii} regions, \citet{Espinosaponce2020} used {\sc pyHIIexplorer}\footnote{
\url{https://github.com/cespinosa/pyHIIexplorerV2}}, a code based on {\sc HIIexplorer}, originally written in Perl \citep{Sanchez2012b}. The {\sc pyHIIexplorer} was written in {\sc Python}, which is a language commonly used in astronomy nowadays, but it essentially performs the same steps as {\sc HIIexplorer}. The detection of ionized regions in {\sc pyHIIexplorer} is based on two initial assumptions: (i) H{\sc ii} regions are isolated structures that have strong emission lines that are clearly above the continuum emission and the average ionized gas emission across each galaxy.;
(ii) H{\sc ii} regions have typical sizes on the order of a few hundred parsecs. After this segregation of the H{\sc ii} regions, a FITS file containing the information of the ionized regions is created, making it possible to obtain the emission line fluxes, as well as information about the stellar populations. Further details of the procedures can be found in \citet{Espinosaponce2020}. 

We have made use of this catalog from \citet{Espinosaponce2020} to obtain the fluxes and $\mbox{EW(H}\alpha)$ of the potential H{\sc ii} regions from the 147 galaxies in our sample. We set limits for the fluxes of the emission lines [O{\sc iii}], H$\beta$, [N{\sc ii}], and H$\alpha$ from the ionized regions based on a Gaussian distribution, excluding those regions whose error in flux exceeded $3\sigma$ of the FWHM from the fitted Gaussian. With these criteria, we obtained a total of 11,410 ionized regions in our sample.
For the correction of interstellar extinction, we adopted the theoretical $\mbox{H}\alpha/\mbox{H}\beta$ ratio as 2.86, considering an electron density $n_{e} = 100 \ \mbox{cm}^{-3}$ and an electron temperature $T_{e} = 10000 \ \mbox{K}$, following the recombination theory \citep{Osterbrock1989}.
The correction in the fluxes were performed as usual and the determination of the E(B-V) is carried out using the extinction curve of \citet{Fitzpatrick1999} as given by the polynomial fit from \citet{cavichia2010}.
The mean value and standard deviation of the extinction E(B-V) are $0.21 \pm 0.33$, with the first, second, and third quartiles corresponding to 0.06, 0.19, and 0.33, respectively.
The physical distances of the ionized regions in galaxies were determined following \citet{Scarano2008} by adopting an intrinsic ellipticity for all galaxies of $q = 0.13$ \citep{Giovanelli1994}.

\subsubsection{Diffuse Ionized Gas (DIG)}
\label{gas.difuso.ionizado}

As explained before, DIG consists of regions with different properties than H{\sc ii} regions, being therefore important to distinguish a H{\sc ii} region from a DIG region. There are studies in the literature that make this distinction based on the surface brightness of $\mbox{H}\alpha$ \cite[e.g.,][]{zurita2000, vogt2017}. However, \citetalias{lacerda2018} argue that using surface brightness of $\mbox{H}\alpha$ is a misleading way to differentiate H{\sc ii} regions from DIG, because if we consider two DIG-dominated regions overlapping along the line of sight, they would together indicate a high surface brightness, causing these DIG-dominated regions be erroneously classified as a single H{\sc ii} region. Therefore, 
\citetalias{lacerda2018} propose an analysis based on $\mbox{EW(H}\alpha)$ using CALIFA data, defining hDIG by $\mbox{EW(H}\alpha) < 3\,\mbox{\AA}$ as regions ionized by HOLMES, mDIG by  $3\,\mbox{\AA} < \mbox{EW(H}\alpha) < 14\,\mbox{\AA}$ as regions ionized by various mixed sources, and SFc by $\mbox{EW(H}\alpha) > 14\,\mbox{\AA}$ as star-forming complexes \footnote{ Regions that contain H{\sc ii} regions, but inevitably mixed with DIG emission in our data. 
As defined by \citetalias{lacerda2018}, SFc are the zones with a larger SF/DIG ratio and are not necessarily dominated by star formation, but simply contain a good proportion of SF-powered line emission.
}.

According to \citetalias{lacerda2018}, early-type spiral galaxies show a high concentration of hDIG compared to late-type galaxies. However, they identified that in early-type galaxies the presence of hDIG is more uniformly distributed in their disks, while in later-type galaxies (Sb or later), hDIG is concentrated in the central regions located within $R < 1$ HLR, where $R$ is the distance to the galaxy center. The identification of hDIG in the central regions of galaxies is interpreted as a higher proportion of old stellar populations in galactic bulges.
It was observed that mDIG can indeed be understood as a mixture of hDIG and SFc since it is primarily located between regions classified as hDIG and SFc. 

\begin{figure}
\begin{center}
\includegraphics[width=0.8\columnwidth]{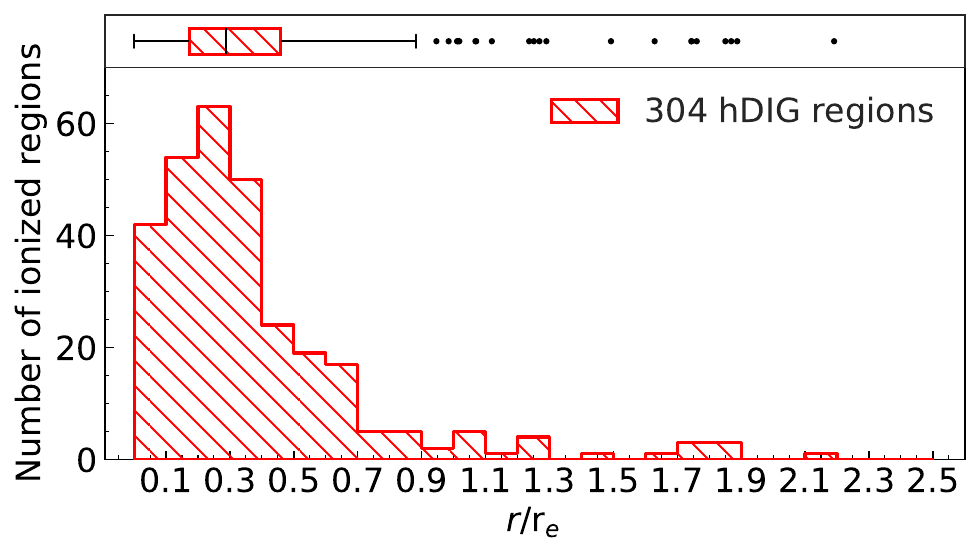}
\includegraphics[width=0.8\columnwidth]{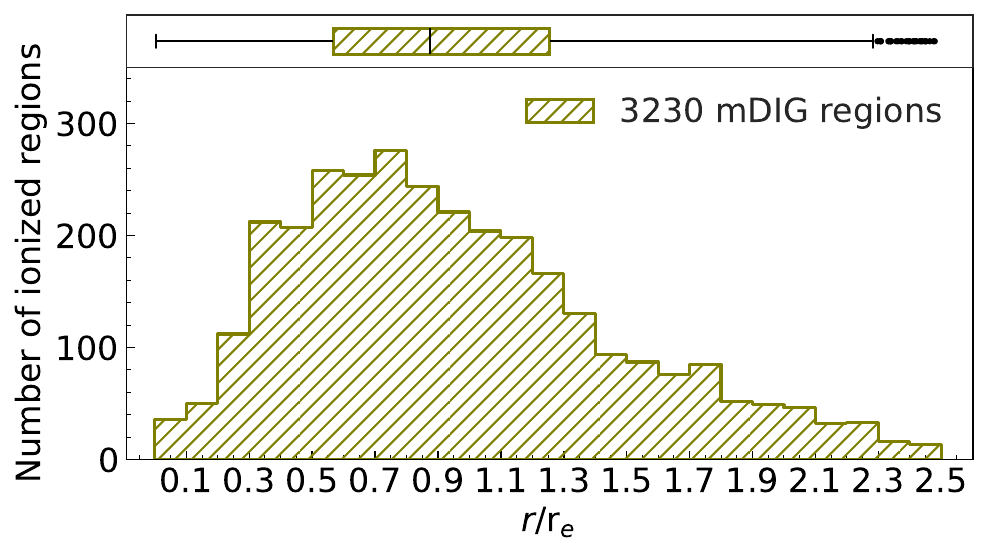}
\includegraphics[width=0.8\columnwidth]{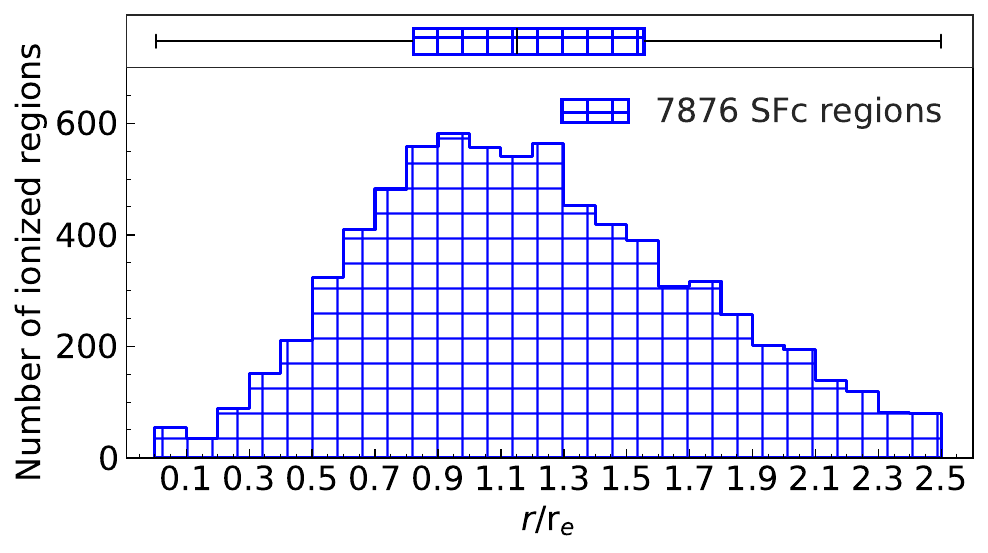}
\caption{Radial distribution of ionized regions in galaxies, separated into hDIG, mDIG, SFc, as shown in the top, middle, and bottom panels, respectively, indicates the number of ionized regions for each classification. The boxplots provide the median of each classification, as well as the points considered outliers.}
\label{dig.todas.fontes}
\end{center}
\end{figure}

We performed an analysis of the distribution of ionized regions from the sample, identified as hDIG, mDIG, and SFc, along the radial distribution of galaxies, as shown in Figure \ref{dig.todas.fontes}. We have limited the radial distribution to $r<2.5$~\re\ for a better visualization.
From Figure~\ref{dig.todas.fontes}, we observe that only a 2.7\% of ionized regions are classified as hDIG and they are predominantly concentrated in the inner regions of galaxies, while SFc tends to move away from the central region and distribute throughout the disk. The mDIG can indeed be understood as a mixture of various sources, as illustrated in Figure~\ref{dig.todas.fontes}, as this distribution is concentrated in the intersection between the peaks of hDIG and SFc.

According to \citet{belfiore.2022}, the main source of ionization of the DIG proceeds from photons leaking from the 
H{\sc ii} regions, capable of ionizing the gas across the disks of the galaxies, where they established that the mean free path of these photons is 1.9\,kpc. Since the average resolution of CALIFA is approximately 1.0\,kpc \citep{Walcher2014}, and considering that H{\sc ii} regions have sizes on the order of a hundred to several hundreds of parsecs \citep{gonzalez_delgado_1997, lopez_2011, Oey_2003}, it is not possible to spatially resolve and separate pure H{\sc ii} regions from other potential ionization sources.

To address the effect of DIG, \citet{Espinosaponce2020} develop a method to subtract DIG contamination in regions identified as H{\sc ii} regions in the CALIFA data cubes. This procedure essentially involves identifying spaxels that were not classified as H{\sc ii} regions by {\sc pyHIIEXPLORER}, calculating the average flux of these spaxels, now referred to as DIG, and subtracting this mean flux from the fluxes of the identified H{\sc ii} regions. Details are available in the work of \citet{Espinosaponce2020}.

Therefore, we can also adopt the catalog of \citet{Espinosaponce2020} with the DIG correction, as this correction is based on spaxels external to those identified by {\sc pyHIIEXPLORER} in H{\sc ii} regions. As the mean free path of the photons extends beyond these regions, it is reasonable to consider these spaxels as sources ionized by leaked photons.
The new initial sample corresponds to the same 147 galaxies, but with the decontamination of the line fluxes performed. Due to our criterion of excluding regions where the line flux errors exceeded $3\sigma$, the new sample contains a slightly smaller number of ionized regions, comprising a total of 10,974 ionized regions. The mean value and standard deviation of the extinction E(B-V) for this sample, excluding DIG contamination, is $0.24 \pm 0.43$, with the first, second, and third quartiles corresponding to 0.07, 0.21, and 0.36, respectively.

Here, we analyze both samples to investigate the effect of performing or not performing DIG decontamination. Throughout the paper, we will refer to the first sample as the ``with DIG" sample, as the line fluxes in this sample have not undergone DIG decontamination. On the other hand, we will refer to the new sample as the ``without DIG" sample, as the line fluxes in this sample have undergone DIG decontamination.


\subsubsection{BPT Diagram}
\label{section.bpt.diagram}

\begin{figure}
\centering
\includegraphics[width=\columnwidth]{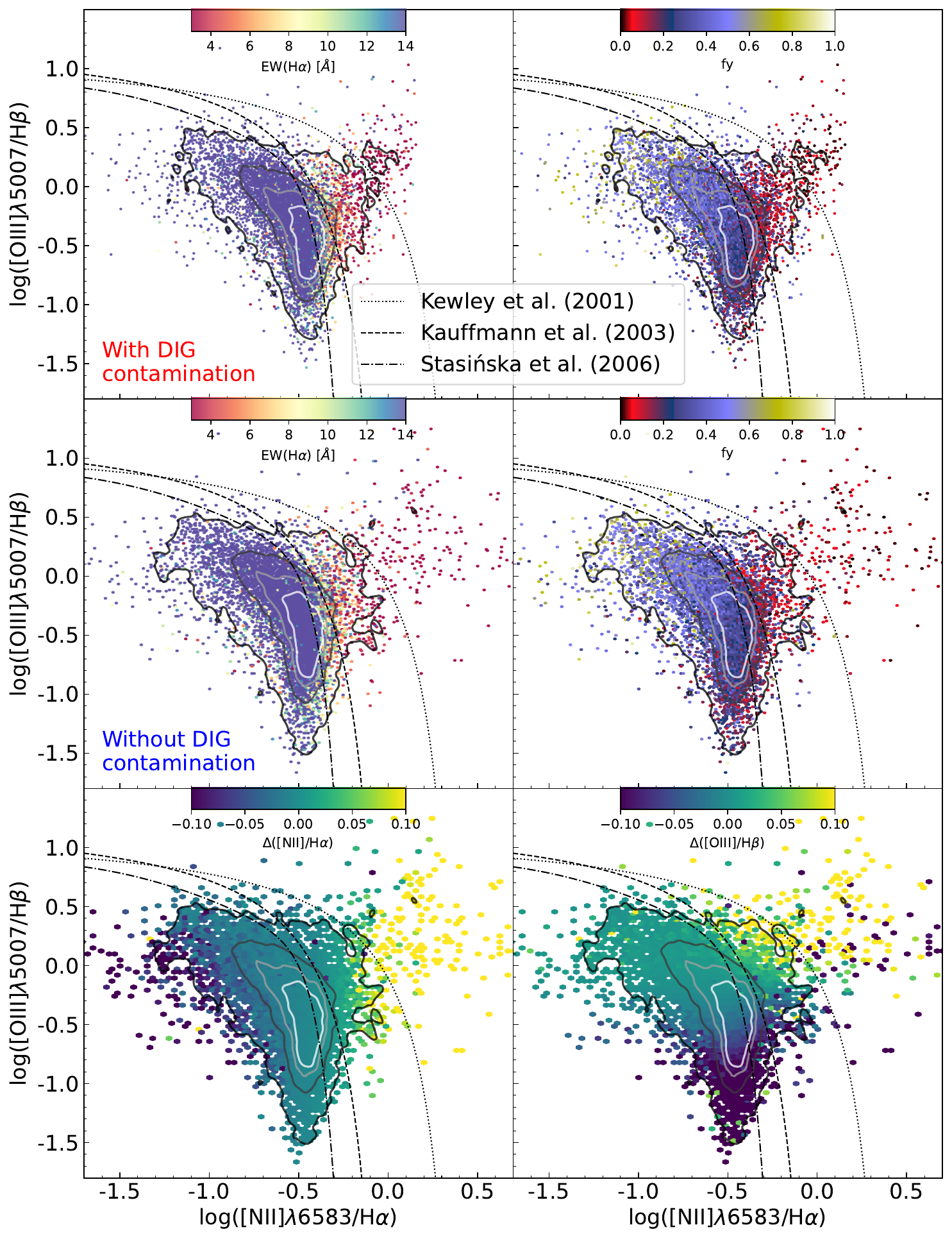}
\caption{The BPT diagrams of the data samples indicate demarcation curves in the separation between H{\sc ii} regions and AGNs as follows: dashed line: \citetalias{Kewley2001}; dash-dotted line: \citetalias{stasinska2006}; dotted line: \citetalias{Kauffmann2003}, as shown in the legend. The top panels, identified as with DIG contamination, correspond to the sample of 11,410 ionized regions. The middle panels, identified as without DIG contamination, correspond to the sample of 10,974 ionized regions. In the top and middle left panels the colors correspond to the $\mbox{EW(H}\alpha)$ proposed by \citetalias{lacerda2018}.
In the top and middle right panels the colors correspond to the fraction of young starlight proposed by \citetalias{Sanchez2014}. 
The bottom panels show the distribution of the H{\sc ii} regions in the sample without contamination and the color diagrams indicate the difference between the fluxes with and without DIG contamination for $[\mbox{\ion{N}{0II}}]/\mbox{H}\alpha$ (left) and $[\mbox{\ion{O}{0III}}]/\mbox{H}\beta$ (right). 
In all panels the contours show the density distribution of the H{\sc ii} regions in the samples, where the outermost contour encloses 95\% of the regions, and each consecutive contour encloses 75\%, 55\%, and 35\% of the points.}
\label{bpt.dados.amostra}
\end{figure}

In the BPT diagram \citep{Baldwin1981}, different curves are proposed to distinguish H{\sc ii} regions from AGNs. Among the most common ones are those from \citet[][hereafter KE01]{Kewley2001}, which uses theoretical photoionization models; from \citet[][hereafter KA03]{Kauffmann2003}, based on the analysis of integrated spectra from the Sloan Digital Sky Survey (SDSS) galaxies; and from \citet[][hereafter ST06]{stasinska2006}, using a photoionization model and a power-law-based approach. The region between the curves of \citetalias{Kewley2001} and \citetalias{Kauffmann2003} remains a subject of debate since, besides other sources of ionization, pure H{\sc ii} regions can be found in this region \citep{Kennicutt1989, Ho1997, perezmontero2009, Sanchez2014}. Therefore, adopting the curve of \citetalias{Kauffmann2003} as a criterion to distinguish H{\sc ii} regions from AGNs may exclude known H{\sc ii} regions.
However, other studies \citep{singh.2013, vogt.2014, belfiore.2015, sanchez.2015.bpt} have explored other demarcation curves in the BPT diagram in investigating ionization sources, as the exact location of the curve separating H{\sc ii} regions from AGNs 
is still a matter of controversy 
\citep{Zinchenko.2016}.
Figure~\ref{bpt.dados.amostra} illustrates the BPT diagram showing the mentioned demarcation curves with the data of our two samples, with and without DIG contamination. The details of the colored diagrams in Figure~\ref{bpt.dados.amostra} are discussed in Sections \ref{gas.difuso.ionizado} and \ref{luminosidade_estrelas_jovens}. It is noticeable in the middle panels of the figure that the ionized regions occupy a larger area of the BPT diagram compared with the top panels, even though they are less numerous.

Additionally, bottom panels of Figure~\ref{bpt.dados.amostra} also show the BPT diagrams for the sample of H{\sc ii} regions without DIG contamination, with the color code representing the difference in the line fluxes involved in this diagram for the samples with and without DIG contamination. It can be observed that the change in these line fluxes occurs mainly at the edges of the distribution along the BPT diagram. For $[\mbox{\ion{N}{0II}}]/\mbox{H}\alpha$, the greatest difference occurs in the right and left corners of the distribution, while for $[\mbox{\ion{O}{0III}}]/\mbox{H}\beta$, the difference appears in the right corner and the lower region.
Furthermore, for the ratio of the line fluxes $[\mbox{\ion{N}{0II}}]/\mbox{H}\alpha$ the difference is less than 10\% for 95\% of the sample, while for $[\mbox{\ion{O}{0III}}]/\mbox{H}\beta$ the difference is less than 10\% for 70\% of the sample. These results are in agreement with the change in the density of points between the distribution with and without DIG contamination (upper and central panels), where a slight variation in the density of points along the distribution can be observed.

\subsubsection{WHAN Diagram}

\begin{figure}
\begin{center}
\includegraphics[width=\columnwidth]{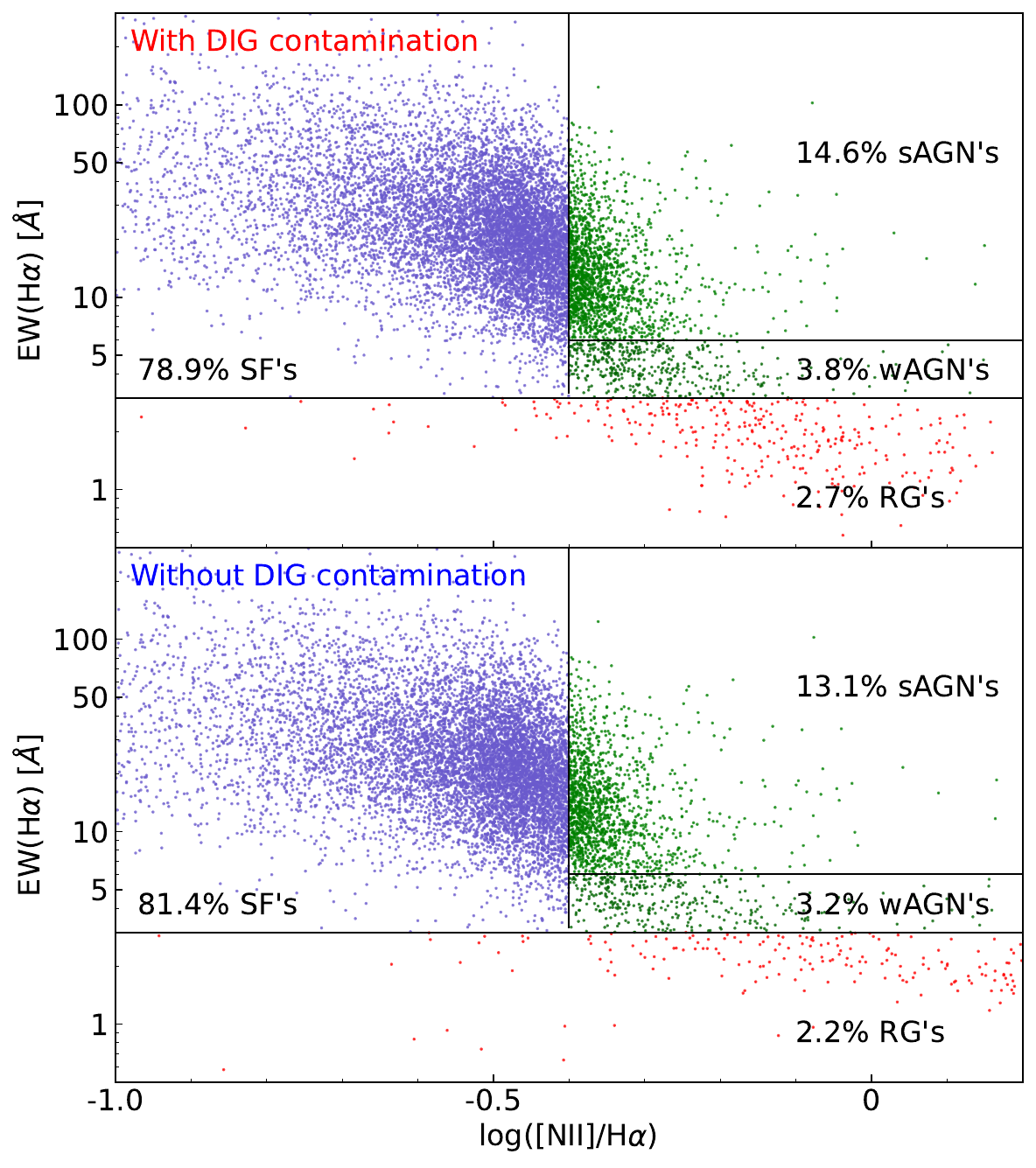}    
\end{center}
\caption{The WHAN diagram from \citetalias{cidfernandes2011} classifying the ionized regions of the samples with and without DIG contamination is shown in the upper and lower panels, respectively, as indicated in each panel. 
The percentage of regions classified as SF (star-forming), wAGN (weak AGN), sAGN (strong AGN), and RG (passive galaxy or retired galaxy) for each sample is displayed in each panel.}
\label{whan.cid.fernandes}
\end{figure}

The WHAN diagram by \citet[hereafter CF11]{cidfernandes2011}, based on the ratio of $[\mbox{\ion{N}{0II}}]/\mbox{H}\alpha$ lines and $\mbox{EW(H}\alpha)$, is another widely used diagram to separate H{\sc ii} regions from AGNs and HOLMES.
Since the diagram only involves strong emission lines, it suffers of less uncertainties than the traditional diagnostic diagrams like the BPT one, that do rely on weaker lines such as $[\mbox{\ion{O}{0III}}]$ and/or H$\beta$. 
Figure~\ref{whan.cid.fernandes} shows the WHAN diagram from \citetalias{cidfernandes2011} used for the classification of different ionization sources in our two samples. For the sample with DIG contamination, out of a total of 11,410 ionized regions, there are 304 RGs (passive galaxies or retired galaxies), 9006 SFs (star-forming), 439 wAGNs (weak AGNs), and 1661 sAGNs (strong AGNs). For the sample without DIG contamination, out of a total of 10,974 ionized regions, there are 245 RGs, 8937 SFs, 354 wAGNs, and 1438 sAGNs.
We can note that, despite the number of ionized regions being lower in the sample without DIG contamination compared to the sample with DIG contamination, proportionally the sample without DIG has a higher number of regions classified as H{\sc ii} regions, while the number of regions classified as RGs, wAGNs, and sAGNs decreases.

\subsubsection{Young Stars Luminosity}
\label{luminosidade_estrelas_jovens}

\citet[][hefereafter SA14]{Sanchez2014} introduce a different method for the identification of H{\sc ii} regions. Their main criterion is based on the fraction of total luminosity ($f_{\rm y}$) in the V band coming from young stars within an ionized region, defining young stars as those with an age of less than 500\,Myr. The luminosity fraction of the different stellar populations (with different ages and metallicities) is derived through the combination of synthetic Simple Stellar Populations (SSPs) obtained from the modeling of the continuous spectrum of the ionized regions. According to \citetalias{Sanchez2014}, a simple analysis of the equivalent width of $\mbox{H}\alpha$ is not sufficient to distinguish between different ionization sources. Figure~\ref{bpt.dados.amostra}, in the upper and middle right panels, shows the distribution of ionized regions in the samples with and without DIG contamination, respectively. The fraction of stellar luminosity is indicated with a color diagram, where points closer to the red color, with $f_{\rm y} < 0.2$, represent regions dominated by old stars, which are not considered H{\sc ii} regions according to the criterion of \citetalias{Sanchez2014}.

From Figure \ref{bpt.dados.amostra}, for the sample with DIG contamination, when we adopt the \citetalias{Sanchez2014} criterion, we obtain 8,345 ionized regions with $f_{\rm y} > 0.2$, corresponding to 73\% of ionized regions classified as H{\sc ii} regions. Of this total, 99.8\% are located below the \citetalias{Kewley2001} curve. Similarly, for the sample without DIG contamination, when we adopt the \citetalias{Sanchez2014} criterion, we obtain 8,156 ionized regions with $f_{\rm y} > 0.2$, corresponding to 74\% of ionized regions classified as H{\sc ii} regions. Of this total, 99.7\% are located below the \citetalias{Kewley2001} curve. Therefore, based on the results of \citetalias{Sanchez2014}, we can conclude that practically all H{\sc ii} regions are located below the curve proposed by \citetalias{Kewley2001}, with a contribution of young star luminosity exceeding a 20\%. 

Table \ref{tab:hii_criterion} summarizes the number statistics obtained considering or not DIG decontamination and the different selection criteria to classify the H{\sc ii} regions, showing the percentage of ionized regions classified as H{\sc ii} regions for each criteria from the total of 11,410 and 10,974 H{\sc ii} regions for the samples with DIG and without DIG, respectively.

\begin{table}
    \caption{Number of galaxies and H{\sc ii} regions per criterion.}
    \label{tab:hii_criterion}
    \begin{footnotesize}
    \begin{center}
    \begin{tabular}{@{\extracolsep{5pt}}lcccccc@{}}
    \hline
    \hline
         & \multicolumn{3}{c}{With DIG} & \multicolumn{3}{c}{Without DIG} \\
       \cline{2-4} 
       \cline{5-7}
       Criterion & $N_{\mbox{gal}}$ &  $N_{\mbox{H{\sc ii}}}$ & \% & $N_{\mbox{gal}}$ & $N_{\mbox{H{\sc ii}}}$ & \% \\
        \hline
       \citetalias{Kewley2001} & 147 & 11,252 & 99 & 145 & 10,752 & 98 \\ 
       \citetalias{Kauffmann2003} & 142 & 10,362 & 91 & 141 & 10,036 & 91 \\
       \citetalias{stasinska2006} & 130 & 8,350 & 73 & 131 & 8,364 & 76 \\
       \citetalias{cidfernandes2011}& 136 & 8,965 & 79 & 135 & 8,887 & 81 \\
       \citetalias{Sanchez2014} & 128 & 8,345 & 73 & 128 & 8,156 & 74 \\
       KE6A & 141 & 10,494 & 92 & 141 & 10,194 & 93 \\
       \hline
    \end{tabular}
    \end{center}
    \end{footnotesize}
    \tablecomments{With DIG and Without DIG mean before and after DIG decontamination. KE01: \citet{Kewley2001}, KA03: \citet{Kauffmann2003}, ST06: \citet{stasinska2006}, CF11: \citet{cidfernandes2011}, SA14: \citet{Sanchez2014}, and KE6A: \citet{Kewley2001} with $\mbox{EW(H}\alpha) > 6\,\mbox{\AA}$. }
    \label{tab:hii_selection}
\end{table}

For the samples with and without DIG contamination, considering the six criteria for selecting H{\sc ii} regions, only 3 galaxies have exactly the requested minimum number of 10 H{\sc ii} regions. However, the distribution of these H{\sc ii} regions extends across the entire galaxy, between 0.5 \re\ and 2.0 \re\ , ensuring the possibility of obtaining the gradient.

\section{Abundance gradients}
\label{sec:abundances.and.fitting}

\subsection{Abundance determinations}

The chemical abundance of ionized gas can be obtained through two main methods: 
(i) the direct method, based on auroral and nebular line ratios, enabling the determination of temperature and electron density, and consequently, the determination of chemical abundance \citep[e.g.][]{peimbert_1969, stasinska2006, pagel_1992, vilchez_1996, izotov_2006}; 
(ii) indirect method, also known as the empirical method, based on theoretical models and calibrations that use strong line ratios in determining chemical abundance. 
The direct method is the most robust way to determine chemical abundances. However, as in the case of H{\sc ii} regions in distant galaxies, these measurements of chemical abundances can be challenging, as the electron temperature decreases with increasing metallicity, making auroral emission lines difficult to measure as they become weak. Therefore, the indirect method should be employed to determine the chemical abundances of these H{\sc ii} regions \citep{pagel1979, Pettini2004}. The O3N2 index was first introduced by \citet{Alloin1979} and uses the ratios of strong emission lines $[\mbox{\ion{O}{0III}}]/\mbox{H}\beta$ and $[\mbox{\ion{N}{0II}}]/\mbox{H}\alpha$. These emission lines in the O3N2 indicator are highly sensitive to the oxygen abundance. Moreover, the advantage of using these lines is that they are close in wavelength, providing the benefit of dust extinction correction not being critical in the determination of oxygen abundance. Additionally, as discussed by \citet{vale.asari.2019} and \citet{kumari.2019}, the O3N2 index is less affected by DIG compared to other indices proposed in the literature, making it more suitable for estimating oxygen abundance in regions with low spatial resolution. For all these reasons, in this study we will adopt a calibration based on the O3N2 index to derive the oxygen abundances, in particular, the one presented in \citet{Pettini2004}. We are aware that a more recent calibrator is proposed by \citet{Marino2013}, however, as discussed in \citet{zurita2021}, the abundances obtained with the calibrator from \citet{Marino2013} are restricted to a limited range, smaller than the one obtained from the Te-based method for the same regions. This fact motivates the use in this study of the PP04 calibrator instead of the more recent one by \citet{Marino2013} for the O3N2 index. 

\subsection{Gradient fitting methodology}
\label{sec:fit.piecewise}

In some cases, as already explained before, the oxygen abundance radial distributions in spiral galaxies do not strictly follow a straight line and can exhibit variations, including breaks in the main gradient \citep{belley.roy.1992, rosales.ortega.2011, Sanchez2012b}. As reported in the studies of \citetalias{Sanchez-Menguiano2016} and \citetalias{Sanchez-Menguiano2018}, these radial distributions can take different profiles, such as a single linear gradient, a broken gradient with an inner drop, a broken gradient with an outer flattening, and a doubly broken gradient with an inner drop and an outer flattening.

In the study by \citetalias{Sanchez-Menguiano2016}, the radial abundance distribution of a large sample of spiral galaxies is investigated, identifying the presence of these deviations from single gradients simply by eye. However, in the study by \citetalias{Sanchez-Menguiano2018}, an automatic procedure without human supervision is implemented, using functions capable of adjusting the inner drops, similarly to this work, where the selection of the best fit was performed by analyzing the residuals using Monte Carlo simulations and fitting Gaussian distributions of these residuals. Details of this procedure can be found in \citetalias{Sanchez-Menguiano2018}.

\begin{figure}
\begin{center}
\includegraphics[width=0.9\columnwidth]{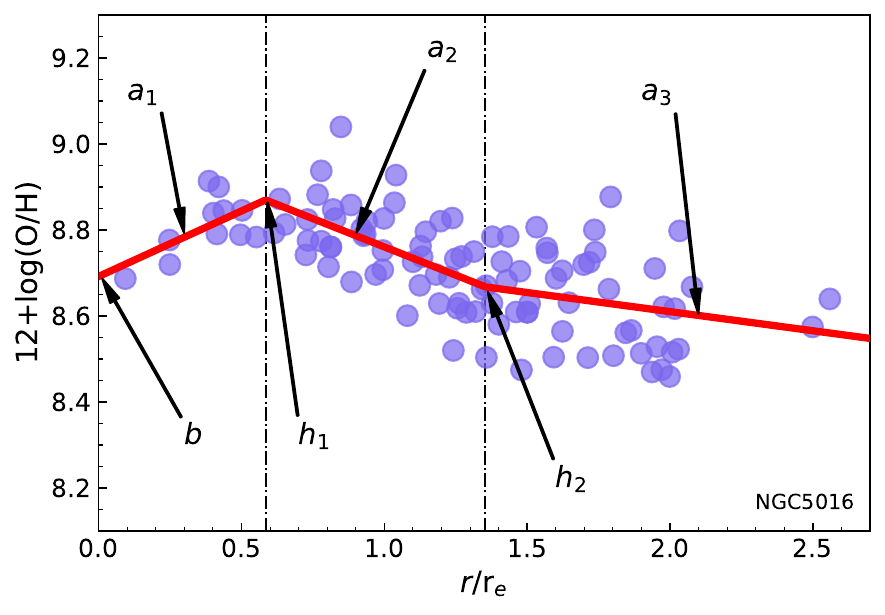}
\caption{Example of a fit performed using the {\sl piecewise regression} from \citet{Pilgrim2021}, illustrating the free parameters of the fit expressed in Equations \ref{fit_1}, \ref{fit_2} and \ref{fit_3}. The blue dots represent the selected H{\sc ii} regions according to the \citet{Kauffmann2003} criterion, with correction for DIG contamination in the galaxy NGC~5016. The solid red line corresponds to the fit of the oxygen abundance radial distribution. The dashed-dotted lines indicate the positions where the inner drop and the outer flattening occur, respectively.} 
\label{exemplo.ajuste}
\end{center}
\end{figure}

Here we will address this problem of the gradient fitting using a more robust unsupervised automatic fitting procedure that employs a bootstrap process \citep{andrea.bootstrap} on the data to escape local minima. In other words, when identifying a breakpoint in the gradient, the algorithm continues processing the data to find another possible breakpoint, where the gradient better fits the data, resulting in the lowest residual sum of squares (RSS).
Thus, the data dispersion may influence the determination of a higher or lower RSS value relative to the fitted profile.
To perform the adjustments of the abundance radial gradient, we used the {\sc Python} package {\sc piecewise regression}\footnote{\url{https://piecewise-regression.readthedocs.io}} developed by \citet{Pilgrim2021}, which implements the iterative algorithm from \citet{muggeo2003} to find breakpoints in the radial distributions. This method simultaneously fits breakpoint positions and the linear models for the different fit segments, and it gives confidence intervals for all the model estimates. The expressions (see below) used are similar to those adopted by \citetalias{Sanchez-Menguiano2018}, and correspond to the functions that fit the gradients for the case of a single linear profile, for the case where the profile has only one breakpoint, either an inner drop or an outer flattening, and for the case where there can be up to two breakpoints, one inner drop and one outer flattening, respectively:

\begin{equation}
    12 + \log(\mbox{O}/\mbox{H}) = a_2r + b,
    \label{fit_1}
\end{equation}
\begin{equation}
    12 + \log(\mbox{O}/\mbox{H}) = a_1r + b + (a_2 - a_1)(r - h_1)\textbf{H}(r - h_1),
    \label{fit_2}
\end{equation}
\begin{eqnarray}
    \nonumber 12 + \log(\mbox{O}/\mbox{H}) &=& a_1r + b + (a_2 - a_1)(r - h_1)\textbf{H}(r - h_1) \\
    &&+ (a_3 - a_2)(r - h_2)\textbf{H}(r - h_2),
\label{fit_3}
\end{eqnarray}

\noindent where $a_1, a_2,$ and $a_3$ are the coefficients of the internal, main, and external radial gradients, respectively; $b$ is the linear coefficient; $h_1$ and $h_2$ are the positions where the inner drop and the outer flattening occur, respectively; and \textbf{H} is the Heaviside function that ensures the change in the gradient slope in the fit. An example of the fit coefficients is shown in Figure~\ref{exemplo.ajuste} for the galaxy NGC~5016.

For each galaxy, each of the three fits was performed and to ensure the attainment of the global minimum that best fits the data, a two thousand bootstrapped resample of the data was implemented, to ensure reliability in the bootstrap confidence limits \citep{davison_1997}.
To determine which of the three fits better captures the data, it is adopted the Akaike Information Criterion \citep[AIC,][]{akaike.test}. The AIC, widely used in statistical inference and discussed in \citet{krishak2020a}, is recommended for cases where the model must fit a larger number of free parameters, making it suitable for our sample. The AIC is based on a probability function, thus providing the probability that a particular model fits the data better than other one according to the model simplicity. In other words, the AIC returns a parameter that compares various fitted models according to their free parameters, and the model with the lowest AIC value has the highest probability of better describing the fitted data. The AIC is calculated as follows:
\begin{equation}
    \mbox{AIC} = 2k - 2\ln(\mathcal{L}),
\end{equation}
where $k$ is the number of model parameters and $\ln(\mathcal{L})$ is the maximized log-likelihood of the model, given by:
\begin{equation}
    \ln(\mathcal{L}) = -\frac{n}{2}\ln(2\pi) -\frac{n}{2} \ln(\hat{\sigma }^{2}) - \frac{\mathrm{RSS}}{2\hat{\sigma }^{2}},
\end{equation}
\noindent where \textit{n} is the number of observed data points, $\hat{\sigma }^{2}=\mathrm {RSS} /n$ is the variance of the model residual distribution and 
$\mbox{RSS} = \sum_{i=1}^{n} \left(y_i - \hat{y}_i\right)^2$ is the residual sum of squares, that is, the sum of squares of the observed data minus the predicted data. Furthermore, according to \citet{akaike}, in cases where the ratio of the number of observed data points to the number of free parameters is less than 40, a correction in determining the AIC is necessary. Therefore, when necessary, this correction was applied, defining the corrected AIC as:
\begin{equation}
    \mbox{AIC}_{c} = \mbox{AIC} + \frac{2k(k+1)}{n-k-1}.
\end{equation}

In the case of a radial distribution with only one break, as suggested by \citetalias{Sanchez-Menguiano2018}, we considered an inner drop when the position of this break was limited to the first half of the radial distribution and the slope of the internal gradient was less negative than the slope of the main negative gradient. The other situation in which a break in the gradient occurs was considered as an outer flattening.

\section{Results}
\label{sec:results}

In Section \ref{sec:hii_region_selection_grad_fit}, we compare results obtained with different approaches for selecting H{\sc ii} regions, also analyzing how contamination by DIG may affect these results for the oxygen abundance gradient fit. Next, Section \ref{sec:inner_drop_physical_properties} present the correlations between the physical parameters of galaxies that present an inner drop.

\subsection{The Influence of the DIG and the Selection Criteria of {\rm H} {\sc ii} Regions}
\label{sec:hii_region_selection_grad_fit}

As discussed in Section \ref{subsec:hii_regions_selection}, several methods are available for selecting star-forming regions in galaxies. We propose to investigate the implications of these criteria and the DIG contamination on the fits of oxygen abundance gradients, as the number of H{\sc ii} regions in the same galaxy can vary significantly depending on the used method in the selection process. Thus, considering the fitting methodology presented in Section \ref{sec:fit.piecewise}, the fits for the oxygen abundance gradients were performed for all criteria presented in Section \ref{subsec:hii_regions_selection}.

\begin{table}[!ht]
\setlength{\tabcolsep}{0pt}
\begin{center}
    \caption{The occurrence of the inner drops.}
    \label{tab:ocurrence_inner_drop}
    \scriptsize
    \begin{tabular}{@{\extracolsep{3pt}}lcc@{\hspace{3pt}}cc@{\hspace{3pt}}cc@{\hspace{3pt}}cc@{\hspace{3pt}}cc@{\hspace{3pt}}cc@{\hspace{3pt}}cc@{}}
    \hline
    \hline
       Galaxy    & \multicolumn{12}{c}{\ion{H}{0II} region selection criteria} & \multicolumn{2}{c}{\% inner} \\
    \cline{2-13}       
     &  \multicolumn{2}{c}{SA14} & \multicolumn{2}{c}{CF11} & \multicolumn{2}{c}{KA03} & \multicolumn{2}{c}{KE01} & \multicolumn{2}{c}{KE6A} & \multicolumn{2}{c}{ST06} & \multicolumn{2}{c}{drop}\\    
    \cline{2-3}
    \cline{4-5}
    \cline{6-7}
    \cline{8-9}
    \cline{10-11}
    \cline{12-13}
    \cline{14-15}
        &C&D&C&D&C&D&C&D&C&D&C&D&C&D\\
\hline   

NGC 7653 & y& y& y& y& y& y& y& y& y& y& y& y& 100.0& 100.0\\ 
NGC 4047 & y& y& y& y& y& y& y& y& y& y& y& y& 100.0& 100.0\\ 
NGC 5406 & y& n& y& y& y& y& y& y& y& y& y& y& 100.0& 83.3\\ 
NGC 0309 & y& n& y& y& y& y& y& y& y& y& y& y& 100.0& 83.3\\ 
NGC 4210 & y& y& y& n& y& y& y& y& y& y& y& y& 100.0& 83.3\\
NGC 5016 & n& n& y& n& y& y& y& y& y& y& y& n& 83.3& 50.0\\ 
NGC 3614 & n& n& y& y& y& n& y& n& y& n& y& y& 83.3& 33.3\\ 
IC 1256 & n& n& y& y& y& y& y& y& y& y& n& y& 66.7& 83.3\\ 
NGC 5378 & n& n& n& n& y& n& y& n& y& n& y& y& 66.7& 16.7\\ 
NGC 0776 & n& y& n& n& y& y& y& y& y& y& n& n& 50.0& 66.7\\ 
NGC 2347 & n& n& n& n& y& y& y& y& y& y& n& y& 50.0& 66.7\\ 
NGC 5533 & y& y& n& y& y& y& n& n& n& n& y& n& 50.0& 50.0\\ 
NGC 6004 & n& n& y& y& n& y& y& n& y& y& n& n& 50.0& 50.0\\ 
NGC 5720 & n& n& n& n& y& y& y& y& n& n& y& n& 50.0& 33.3\\ 
NGC 4185 & n& n& y& y& y& y& n& y& n& y& n& y& 33.3& 83.3\\ 
NGC 5056 & n& n& n& n& n& n& y& y& y& y& n& n& 33.3& 33.3\\ 
NGC 6941 & n& n& n& n& y& y& n& n& n& n& y& y& 33.3& 33.3\\ 
NGC 7782 & n& n& y& n& n& n& y& y& n& n& n& n& 33.3& 16.7\\ 
NGC 0214 & n& --& n& --& y& --& n& --& n& --& y& --& 33.3& --\\ 
NGC 0257 & n& --& y& --& n& --& n& --& n& --& y& --& 33.3& --\\ 
NGC 7631 & n& n& n& n& y& y& y& n& n& n& n& n& 33.3& 16.7\\ 
NGC 5957 & n& n& n& y& y& y& n& y& n& y& n& n& 16.7& 66.7\\ 
NGC 1667 & n& n& n& y& n& y& n& n& n& n& y& y& 16.7& 50.0\\ 
NGC 5267 & y& y& n& y& n& n& n& y& n& n& n& n& 16.7& 50.0\\ 
NGC 3687 & n& y& n& n& n& n& y& y& n& n& n& n& 16.7& 33.3\\ 
NGC 4644 & n& n& y& y& n& n& n& y& n& n& n& n& 16.7& 33.3\\ 
NGC 2805 & y& y& n& n& n& n& n& n& n& n& n& n& 16.7& 16.7\\ 
NGC 2916 & n& n& n& n& n& n& y& y& n& n& n& n& 16.7& 16.7\\ 
NGC 7819 & n& --& y& --& n& --& n& --& n& --& n& --& 16.7& --\\ 
NGC 5735 & n& n& n& n& n& n& y& y& n& n& n& n& 16.7& 16.7\\ 
UGC 04195 & n& n& n& n& n& n& y& y& n& n& n& n& 16.7& 16.7\\ 
MCG-01... & y& --& n& --& n& --& n& --& n& --& n& --& 16.7& --\\ 
NGC 0036 & n& --& n& --& n& --& y& --& n& --& n& --& 16.7& --\\ 
NGC 3811 & n& --& n& --& n& --& y& --& n& --& n& --& 16.7& --\\ 
NGC 5622 & y& --& n& --& n& --& n& --& n& --& n& --& 16.7& --\\ 
NGC 5157 & n& --& n& --& n& --& y& --& n& --& n& --& 16.7& --\\ 
NGC 5656 & n& --& n& --& n& --& y& --& n& --& n& --& 16.7& --\\ 
NGC 7787 & --& n& --& y& --& n& --& n& --& n& --& n& --& 16.7\\ 
NGC 7716 & --& n& --& n& --& n& --& y& --& n& --& n& --& 16.7\\ 
NGC 5376 & --& n& --& n& --& n& --& y& --& n& --& n& --& 16.7\\ 
NGC 5205 & --& y& --& n& --& n& --& n& --& n& --& n& --& 16.7\\
\hline
$N_{\mbox{gal}}$ &10&9&14&14&18&17&24&22&13&13&14&12& 37  & 32  \\
$\%_{\mbox{gal}}$ &8&7&10&10&13&12&16&15&9&9&11&9& 25 & 22  \\ 
\hline
\end{tabular}
\end{center}
\tablecomments{
Occurrence of the inner drops for different \ion{H}{0II} regions selection criteria for the DIG contaminated sample (C, left columns) and for the DIG decontaminated sample (D, right columns).}

\end{table}

Table \ref{tab:ocurrence_inner_drop} summarizes the results of this analysis, indicating the frequency of the inner drop for a given galaxy and adopted H{\sc ii} region selection criterion. For each H{\sc ii} region selection criterion, we show in the left the results with DIG contamination (C) and in the right without DIG contamination (D).
The column ``Galaxy'' corresponds to the name of the galaxy. 
The column ``\% inner drop'' corresponds to the frequency with which a particular galaxy has an inner drop, considering all the six H{\sc ii} region selection criteria, where (C) and (D) also correspond to the samples with and without DIG, respectively. For instance, a galaxy that has an inner drop detected in all the H{\sc ii} region selection criteria after DIG decontamination is marked as 100\% in the column ``D'' inside the ``\% inner drop'' column. The other columns are identified according to the H{\sc ii} region selection criterion discussed in Section \ref{subsec:hii_regions_selection}, where ``y" indicates the presence of an inner drop, and ``n" indicates that there is no inner drop in that criterion. The empty cells in the table, identified by ``--'', indicate that there was no occurrence of an inner drop in all of the H{\sc ii} region selection criteria, considering  the DIG contaminated and decontaminated samples separately. The rows ``$N_{\mbox{gal}}$'' and ``$\%_{\mbox{gal}}$'' correspond to the total number of galaxies with an inner drop and the percentage relative to the number of galaxies in each sample. 

From the initial sample of 147 galaxies with DIG, 37 galaxies showed an inner drop in at least one of the H{\sc ii} region selection criteria, corresponding to 25\% of the sample. On the other hand, for the sample of 147 galaxies without DIG, 32 showed an inner drop in at least one of the criteria, corresponding to 22\% of the sample. These numbers are given in the two last rows of the column ``\% inner drop'' in Table \ref{tab:ocurrence_inner_drop}.
For each H II region selection criterion, the number of galaxies that exhibit an inner drop is also given in the last two rows of Table \ref{tab:ocurrence_inner_drop}, for both samples.

Considering the sample with DIG contamination, we can see from Table \ref{tab:ocurrence_inner_drop} that only 5 galaxies (NGC~7653, NGC~4047, NGC~5406, NGC~0309, NGC~4210) exhibit an inner drop in all H{\sc ii} region selection criteria, corresponding to only 3.4\% of the total initial, with only 14 galaxies showing an inner drop in at least three methods (9.5\%). It is observed that using the criterion of \citetalias{Kewley2001} results in a higher number of galaxies with an inner drop, while using the criterion of \citetalias{Sanchez2014} leads to a lower number of galaxies with an inner drop. This is because this last method excludes more H{\sc ii} regions in the center of the galaxy compared to other methods and, as there are fewer identified H{\sc ii} regions in the inner regions, there is consequently a lower number of galaxies exhibiting the inner drops.  For the sample without DIG contamination, only 2 galaxies (NGC~7653, NGC~4047) exhibited an inner drop in all H{\sc ii} region selection criteria, corresponding to 1.4\% of the total sample, with only 15 galaxies showing an inner drop in at least three methods (10.2\%). Once again, the \citetalias{Kewley2001} criterion presents the highest number of galaxies with an inner drop, while the \citetalias{Sanchez2014} criterion presents the lowest number.
Moreover, it is evident that, although the difference is small, the sample without DIG has a larger number of galaxies showing an inner drop in at least three of the HII region selection criteria.

\begin{figure*}
\figurenum{6}
\plotone{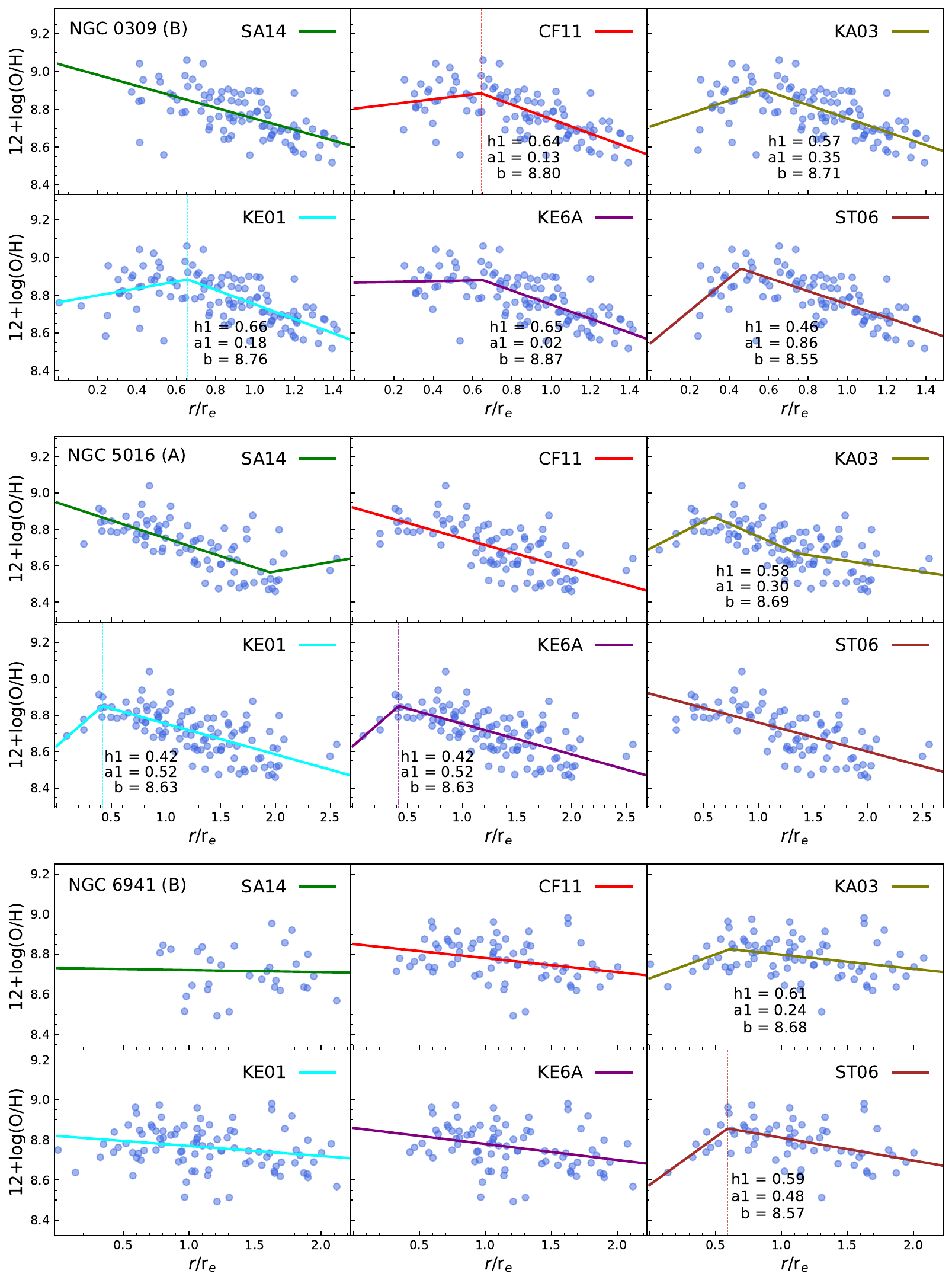}
\caption{Profiles of abundance gradients without DIG contamination. In each panel in the graphs, the blue points correspond to the H{\sc ii} regions selected by each criterion. The colored solid curve corresponds to the fit of that specific criterion, indicated in the upper right corner of each panel. The dashed line, when present, indicates the position of the inner drop, with the color corresponding to the legend. The gray dashed vertical line, when present, represents the position where the external flattening occurs. The name of each galaxy is presented in the upper left corner of each panel followed by the classification of barred (B), unbarred (A) or intermediate (AB). The complete figure set (11 images) is available in Appendix \ref{ap02}.}
\label{fig:profiles_grad_with_dig}
\end{figure*}


The results presented above show that the radial profile of oxygen abundances may vary depending on the adopted criterion to select H{\sc ii} regions, regardless of whether the sample is contaminated by DIG or not. In Figure \ref{fig:profiles_grad_with_dig}, the fittings for each criterion in three galaxies from the DIG decontaminated sample are presented to illustrate this changing effect on the gradient profile. In the Online Supporting Material (available in the online journal) the fits of the abundance gradients for all H{\sc ii} region selection criteria are shown for the 32 galaxies that present an inner drop in at least one criterion. The galaxy NGC~0309 did not exhibit an inner drop in the \citetalias{Sanchez2014} criterion but it does in the other five criteria. When analyzing the position of the inner drop in these last cases, we see that for the \citetalias{cidfernandes2011}, \citetalias{Kewley2001}, and KE6A criteria, the positions are practically the same. 
However, it is possible to notice that the inclinations of the internal radial gradients in these criteria vary considerably. In the KE6A criterion, this gradient is practically flat, while in the \citetalias{Kewley2001} criterion, it is steeper. For the \citetalias{Kauffmann2003} and \citetalias{stasinska2006} criteria, the positions of the inner drops are closer to the center of the galaxy than in the other criteria, also presenting steeper internal radial gradients, indicating lower oxygen abundances in the center of the galaxy. 

In the case of the galaxy NGC~5016, the presence of the inner drop was identified in the \citetalias{Kauffmann2003}, \citetalias{Kewley2001}, and KE6A criteria, with the \citetalias{Kewley2001} and KE6A criteria showing the same values for both the positions and inclinations of the radial gradients. However, for the \citetalias{Kauffmann2003} criterion, the position of the inner drop is different, occurring farther from the center of the galaxy, and it also presents a less steep internal radial gradient, indicating higher abundance in the center of the galaxy. For the galaxy NGC~6941, an inner drop was identified only with the \citetalias{Kauffmann2003} and \citetalias{stasinska2006} criteria, with the position being quite similar for both of them. However, the internal radial gradient is steeper in the \citetalias{stasinska2006} criterion, indicating a lower oxygen abundance in the inner region of the galaxy. In the case of \citetalias{Kewley2001}, the dispersion of the data around the linear fit is higher and the inner drop was not fitted. These results show that both the positions of the inner drop and the profiles of the radial gradients can undergo changes depending on the adopted criteria for H{\sc ii} region selection.

\begin{figure}
\figurenum{7}
\centering
\includegraphics[width=\columnwidth]{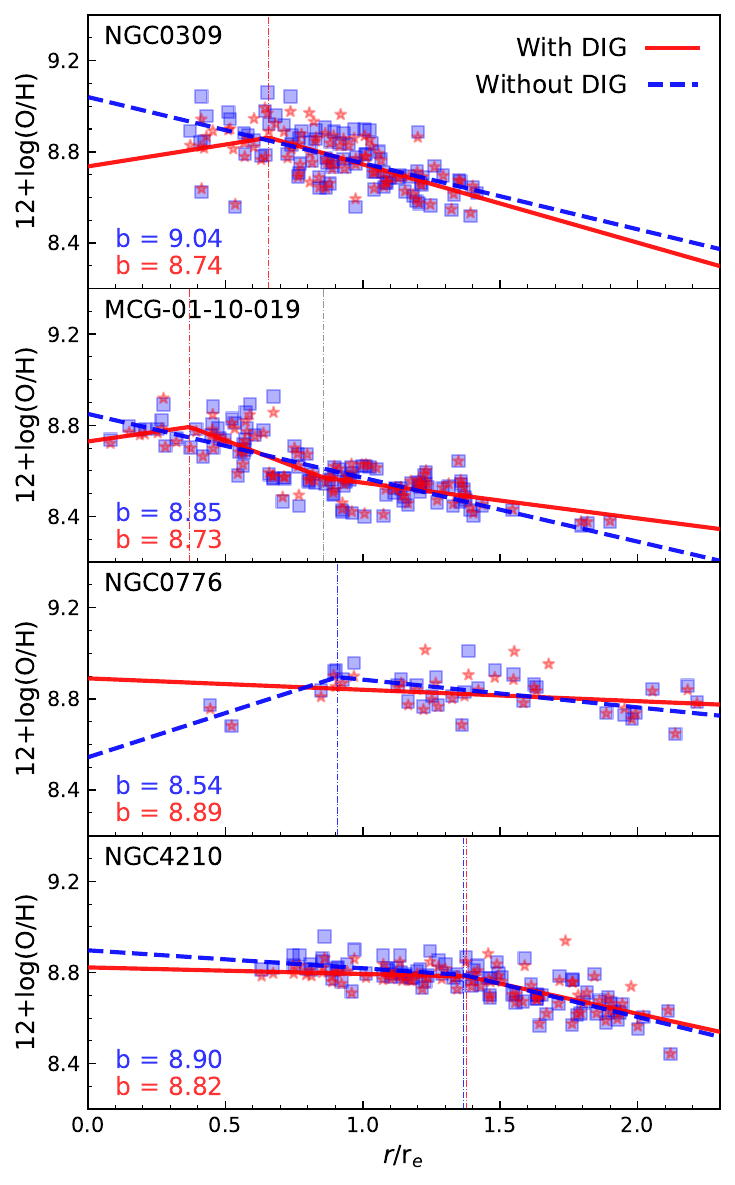}
\caption{Profiles of oxygen gradients of four galaxies from the sample with and without DIG contamination, represented by red stars and blue squares, respectively. The H{\sc ii} regions were selected according to \citetalias{Sanchez2014} criterion. The solid red and dashed blue lines correspond to the fit of the gradient profile for the oxygen abundances derived with and without DIG contamination, respectively. The dash-dotted blue and red vertical lines correspond to the positions of the inner drops. The gray dash-dotted vertical line corresponds to the position of the outer flattening. The abundances at $r=0$, represented by the value \textit{b}, are shown in the bottom left corner of each panel.}
\label{comp.dig.corrected.inner.drop}
\end{figure}

In addition to the method of selecting H{\sc ii} regions influencing the abundance gradient profiles, it is also evident from Table~\ref{tab:ocurrence_inner_drop} that DIG contamination also influences the profiles, as the number of galaxies present an inner drop varies before and after DIG decontamination, as reported in the last two rows of the table. We can notice that some galaxies that exhibited an inner drop in the sample with DIG contamination no longer show this drop, and vice versa. To assess this effect, we analyzed the gradient profiles of four galaxies, considering the \citetalias{Sanchez2014} method for selecting H{\sc ii} regions
just to highlight this effect, 
as shown in Figure~\ref{comp.dig.corrected.inner.drop}. We can observe that performing DIG decontamination leads to higher oxygen abundances, especially in the inner regions of the galaxies, within 1~\re, which are the regions where an inner drop tends to occur. Furthermore, for some galaxies that previously showed an inner drop before DIG decontamination, they no longer exhibit it, as is the case with NGC~0309 and MCG-01-10-019. We can also observe that NGC~0776 was a galaxy that did not show an inner drop, but after DIG decontamination, it presented an inner drop in the gradient.
These effects occur due to the fitting methodology, which also considers the dispersion of data across the entire radial distribution of the galaxy. That is, this slight variation in abundances, although visually minimal, directly impacts the calculation of the RSS of the fitted gradient and, consequently, affects the AIC values of the models. This causes changes in the gradient profiles, altering the presence or absence of the inner drop. In the case of NGC 0776, it seems that the dispersion is slightly lower and the inner drop is fitted in this case. We should keep in mind that this prevents fitting false drops in the radial distribution.

Regarding NGC~4210, we see that regardless of DIG decontamination, this galaxy showed an inner drop; however, the profile of the inner radial gradient was altered. This effect on the alteration of the inner radial gradient was visually verified in almost all galaxies of the sample and also occurred for all criteria of selecting H{\sc ii} regions.

\begin{figure}
\figurenum{8}
\includegraphics[width=\columnwidth]{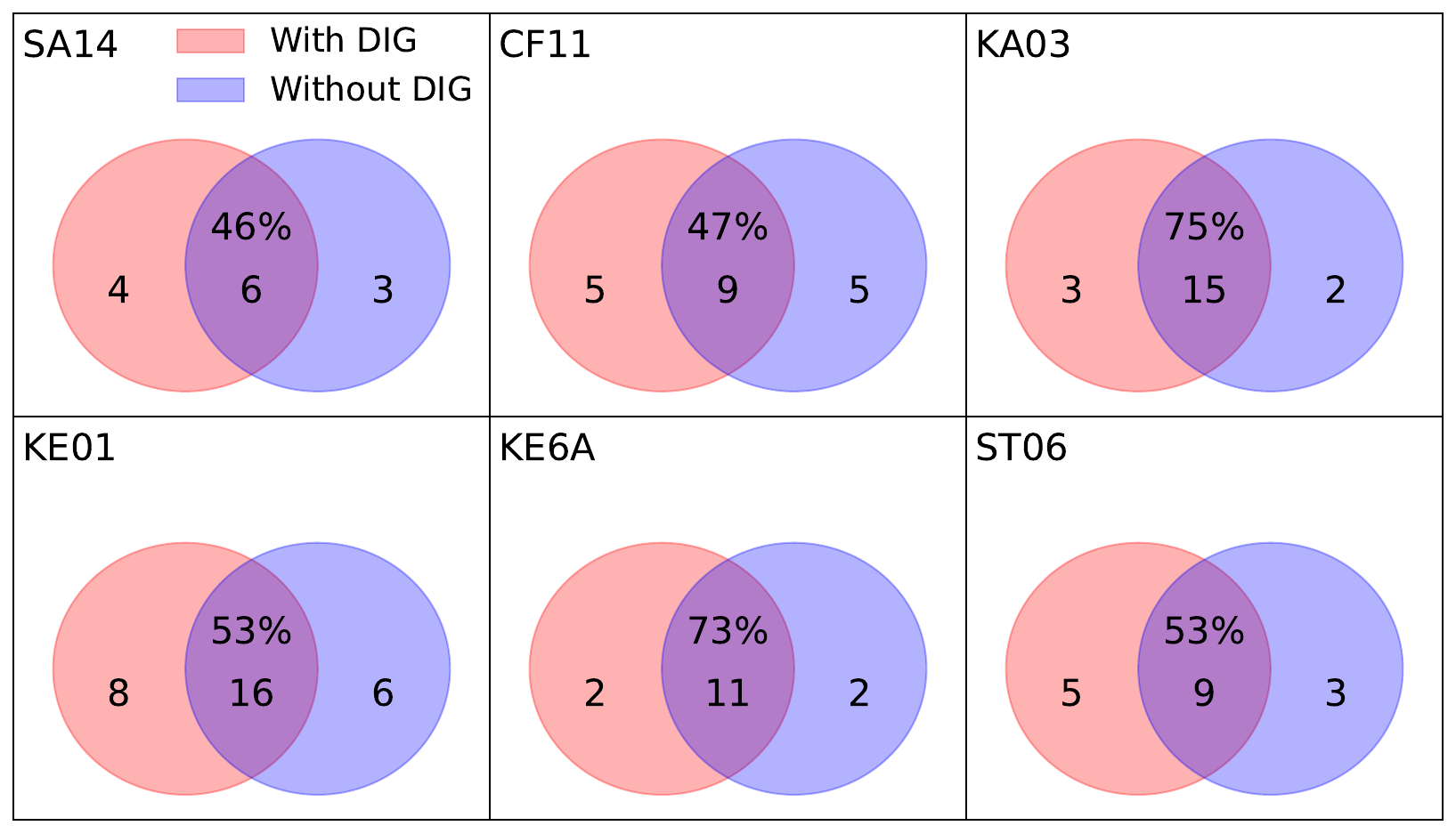}
\caption{Diagrams depicting the number of galaxies with an inner drop using different H{\sc ii} region selection methods, with and without DIG contamination, differentiated by red and blue colors as indicated in the legend. At the center of each panel is the intersection indicating the percentage of the total number of galaxies with an inner drop relative to the complete set, considering DIG contamination or not.}
\label{comp.quebra.com.sem.dig}
\end{figure}

Another result that can be observed in Table \ref{tab:ocurrence_inner_drop} is that the 14 galaxies that exhibit an inner drop in the DIG contaminated and the 15 galaxies from the sample without DIG contamination are not exactly the same. In this way, we conducted a comparison to analyze the number of galaxies exhibiting an inner drop with and without DIG contamination, as shown in Figure \ref{comp.quebra.com.sem.dig}. We can observe that the criteria of KE6A and \citetalias{Kauffmann2003} are the ones that maintained the highest number of galaxies showing an inner drop regardless of DIG correction (highest number of galaxies at the intersection of the diagrams). This can be explained by the fact that using the $\mbox{EW(H}\alpha) > 6 \ \mbox{\AA}$ criterion in the \citetalias{Kewley2001} demarcation curve eliminates HOLMES according to \citetalias{lacerda2018}, significantly reducing hDIG. Additionally, as discussed in \citet{zinchenko.2019}, \citet{pilyugin.2018} also show that applying the demarcation curve proposed by \citetalias{Kauffmann2003} eliminates ionized regions that are contaminated by DIG. However, as Figure \ref{comp.quebra.com.sem.dig} shows, both criteria do not completely eliminate the ionized regions that are contaminated by DIG.

As it is evident that the presence of DIG directly influences the gradient profiles and the occurrence of the inner drop, the subsequent analysis is based solely on the sample without DIG contamination. For the purpose of comparison, we performed a fit on the sample without DIG contamination according to the calibrators proposed by \citet{dopita2016} and \citet{Marino2013} for the O3N2 index, as well as the calibrator proposed by \citet{Pettini2004} using the N2 index \citep{storchi1994}. In the case of the \citet{dopita2016} calibrator, only 6 galaxies exhibited an inner drop in at least one method of H{\sc ii} region selection, with only 4 galaxies showing an inner drop in at least three criteria. For the \citet{Marino2013}, 34 galaxies presented an inner drop with 14 galaxies in at least three criteria. In the case of the \citet{Pettini2004} calibrator with the N2 index, 29 galaxies exhibited an inner drop in at least one of the H{\sc ii} region selection criteria, with 11 galaxies showing an inner drop in at least three criteria.

\subsection{The presence of the Inner Drop and the Physical Properties of the Galaxies \label{sec:inner_drop_physical_properties}}

Throughout this work we discussed the importance of a careful selection of H{\sc ii} regions in galaxies, noting that the possibility of an inner drop in oxygen abundance gradients may be related to this selection. Therefore, in our next analysis, we restrict the analysis to the galaxies presenting a higher probability of displaying an inner drop according to the different selection criteria, selecting the 15 galaxies that consistently showed an inner drop in at least three criteria after DIG decontamination. This approach aims to ensure greater reliability in identifying galaxies that genuinely exhibit an inner drop. Table \ref{mean.std.coeff.hii} displays the coefficients $a_1$, $a_2$ and $h_1$ for each H{\sc ii} region selection method in the 15 galaxies aforementioned. The two last lines in this table show the mean and the standard deviation for $a_1$, $a_2$ and $h_1$ for all the galaxies considering the same H{\sc ii} region selection method. We note that the average values for the position of the inner drop converge to 0.8 -- 0.9\,\re, for all the H{\sc ii} region selection criterion. Also, they are higher than the value of $0.5 \pm 0.2$\,\re \, from \citetalias{Sanchez-Menguiano2018}, although they are compatible considering the errors in the measurements. Considering the average standard deviations for $h_1$ for all the galaxies in a given H{\sc ii} region selection criterion in this table, they are in the range $\sim$0.20 -- 0.30\,\re\ and are consistent with the errors in the mean positions obtained in the work of \citetalias{Sanchez-Menguiano2018}. Additionally, we observe that for \citetalias{Kauffmann2003} and \citetalias{cidfernandes2011}, the standard deviations have the smallest values, indicating less dispersion in the mean positions of inner drops for these criteria. 

\begin{table*}
\setlength\tabcolsep{0.02cm}
\centering
\caption{Coefficients $a_1$, $a_2$  and $h_1$ from profiles of the radial abundance gradients with an inner drop.}
\label{mean.std.coeff.hii}
\begin{scriptsize}
 \begin{tabular}{@{\extracolsep{7pt}}lccc@{\hspace{7pt}}ccc@{\hspace{7pt}}ccc@{\hspace{7pt}}ccc@{\hspace{7pt}}ccc@{\hspace{7pt}}ccc@{}}
\hline \hline
 Galaxy &  \multicolumn{18}{c}{ H{\sc ii} region selection criteria\tablenotemark{\scriptsize a}}\\
      \cline{2-19} 
  & \multicolumn{3}{c}{\citetalias{Sanchez2014}} & \multicolumn{3}{c}{\citetalias{cidfernandes2011}} & \multicolumn{3}{c}{\citetalias{Kauffmann2003}} & \multicolumn{3}{c}{\citetalias{Kewley2001}} & \multicolumn{3}{c}{KE6A} & \multicolumn{3}{c}{\citetalias{stasinska2006}} \\
 \cline{2-4}
 \cline{5-7}
 \cline{8-10}
 \cline{11-13}
 \cline{14-16}
 \cline{17-19}
   & $a_1$ & $a_2$ & $h_1$ & $a_1$ & $a_2$ & $h_1$ & $a_1$ & $a_2$ & $h_1$ & $a_1$ & $a_2$ & $h_1$  & $a_1$ & $a_2$ & $h_1$ & $a_1$ & $a_2$ & $h_1$ \\  
  \hline
    NGC~7653 	&	-0.01	& -0.15	& 	0.56	&	-0.08	& -0.26	& 	0.91	&	-0.08	& -0.26	& 	0.91	&	-0.08	& -0.25	& 		0.91	&	-0.09	& -0.25	& 	0.92	&	-0.08	& -0.26	& 	0.92	\\
    NGC~4047 	&	0.08	& -0.24	&	0.67	&	0.08	& -0.23	&	0.66	&	0.08	& -0.21	&	0.64	&	0.08	& -0.21	&		0.64	&	0.08	& -0.21	&	0.64	&	0.08	& -0.21	&	0.64	\\
    NGC~5406 	&	 -- 	& -0.22	&	 -- 	&	0.09	& -0.35	&	1.01	&	0.10	& -0.35	&	1.01	&	0.10	& -0.35	&		1.01	&	0.10	& -0.35	&	1.01	&	0.10	& -0.31	&	1.00	\\
    NGC~0309 	&	 -- 	& -0.29	&	 -- 	&	0.13	& -0.38	&	0.64	&	0.35	& -0.35	&	0.57	&	0.18	& -0.38	&		0.66	&	0.02	& -0.37	&	0.65	&	0.86	& -0.35	&	0.46	\\
    NGC~4210 	&	-0.08	& -0.29	&	1.37	&	 -- 	& -0.18	&	 -- 	&	0.13	& -0.22	&	0.87	&	0.00	& -0.29	&		1.27	&	-0.06	& -0.29	&	1.31	&	-0.04	& -0.28	&	1.30	\\
    NGC~5016 	&	 -- 	& -0.20	&	 -- 	&	--	& -0.17	&	 -- 	&	0.30	& -0.26	&	0.58	&	0.52	& -0.17	&		0.42	&	0.52	& -0.17	&	0.42	&	 -- 	& -0.16	&	 -- 	\\
    IC~1256 	    &	 -- 	& -0.28	&	 -- 	&	0.06	& -0.28	&	0.78	&	0.04	& -0.28	&	0.81	&	0.04	& -0.28	&		0.81	&	0.04	& -0.28	&	0.81	&	0.06	& -0.27	&	0.79	\\
    NGC~0776 	&	0.39	& -0.12	&	0.91	&	 -- 	& -0.13	&	 -- 	&	0.12	& -0.12	&	1.03	&	0.11	& -0.12	&		1.02	&	0.11	& -0.12	&	1.02	&	 -- 	& -0.11	&	 -- 	\\
    NGC~2347 	&	 -- 	& -0.21	&	 -- 	&	--	& -0.21	&	 -- 	&	0.09	& -0.28	&	0.7	    &	0.09	& -0.28	&		0.70	&	0.09	& -0.28	&	0.70	&	0.09	& -0.28	&	0.68    \\
    NGC~5533 	&	1.55	& -0.17	&	0.34	&	0.32	& -0.20	&	0.62	&	0.26	& -0.21	&	0.62	&	 -- 	& -0.02	&		 -- 	&	--	& -0.17	&	 -- 	&	 -- 	& -0.12	&	 -- 	\\
    NGC~6004 	&	 -- 	& -0.11	&	 -- 	&	0.12	& -0.17	&	0.67	&	0.12	& -0.15	&	0.68	&	 -- 	& -0.09	&		 -- 	&	0.08	& -0.17	&	0.75	&	 -- 	& -0.10	&	 -- 	\\
    NGC~4185 	&	 -- 	& -0.27	&	 -- 	&	0.17	& -0.22	&	0.71	&	0.16	& -0.24	&	0.72	&	0.14	& -0.24	&		0.71	&	0.20	& -0.23	&	0.68	&	0.16	& -0.24	&	0.73	\\
    NGC~5957 	&	 -- 	& -0.19	&	 -- 	&	0.18	& -0.26	&	1.01	&	0.21	& -0.18	&	0.87	&	0.24	& -0.17	&		0.90	&	-0.04	& -0.26	&	1.19	&	 -- 	& -0.13	&	 -- 	\\
    NGC~1667 	&	 -- 	& -0.06	&	 -- 	&	0.13	& -0.12	&	0.98	&	0.04	& -0.09	&	1.17	&	 -- 	& -0.06	&		 -- 	&	--	& -0.06	&	 -- 	&	0.04	& -0.14	&	1.41	\\		
    NGC~5267 	&	0.21	& -0.07	&	1.28	&	0.13	& -0.09	&	1.40	&	 -- 	& -0.01	&	 -- 	&	0.12	& -0.06	&		1.38	&	 -- 	& -0.02	&	 -- 	&	--	& -0.03	&	 -- 	\\
    \hline
    Mean & 0.36 & -0.19 & 0.85 & 0.12 & -0.22 & 0.85 & 0.14 & -0.21 & 0.80 & 0.13 & -0.20 & 0.87 & 0.09 & -0.22 & 0.84 & 0.14 & -0.20 & 0.88 \\
    Std & 0.56 & 0.07 & 0.37 & 0.09 & 0.08 & 0.23 & 0.11 & 0.09 & 0.18 & 0.14 & 0.11 & 0.26 & 0.15 & 0.10 & 0.24 & 0.26 & 0.09 & 0.29\\ 
    \hline
\end{tabular}
\tablecomments{The coefficients $a_1$ and $a_2$ are in dex$/$\re\ and $h_1$ in $r/$\re\, and are for each H{\sc ii} region selection criterion for galaxies with an inner drop and after DIG decontamination. The last two rows display the mean and standard deviation of the coefficients.}
\tablenotetext{a}{As in Table \ref{tab:hii_selection}.}
\end{scriptsize}
\end{table*}

The optimal position for the inner drop for each one of the 15 galaxies was determined by calculating the average of positions obtained from the different H{\sc ii} region selection criteria in Table \ref{mean.std.coeff.hii}, weighted by the individual errors of the coefficients obtained from the fitting procedure. The associated error of the averaged value was determined by propagating the errors from the individual coefficients. The final result can be seen in Table \ref{mean.std.coeff.a1.a2.h1}. In this table we also show the average values for the coefficients $a_1$ and $a_2$ for each galaxy. Therefore, all the results presented from now on are based on the mean values of the slopes and positions of the inner drops as given in Table \ref{mean.std.coeff.a1.a2.h1}.

\begin{table}
\centering
\caption{Average coefficients $a_1$, $a_2$ and $h_1$ from profiles of the radial gradient in the 15 galaxies that presented an inner drop without DIG contamination.}
\label{mean.std.coeff.a1.a2.h1}
\begin{tabular}{lccc}
\hline \hline
Galaxy & $a_1$ & $a_2$ & $h_1$ \\
       & (dex$/$\re) & (dex$/$\re) & ($r/$\re) \\
\hline
    NGC~7653 	&	-0.08	$\pm$ 0.03	& 	-0.255 
    $\pm$ 0.009	&	0.90	$\pm$ 0.06 \\
    NGC~4047 	&	0.08	$\pm$ 0.05	&	-0.218 
    $\pm$ 0.008	&	0.65	$\pm$ 0.05 \\
    NGC~5406 	&	0.10	$\pm$ 0.09	&	-0.294  
    $\pm$ 0.032	&	1.01	$\pm$ 0.06 \\
    NGC~0309 	&	0.17	$\pm$ 0.07	&	-0.347 
    $\pm$ 0.019	&	0.55	$\pm$ 0.03 \\
    NGC~4210 	&	-0.04	$\pm$ 0.02	&	-0.214 
    $\pm$ 0.008	&	1.19	$\pm$ 0.04 \\
    NGC~5016 	&	0.40	$\pm$ 0.16	&	-0.173  
    $\pm$ 0.009	&	0.46	$\pm$ 0.06 \\
    IC~1256 	&	0.04	$\pm$ 0.05	&	-0.278
    $\pm$ 0.008	&	0.80	$\pm$ 0.05 \\
    NGC~0776 	&	0.13	$\pm$ 0.04	&	-0.119
    $\pm$ 0.013	&	0.97	$\pm$ 0.06 \\
    NGC~2347 	&	0.09	$\pm$ 0.06	&	-0.218 
    $\pm$ 0.007	&	0.69	$\pm$ 0.06 \\
    NGC~5533 	&	0.30	$\pm$ 0.09	&	-0.106 
    $\pm$ 0.019	&	0.45	$\pm$ 0.03 \\
    NGC~6004 	&	0.09	$\pm$ 0.07	&	-0.115
    $\pm$ 0.011	&	0.71	$\pm$ 0.07 \\
    NGC~4185 	&	0.16	$\pm$ 0.04	&	-0.236 
    $\pm$ 0.016	&	0.72	$\pm$ 0.04 \\
    NGC~5957 	&	0.12	$\pm$ 0.04	&	-0.183 
    $\pm$ 0.018	&	0.95	$\pm$ 0.03 \\
    NGC~1667 	&	0.05	$\pm$ 0.04	&	-0.074 
    $\pm$ 0.005	&	1.19	$\pm$ 0.13 \\
    NGC~5267 	&	0.13	$\pm$ 0.04	&	-0.024
    $\pm$ 0.006	&	1.35	$\pm$ 0.15 \\
    \hline
\end{tabular}
\end{table}

\begin{figure}
\figurenum{9}
\includegraphics[width=\columnwidth]{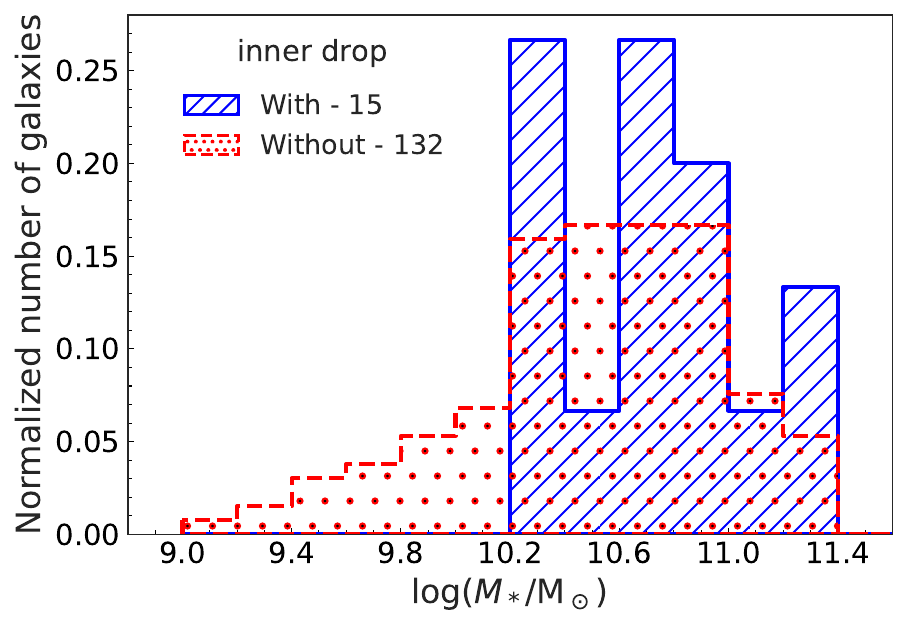}
\caption{Distribution of the number of galaxies that exhibit an inner drop (in blue) compared to galaxies without an inner drop (in red), considering the 15 galaxies that showed an inner drop in at least three H{\sc ii} region selection criteria, as shown in Table \ref{mean.std.coeff.a1.a2.h1}.}
\label{mass.inner.drop}
\end{figure}

Similarly to the work from \citetalias{Sanchez-Menguiano2016} and \citetalias{Sanchez-Menguiano2018}, we also identified that the presence of the inner drop occurs in the most massive galaxies, as can be seen in Figure ~\ref{mass.inner.drop}. However,
contrary to previous study by SM18, 
we only identified inner drops in galaxies with $\log(M_*/\mbox{M}_\sun) > 10.2$, that is, the presence of an inner drop in the metallicity gradient is exclusive to the most massive galaxies in our sample.

\begin{figure}
\figurenum{10}
\includegraphics[width=\columnwidth]{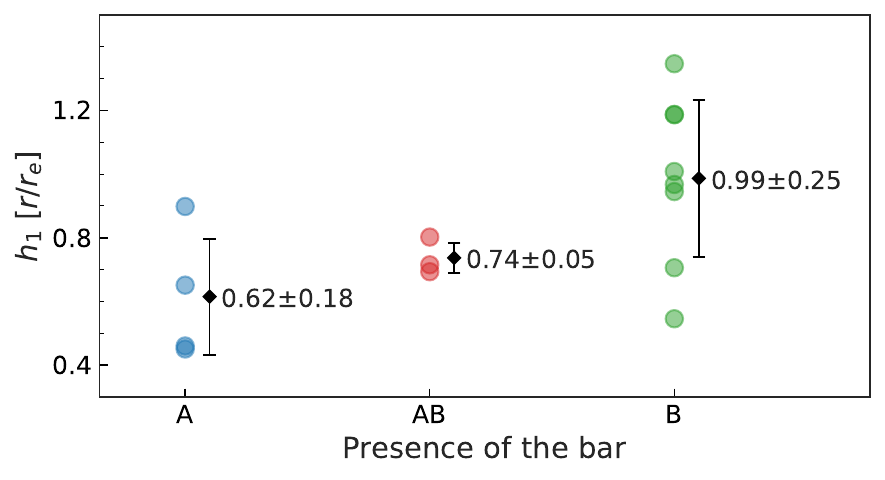} \\
\includegraphics[width=\columnwidth]{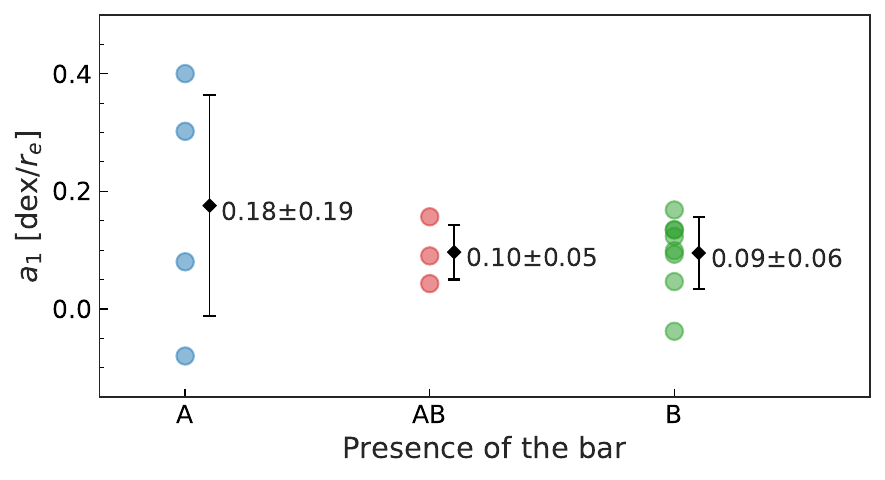} \\
\includegraphics[width=\columnwidth]{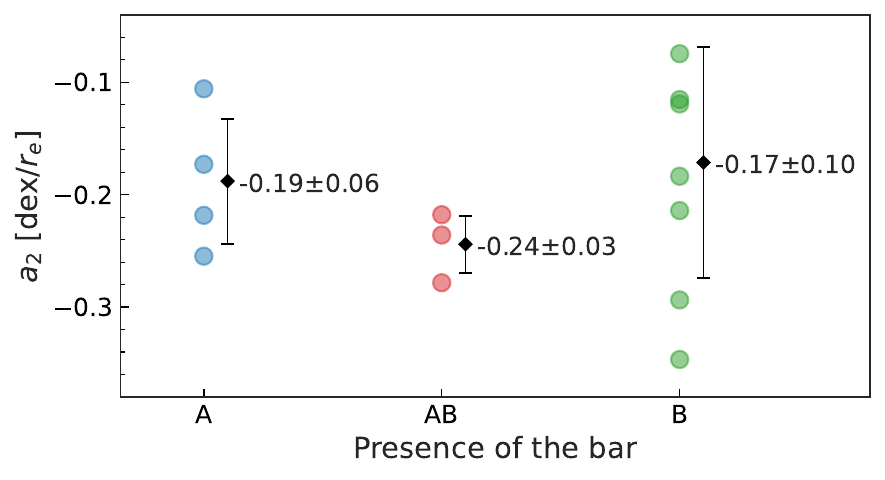}
\caption{Relation of the coefficients of the fit with the presence or absence of the bar for the 15 galaxies that exhibited an inner drop in at least three methods of selection of H{\sc ii} regions, as shown in Table \ref{mean.std.coeff.a1.a2.h1}, without DIG contamination. The top, middle, and bottom panels correspond to the position where the inner drop occurs, the coefficient $a_1$ related to the inner radial gradient, and the coefficient $a_2$ related to the main negative gradient, respectively. The indices ``A'', ``AB'', and ``B'' correspond, respectively, to unbarred galaxies, galaxies that may or may not have bars, and barred galaxies, according to the Hubble classification. For each classification the mean and the standard deviation values are given, and also represented as black diamonds with error bars, respectively.}
\label{morph.bar.inner.drop}
\end{figure}

In the same way, as reported in previous works from \citetalias{Sanchez-Menguiano2016, Sanchez-Menguiano2018}, the inner drops are detected in barred and unbarred galaxies, as seen in Figure~\ref{morph.bar.inner.drop}. In this figure, the top, middle and bottom panels show the relation of the position of the inner drop ($h_1$), slope of the inner radial abundance gradient ($a_1$) and the slope of the main abundance gradient ($a_2$), with the presence of the bar, respectively. We can observe in the top panel that the average value and the dispersion for the barred galaxies are higher than for unbarred galaxies. We also note that only 4 unbarred galaxies showed an inner drop, while 8 barred galaxies and 3 intermediate galaxies presented the inner drop. It is also possible to observe that barred galaxies exhibit a greater dispersion in the slope of the main negative gradient, while the dispersion in the inner radial gradient is smaller.
When the inner drop exists, and considering all the morphological types in Table~\ref{mean.std.coeff.a1.a2.h1}, the main abundance gradient has an average slope of $ \left<a_2\right> = -0.19 \pm 0.09$\,dex\,\re$^{-1}$, which is steeper than the value of $-0.08\pm 0.09$\,dex\,\re$^{-1}$, for the common radial gradient for spiral galaxies in previous works using CALIFA data \citepalias{Sanchez-Menguiano2016}. This result was also noted by previous works from \citetalias{Sanchez-Menguiano2016,Sanchez-Menguiano2018}. The average values and the standard deviations for the slope of the inner radial gradient and the position of the inner drop are $\left<a_1\right> = 0.12\pm 0.11$\, dex\,\re$^{-1}$, and $\left<h_1\right> = 0.84\pm 0.26$\,\re, respectively.

\begin{figure*}
\figurenum{11}
\begin{center}
\includegraphics[width=2.0\columnwidth]{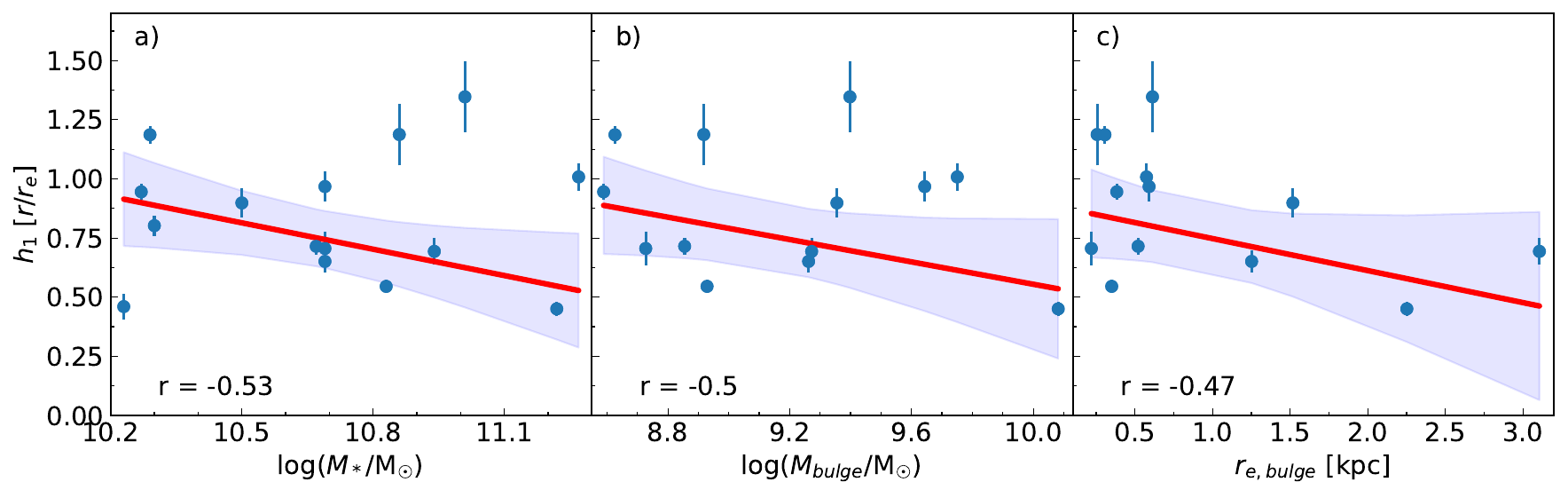}
\includegraphics[width=2.0\columnwidth]{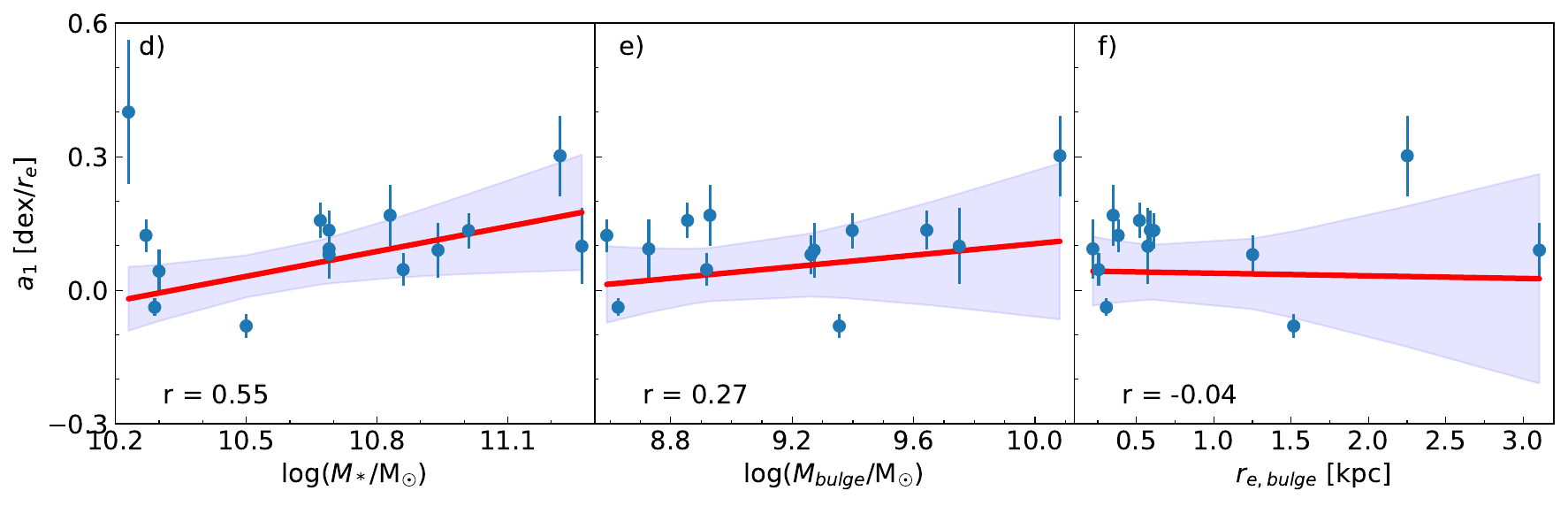}
\end{center}
\caption{Relationships between the mean position where the inner drop occurs ($h_1$) with the galaxy and bulge masses and and the bulge effective radius presented in panels a) to c). Panels d) to f) show the same relations but for the slope of the inner radial abundance gradient ($a_1$). The blue points with error bars correspond to the averages values with their respective errors for each galaxy, as given by Table \ref{mean.std.coeff.a1.a2.h1}. The red continuous lines are the weighted linear fit to the data and the blue shaded areas the 95\% confidence interval. The ``r'' index in the bottom left corner of each panel corresponds to the weighted Pearson correlation coefficient \citep{pearson.coef}.
}
\label{analise_galaxy}
\end{figure*}

In an attempt to study possible relationships between the presence of the inner drop and the physical properties of galaxies not yet explored in the literature, as e.g. the physical parameters of the bulges, we conducted analyses using the positions where the inner drop occurs and we also analyzed the slope of the inner radial gradient profile, identified by the coefficients $h_1$ and $a_1$ in Figure~\ref{exemplo.ajuste}, respectively. Figure~\ref{analise_galaxy} shows the relationship between the mean positions of the inner drop and the mean slope of the inner radial abundance gradient with the galaxy and bulge masses and also the bulge effective radius for the 15 galaxies that presented the inner drop without DIG contamination. For all the relations the Pearson correlation coefficient \citep{pearson.coef} was calculated weighted by the errors of the individual points as given in Table \ref{mean.std.coeff.a1.a2.h1}. In the case of the position of the inner drop $h_1$ (panels a) to c)) the Pearson coefficients point to negative correlations with all the parameters analyzed, although the correlation with the mass of the galaxies is slightly higher. 

In Figure \ref{analise_galaxy} panels b) and c) show the relationships between the mass and effective radius of the bulges with the mean positions of the inner drop. For the bulge parameters, the galaxies IC~1256 and NGC~5016 were excluded from the analysis as there is no information on the bulge parameters for these galaxies. A moderate negative correlation between the masses and the effective radii of the bulges with the positions of the inner drop is observed. Therefore, we can expect that a galaxy with an inner drop closer to the center will tend to have a more massive bulge, with larger bulge effective radius, and higher surface brightness at the effective radius of the bulge
\footnote{The effective radii and the surface brightness at the effective radii are correlated following the Kormendy relation \citep{kormendy77}.}.

Figure \ref{analise_galaxy} panels d) to f) show the same relations but for the slope of the inner radial gradient $a_1$ with the galaxy and bulge masses and the bulge effective radius. In the case of the bulge effective radius, the result pointed to a non-existence of correlation. On the other hand, there are positive correlations with the galaxy and bulge masses, although the first one is stronger. The result that the most massive galaxies tend to have the steepest internal radial gradients was already identified by \citetalias{Sanchez-Menguiano2016} and \citetalias{Sanchez-Menguiano2018}, but now it is quantified in our Figure~\ref{analise_galaxy}. 

\section{Discussion \label{sec:discussion}} 

The explanation about the existence of an inner drop in the radial abundance gradient of spiral galaxies is still an unsolved problem. Previous works from \citetalias{Sanchez-Menguiano2016,Sanchez-Menguiano2018} have made progress on this topic, finding that the average slope of the main radial gradient for O/H of $-0.10 \pm 0.04$~dex/\re\ is slightly steeper than the value of $-0.06\pm0.05$~dex/\re\ derived for the galaxies without evidence of this inner drop in the abundances. With our improved method to fit the abundance gradients, 
we confirmed this result: $-0.19 \pm 0.09$~dex/\re\, and  $-0.08 \pm 0.04$~dex/\re for galaxies with and without an inner drop, respectively.
\citetalias{Sanchez-Menguiano2016,Sanchez-Menguiano2018} also find that the galaxies displaying the strongest oxygen abundance inner drop are the most massive
ones, suggesting that stellar mass plays a key role in shaping the inner abundance profiles. Our detailed analysis identified a correlation between the slope of the inner drop gradient and the galaxy stellar mass, but we also found that the bulge mass presents a correlation with the slope of the inner radial gradient. Nonetheless, given the present uncertainties in the slope of the inner radial gradient, it is not possible to point which correlation is higher. On the other hand, our analysis pointed that the inner drop occurs exclusively for galaxies with $\log(M_*/\mbox{M}_\sun) > 10.2$. 

From the theoretical point of view, the presence of the bar in the galactic center and the associated secular evolution of the disk could be an explanation to the effect of the inner drop in the radial abundance distribution. Models and simulations show that the non-axisymmetric bar gravitational potential drives gas flow toward the galaxy central region along the bar dust lane \citep[e.g.][and references therein]{fragkoudi16}, and there are observational results confirming bar-driven gas transport \citep[e.g.][and references therein]{lopez-coba22}. According to the chemical evolution model (CEM) of the Milky Way by \citet{Cavichia2014}, the presence of the bar induces radial gas flows within the corotation radius, being a mechanism for the movement and mixing of gas in barred spiral galaxies, causing the increase of the gas density and decreasing the abundances by dilution and, simultaneously, provoking the possibility of creating new stars, by increasing the star formation rate. Consequently, it is expected the effect of the bar might play an important role in the profiles of abundance gradients. 
However, as identified in the present work and in those from \citetalias{Sanchez-Menguiano2016,Sanchez-Menguiano2018}, the occurrence of the inner drop was not only associated with the presence of a bar. In spite of that, in our analysis the inner drop in barred galaxies tend to occur at positions furthest from the center in the disk, and also the dispersion of the values is higher. Therefore, there is an evidence that bars can be related with the phenomenon of the inner drop. Although bars are not the only causes of the inner drop, they may amplify it when a bar is present in the central region of the galaxy. From visual inspection of $r$-band images from SDSS and also from the morphological classifications, we also identified that 12 out of the 15 galaxies (80\%) that present an inner drop have some asymmetry in the center, either a bar or a circumnuclear star-forming ring. The possibility that the inversion of the gradient in the central region of spiral galaxies is related with a star-forming ring was also considered in previous works \citepalias[][and references therein]{Sanchez-Menguiano2016}. 

\begin{figure}
    \figurenum{12}
    \centering
    \includegraphics[width=1.0\linewidth]{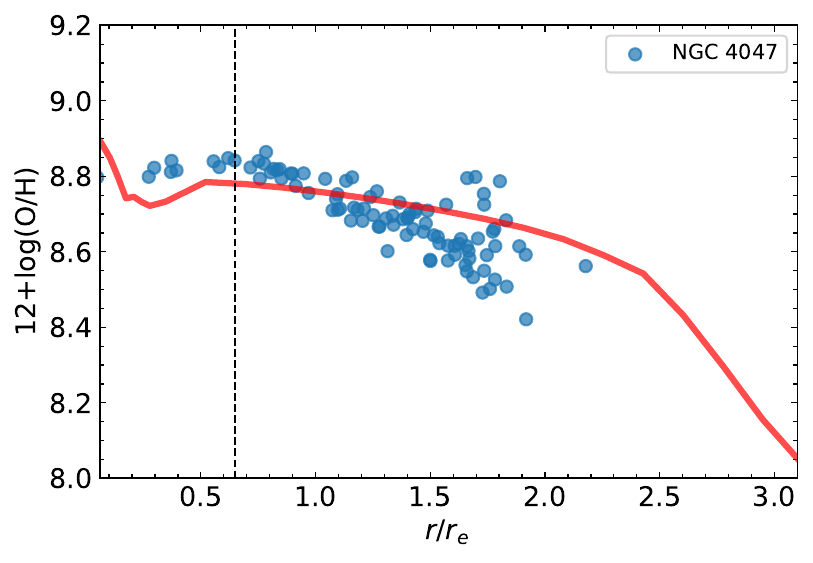}
    \caption{Oxygen abundance radial profile for NGC~4047 (blue filled circles) obtained using KA03 \ion{H}{0II} region selection method. The dashed vertical line marks the position 0.64~$r/r_e$ where an inner drop is detected for this galaxy. The red continuous line corresponds to preliminary results for the Milky Way CEM from \citet{cavichia2023} adapted for a galaxy with similar mass to NGC~4047.}
    \label{fig:cem}
\end{figure}

Bulges are another important subject of study in spiral galaxies, as they can provide important information regarding chemical abundances in the central region of galaxies. Using a CEM, \citet{cavichia2023} propose a new model for the formation of the Galactic bulge, where it is formed from inside-out by gas accretion from the halo. The shortest collapse timescales in the central regions of the Galaxy produces a drop of chemical abundances near the bulge-disk interface at $\sim 3$\,kpc (see their figures 5 and 6), similar to the inner drop in the radial abundance gradients in the present work. For the Milky Way, using the average effective radius of $5.07\pm0.93$\,kpc from \citet{molla2019}, the inner drop for the O/H radial abundance gradient is located at 0.6~\re\ in the work from \citet{cavichia2023}. Thus, a possible relationship between the inner drop and the bulge formation/evolution may be motivated. For a comparison purpose, in Figure \ref{fig:cem} the oxygen radial profile for NGC~4047 -- a galaxy that present an inner drop in all the \ion{H}{0II} region selection method in our analysis -- is compared with the results of the CEM from \citet{cavichia2023} developed for the Milky Way and adapted in this work for a spiral galaxy similar to  NGC~4047, with virial mass $\log(M_{\rm vir})$ = 11.89, which corresponds to a stellar mass $M_* = 10^{10.68} \,\text{M}_{\sun}$, and an efficiency $N = 5$ for an Sbc spiral galaxy. We refer to \citet{cavichia2023} and references therein for more details about the CEM. The CEM presents a drop near the bulge-disk interface because the timescales for the bulge and disk are different in the model, as a result of the inside-out formation of the bulge. Although the result is still preliminary, the model is promising in explaining the inner drop in the chemical abundances radial profiles of the spiral galaxies. To confirm this we need observations with higher spatial resolution to obtain data for $r < 0.5 \, r_e$ for a larger number of galaxies. However we should note that these are still preliminary results and detailed CEM for spiral galaxies with different masses will be published in a forthcoming work.

Our analysis pointed to a negative correlation between the position of the inner drop and the masses and effective radii of the bulges. Thus, the higher the values of these parameters, the lower the positions where the inner drops occur. The stellar masses of the galaxies may as well as be correlated with the positions of the inner drops, given the uncertainties in the determination of the parameters. 

Considering the DIG contamination and all the six different criteria for selecting H{\sc ii} regions, it is evident that, in some cases, the change in the gradient profile experiences a considerable alteration depending on the criterion. In addition to DIG influencing the profile of the radial gradients, we verified that the choice of different methods for selecting H{\sc ii} regions changes the profile of the oxygen abundance distribution, as can be seen in Figure \ref{fig:profiles_grad_with_dig}, where the fits of the abundance gradients for all H{\sc ii} region selection criteria are shown for the 32 galaxies that present an inner drop in at least one criterion with this new sample. In some cases, the method to select H{\sc ii} regions removed the data in the inner regions of the galaxies and the radial distribution of abundances could not be characterized in these central regions. On the other hand, there were situations where the radial distribution of the chemical abundances along the disk changed, depending on the H{\sc ii} regions selection method, increasing the dispersion of the data and the inner drop, that was detected in a different method, was not detected anymore. It is also clear from the analysis presented in Section \ref{sec:hii_region_selection_grad_fit} that the presence of the inner drop was also sensitive on the adopted method to derive the abundances. In the case of the \citet{dopita2016} calibration based on the H$\alpha$, \ion{N}{0II} and \ion{S}{0II} lines, it was observed the largest difference in the abundance gradient fitting and the detection of the inner drops. On the other hand, the calibrators for the O3N2 index, based on the ratios of strong emission lines $[\mbox{\ion{O}{0III}}]/\mbox{H}\beta$ and $[\mbox{\ion{N}{0II}}]/\mbox{H}\alpha$ from \citet{Pettini2004} and \citet{Marino2013}, produced similar results, detecting a large number of inner drops in the radial abundance distributions.

In order to draw firmer conclusions about the correlations obtained in this work between the position where the inner drop occurs and the slope of the inner radial O/H gradient, we need to increase the sample of galaxies that present an inner drop in the radial abundance gradient. We also need more observations of the gas motions in the inner regions of the spiral galaxies to better understand the relation of the gas motions and the inner radial abundance gradients. 

\section{Conclusions \label{sec:conclusions}}

This study provided a comprehensive investigation of oxygen abundance gradients and, consequently, star-forming regions in galaxies, employing different methods for the selection of the H{\sc ii} regions. The results obtained highlight the intrinsic complexity of the chemical evolution of galaxies.

One of the most significant findings was the sensitivity of abundance gradients to different criteria for selecting H{\sc ii} regions. It was observed that the choice of a specific method can lead to substantial variations in the profiles of abundances, with marked changes in the positions of breaks and the slopes of the gradients in several galaxies. The selection of different methods implies the exclusion of more or fewer H{\sc ii} regions in the inner region, causing the exclusion of these regions to affect the number of galaxies that may exhibit an inner drop. This result highlights the critical importance of adopting robust and consistent criteria when investigating the properties of galaxies.

We also observed that the DIG can impact the results by altering the profile of the gradients. This alteration in the profiles can be more or less significant depending on the selection criteria of H{\sc ii} regions, as shown in Figure \ref{comp.quebra.com.sem.dig}. In other words, depending on the adopted criterion, galaxies that previously showed an inner drop may no longer present this drop, and vice versa. This may be associated with the determination of oxygen abundance in spaxels which emission might not come exclusively from the H{\sc ii} regions. This may lead to wrong determination of abundances when using a calibrator that is not suitable for determining abundances in regions not strictly ionized by young stars. However, we noted that the criteria of \citet{Kauffmann2003} and \citet{Kewley2001} with $\mbox{EW(H}\alpha) > 6 \ \mbox{\AA}$ for the selection of H{\sc ii} regions stand out for their robustness regarding DIG contamination.
As discussed earlier, these two criteria maintained about 70\% of the number of galaxies exhibiting a break, regardless of DIG decontamination, despite it being clear that DIG directly affects the distribution irrespective of the adopted criteria.

Another highlight of this study was the observed trend in the relationship between the position of the inner drop and the galaxy stellar mass. Despite the small number of galaxies, the robustness of the adopted criterion, which considered galaxies with breaks in at least three different criteria, ensured the reliability of the observed trend. We found that galaxies that present larger masses tend to have an inner drop closer to the center of the galaxy. Similar relations were also found for the bulges masses and effective radii, although the correlations were slightly weaker. In other words, the inner drop tends to occur closer to the center of the galaxy in galaxies with larger effective bulge radii and, consequently, higher surface brightness at the effective radius of the bulge.

Besides the positions where the inner drops occur, we also analyzed the slope of the inner radial gradient, identified by the coefficient $a_1$. Correlations were also identified between the slope of the inner radial gradient and the galaxy stellar mass and the mass of the bulge, where more massive galaxies and more massive bulges tend to present steeper inner radial gradients.

A more robust approach to clarify the obtained results involves acquiring data with higher spatial resolution, aiming to more effectively eliminate DIG contamination. Additionally, investigating the radial velocities of gas in galaxies can provide clues about the presence of inner drops, as the gas radial movements have the potential to cause significant changes in the distribution of gas clouds. Therefore, obtaining data with high spatial resolution becomes crucial, allowing for the effective and reliable distinction and separation of star-forming regions from regions ionized by different types of sources. This is essential for deepening the study of chemical abundance gradients.

\begin{acknowledgments}
We thank the anonymous referee for their suggestions, which improved this paper.

AFSC acknowledges the scholarship from FAPEMIG.
OC acknowledges funding support from FAPEMIG grant APQ-00915-18. 
OC and MM have been funded through grants PID2019-107408GB-C41 and
PID2022-136598NB-C33 by MCIN/AEI/10.13039/501100011033 and by “ERDF A way of making Europe”.

This study uses data provided by the CALIFA survey (\url{http://califa.caha.es/}); it is based on observations collected at the Centro
Astron\'omico Hispano Alem\'an (CAHA) at Calar Alto, operated
jointly by the Max-Planck-Institut f\"ur Astronomie and the Instituto
de Astrof\'isica de Andaluc\'ia (CSIC). 

This work has made use of the computing facilities available at the Laboratory of Computational Astrophysics of the Universidade Federal de Itajub\'a (LAC-UNIFEI). The LAC-UNIFEI is maintained with grants from CAPES, CNPq and FAPEMIG.

\end{acknowledgments}

%







\section*{}

\appendix

\restartappendixnumbering 

\onecolumngrid

\section{General Properties of Sample Galaxies}
\label{ap01}

Table \ref{tab:sample_selection} below contains information on the main properties of the galaxies and bulges in the sample. The columns of the table correspond to the following identifications:
    (a) galaxy name; (b) right ascension coordinate in hours; (c) declination coordinate in degrees; (d) morphological type from the Hubble classification, indicating barred galaxies (B), non-barred galaxies (A), and intermediate galaxies that may or may not have a bar (AB); (e) integrated stellar mass of the galaxy in $\log(\mbox{M}_{\sun})$; (f) effective radius of the galaxy in units of kpc; (g) mass of the bulge in $\log(\mbox{M}_{\sun})$; (h) effective radius of the bulge in units of kpc; (i) surface brightness at the effective radius of the bulge in units of mag/\arcsec$^2$; (j) redshift; (k) galaxy distance in Mpc; (l) position angle of the galaxy disk in degrees; (m) ratio between the semi-minor and semi-major axes of the galaxy; (n) inclination of the galaxy with respect to the line of sight in degrees.

\begin{center}
    \setlength\tabcolsep{0.12cm}
    \tabletypesize{\scriptsize}
    \begin{longtable}{ccclccccccrrcc}
    \caption{Fundamental properties of the galaxies in the sample. \label{tabela.ap01}}\\
    \hline
    \hline
    Name & R.A. & Decl. & Morph & $\log(M_{*})$ & $r_{e,*}$ & $\log(M_{\text{bulge}})$ & $r_{\text{e,bulge}}$ & $\mu_{\text{bulge}}$ & $z$ & Dist & PA & $b/a$ & $i$ \\
     & (h) & ($\deg$) & type & (M$_{\sun}$) & (kpc) & (M$_{\odot}$) & (kpc) & (mag/$\arcsec^2$) & & (Mpc) & ($\deg$) & & ($\deg$) \\
    (a) & (b) & (c) & (d) & (e) & (f) & (g) & (h) & (i) & (j) & (k) & (l) & (m) & (n) \\
    \hline
    \endfirsthead
    \multicolumn{14}{c}%
    {\tablename\ \thetable\ -- \textit{Continuation}} \\
    \hline
    \hline
    Name & R.A. & Decl. & Morph & $\log(M_{*})$ & $r_{e,*}$ & $\log(M_{\text{bulge}})$ & $r_{\text{e,bulge}}$ & $\mu_{\text{bulge}}$ & $z$ & Dist & PA & $b/a$ & $i$ \\
     & (h) & ($\deg$) & type & (M$_{\sun}$) & (kpc) & (M$_{\odot}$) & (kpc) & (mag/$\arcsec^2$) & & (Mpc) & (deg) & & (deg) \\
    (a) & (b) & (c) & (d) & (e) & (f) & (g) & (h) & (i) & (j) & (k) & (l) & (m) & (n) \\
    \hline
    \endhead
    \hline \multicolumn{14}{r}{\textit{Continues on the next page}} \\
    \endfoot
    \hline \multicolumn{14}{r}{\textit{End of the table}} \\
    \endlastfoot
	IC~0159 & 01.77 & -08.64 & SBdm & 9.8 & 5.7 & 9.2 & 0.9 & 21.1 & 0.013 & 57 & 16 & 0.78 & 40 \\
	IC~0674 & 11.19 & +43.63 & SBab & 10.9 & 10.9 & 9.7 & 2.1 & 21.4 & 0.026 & 107 & 125 & 0.65 & 50 \\
	IC~0776 & 12.32 & +08.86 & SAdm & 9.3 & 7.0 & 8.7 & 1.4 & 23.1 & 0.010 & 35 & 85 & 0.56 & 56 \\
	IC~1151 & 15.98 & +17.44 & SBcd & 9.8 & 4.7 & 8.1 & 0.3 & 21.2 & 0.009 & 31 & 39 & 0.49 & 61 \\
	IC~1256 & 17.40 & +26.49 & SABb & 10.3 & 6.0 & - & - & - & 0.018 & 66 & 90 & 0.59 & 54 \\
	IC~1683 & 01.38 & +34.44 & SABb & 10.6 & 5.0 & 9.4 & 0.4 & 19.0 & 0.016 & 71 & -14 & 0.59 & 54 \\
	IC~4566 & 15.61 & +43.54 & SBb & 10.9 & 9.1 & 9.3 & 0.5 & 19.7 & 0.021 & 80 & 148 & 0.69 & 47 \\
	IC~5309 & 23.32 & +08.11 & SABc & 10.3 & 3.5 & 9.4 & 1.4 & 21.8 & 0.014 & 57 & 21 & 0.49 & 61 \\
	MCG-01-10-019 & 03.68 & -06.42 & SABbc & 10.2 & 12.4 & 9.2 & 1.1 & 21.4 & 0.017 & 75 & 12 & 0.58 & 55 \\
	NGC~0001 & 00.12 & +27.71 & SAbc & 10.8 & 6.2 & 9.9 & 2.2 & 21.3 & 0.015 & 66 & 108 & 0.80 & 37 \\
	NGC~0036 & 00.19 & +06.39 & SBb & 10.9 & 15.3 & 9.6 & 4.1 & 22.9 & 0.020 & 88 & 77 & 0.65 & 50 \\
	NGC~0160 & 00.60 & +23.96 & SAa & 11.0 & 11.7 & 10.2 & 2.3 & 20.9 & 0.017 & 75 & 49 & 0.63 & 52 \\
	NGC~0165 & 00.61 & -10.11 & SBb & 10.6 & 13.3 & 9.6 & 0.8 & 20.1 & 0.019 & 84 & 88 & 0.82 & 35 \\
	NGC~0180 & 00.63 & +08.64 & SBb & 10.9 & 13.1 & 9.4 & 0.6 & 19.9 & 0.017 & 75 & 159 & 0.64 & 51 \\
	NGC~0214 & 00.69 & +25.50 & SABbc & 10.8 & 6.9 & 9.3 & 0.5 & 19.5 & 0.015 & 66 & 62 & 0.66 & 49 \\
	NGC~0234 & 00.73 & +14.34 & SABc & 10.7 & 6.3 & 9.2 & 0.5 & 20.0 & 0.015 & 62 & 75 & 0.85 & 32 \\
	NGC~0237 & 00.72 & -00.12 & SBc & 10.3 & 4.2 & 9.4 & 1.0 & 21.1 & 0.014 & 57 & 144 & 0.57 & 56 \\
	NGC~0257 & 00.80 & +08.30 & SAc & 10.8 & 8.9 & 9.6 & 0.8 & 20.0 & 0.017 & 75 & 92 & 0.58 & 56 \\
	NGC~0309 & 00.95 & -09.91 & SBcd & 10.8 & 13.5 & 8.9 & 0.4 & 19.7 & 0.018 & 80 & 108 & 0.84 & 33 \\
	NGC~0447 & 01.26 & +33.07 & SBa & 11.1 & 13.8 & 9.9 & 1.2 & 20.2 & 0.018 & 80 & 20 & 0.60 & 54 \\
	NGC~0477 & 01.36 & +40.49 & SABbc & 10.5 & 14.5 & 9.3 & 0.8 & 20.7 & 0.019 & 84 & -38 & 0.66 & 50 \\
	NGC~0496 & 01.39 & +33.53 & SAcd & 10.4 & 11.4 & 8.2 & 0.3 & 20.5 & 0.020 & 88 & 38 & 0.58 & 55 \\
	NGC~0570 & 01.48 & -00.95 & SBb & 11.0 & 9.0 & 9.8 & 1.1 & 20.2 & 0.018 & 80 & 102 & 0.70 & 46 \\
	NGC~0681 & 01.82 & -10.43 & SABa & 10.5 & 3.7 & - & - & - & 0.006 & 22 & 69 & 0.65 & 50 \\
	NGC~0716 & 01.88 & +12.71 & SABb & 10.6 & 5.8 & - & - & - & 0.015 & 66 & 57 & 0.64 & 51 \\
	NGC~0768 & 01.98 & +00.53 & SBc & 10.5 & 10.9 & 9.1 & 2.2 & 22.8 & 0.023 & 102 & -65 & 0.55 & 57 \\
	NGC~0776 & 02.00 & +23.64 & SBb & 10.7 & 7.1 & 9.6 & 0.6 & 19.3 & 0.016 & 71 & 141 & 0.69 & 47 \\
	NGC~0787 & 02.01 & -09.00 & SAa & 11.0 & 6.8 & 9.7 & 0.7 & 19.5 & 0.015 & 66 & 81 & 0.81 & 37 \\
	NGC~0873 & 02.28 & -11.35 & SAcd & 10.4 & 4.5 & 9.1 & 0.2 & 18.2 & 0.013 & 57 & 122 & 0.86 & 32 \\
	NGC~0941 & 02.47 & -01.15 & SAcd & 9.3 & 2.8 & - & - & - & 0.005 & 22 & 156 & 0.87 & 30 \\
	NGC~0991 & 02.59 & -07.15 & SABcd & 9.6 & 3.8 & - & - & - & 0.005 & 22 & 85 & 0.92 & 23 \\
	NGC~1056 & 02.71 & +28.57 & SAa & 10.0 & 3.4 & - & - & - & 0.005 & 22 & -24 & 0.57 & 56 \\
	NGC~1070 & 02.72 & +04.97 & SAb & 10.9 & 7.2 & 9.2 & 0.4 & 19.3 & 0.013 & 57 & 182 & 0.82 & 35 \\
	NGC~1093 & 02.80 & +34.42 & SBbc & 10.5 & 7.4 & 9.6 & 1.6 & 21.5 & 0.017 & 75 & 109 & 0.62 & 52 \\
	NGC~1094 & 02.79 & -00.29 & SABb & 10.7 & 7.8 & 9.4 & 0.4 & 18.9 & 0.021 & 93 & 90 & 0.71 & 46 \\
	NGC~1659 & 04.77 & -04.79 & SABbc & 10.5 & 6.4 & 9.9 & 0.2 & 15.9 & 0.015 & 64 & 45 & 0.59 & 55 \\
	NGC~1667 & 04.81 & -06.32 & SBbc & 10.9 & 5.1 & 8.9 & 0.3 & 19.1 & 0.015 & 66 & -7 & 0.64 & 51 \\
	NGC~2253 & 06.73 & +65.21 & SBbc & 10.5 & 2.3 & 9.0 & 0.2 & 18.6 & 0.013 & 48 & 137 & 0.87 & 30 \\
	NGC~2347 & 07.27 & +64.71 & SABbc & 10.9 & 5.9 & 9.3 & 3.1 & 23.0 & 0.015 & 62 & 8 & 0.64 & 51 \\
	NGC~2449 & 07.79 & +26.93 & SABab & 10.9 & 5.7 & 7.7 & 1.5 & 25.3 & 0.017 & 71 & 210 & 0.50 & 61 \\
	NGC~2487 & 07.97 & +25.15 & SBb & 10.8 & 11.8 & 9.2 & 0.6 & 20.1 & 0.017 & 71 & 42 & 0.67 & 48 \\
	NGC~2530 & 08.13 & +17.82 & SABd & 10.2 & 9.2 & 9.0 & 0.8 & 20.9 & 0.017 & 71 & 107 & 0.80 & 38 \\
	NGC~2540 & 08.21 & +26.36 & SBbc & 10.5 & 7.9 & - & - & - & 0.022 & 89 & 131 & 0.71 & 45 \\
	NGC~2543 & 08.22 & +36.25 & SBbc & 10.3 & 9.5 & 8.9 & 0.3 & 19.1 & 0.009 & 39 & 48 & 0.60 & 54 \\
	NGC~2558 & 08.32 & +20.51 & SABb & 10.8 & 9.1 & 9.7 & 0.6 & 19.3 & 0.017 & 75 & -21 & 0.63 & 51 \\
	NGC~2565 & 08.33 & +22.03 & SBb & 10.7 & 5.6 & 9.9 & 0.4 & 17.9 & 0.013 & 48 & -13 & 0.46 & 63 \\
	NGC~2595 & 08.46 & +21.48 & SABc & 10.6 & 12.4 & 9.9 & 1.0 & 20.1 & 0.015 & 66 & -9 & 0.70 & 46 \\
	NGC~2604 & 08.56 & +29.54 & SBd & 9.7 & 3.7 & - & - & - & 0.008 & 31 & -9 & 0.88 & 29 \\
	NGC~2638 & 08.71 & +37.22 & SAb & 10.8 & 5.0 & - & - & - & 0.014 & 58 & 64 & 0.49 & 62 \\
	NGC~2639 & 08.73 & +50.21 & SAa & 11.2 & 4.2 & 9.9 & 1.0 & 19.7 & 0.012 & 44 & -46 & 0.51 & 60 \\
	NGC~2730 & 09.04 & +16.84 & SBcd & 10.1 & 7.0 & 8.7 & 0.7 & 21.7 & 0.014 & 53 & 65 & 0.64 & 51 \\
	NGC~2805 & 09.34 & +64.10 & SAc & 10.1 & 5.8 & 9.1 & 2.7 & 23.7 & 0.007 & 22 & 18 & 0.76 & 41 \\
	NGC~2906 & 09.54 & +08.44 & SAbc & 10.4 & 3.0 & 8.6 & 0.2 & 19.1 & 0.008 & 31 & 87 & 0.51 & 60 \\
	NGC~2916 & 09.58 & +21.71 & SAbc & 10.8 & 7.3 & 9.3 & 0.7 & 20.3 & 0.014 & 53 & 13 & 0.59 & 55 \\
	NGC~3057 & 10.09 & +80.29 & SBdm & 9.1 & 4.9 & - & - & - & 0.006 & 22 & 182 & 0.58 & 55 \\
	NGC~3106 & 10.07 & +31.19 & SAab & 11.2 & 12.5 & 9.7 & 2.5 & 21.5 & 0.022 & 89 & 141 & 0.93 & 22 \\
	NGC~3381 & 10.81 & +34.71 & SBd & 9.7 & 2.2 & 7.7 & 0.1 & 20.5 & 0.007 & 22 & 48 & 0.71 & 46 \\
	NGC~3614 & 11.31 & +45.75 & SABbc & 10.2 & 7.3 & 8.3 & 1.2 & 23.5 & 0.009 & 31 & 98 & 0.72 & 44 \\
	NGC~3687 & 11.47 & +29.51 & SBb & 10.3 & 3.9 & 8.5 & 0.2 & 19.2 & 0.010 & 35 & 156 & 0.92 & 24 \\
	NGC~3811 & 11.69 & +47.69 & SBbc & 10.4 & 4.6 & 9.3 & 0.3 & 18.4 & 0.012 & 44 & 13 & 0.62 & 52 \\
	NGC~3994 & 11.96 & +32.28 & SABbc & 10.4 & 3.4 & 9.0 & 0.3 & 18.8 & 0.012 & 44 & 9 & 0.47 & 63 \\
	NGC~4047 & 12.05 & +48.64 & SAbc & 10.7 & 4.4 & 9.3 & 1.3 & 21.7 & 0.013 & 48 & 99 & 0.79 & 38 \\
	NGC~4185 & 12.22 & +28.51 & SABbc & 10.7 & 8.6 & 8.9 & 0.5 & 20.7 & 0.015 & 53 & 169 & 0.64 & 51 \\
	NGC~4210 & 12.25 & +65.99 & SBb & 10.3 & 3.9 & 8.6 & 0.3 & 20.1 & 0.011 & 39 & 91 & 0.73 & 44 \\
	NGC~4470 & 12.49 & +07.82 & SAc & 10.0 & 2.6 & - & - & - & 0.009 & 31 & 3 & 0.66 & 50 \\
	NGC~4644 & 12.71 & +55.15 & SAb & 10.4 & 6.6 & 9.1 & 1.2 & 21.9 & 0.018 & 71 & 48 & 0.45 & 64 \\
	NGC~4961 & 13.10 & +27.73 & SBcd & 9.7 & 3.1 & 8.0 & 0.2 & 21.1 & 0.010 & 35 & 107 & 0.66 & 50 \\
	NGC~5000 & 13.16 & +28.91 & SBbc & 10.7 & 7.3 & 8.8 & 0.5 & 20.4 & 0.021 & 80 & 82 & 0.60 & 53 \\
	NGC~5016 & 13.20 & +24.10 & SAbc & 10.2 & 3.6 & - & - & - & 0.011 & 35 & 58 & 0.73 & 44 \\
	NGC~5056 & 13.27 & +30.95 & SABc & 10.5 & 7.8 & 8.9 & 1.1 & 22.0 & 0.021 & 80 & 180 & 0.59 & 55 \\
	NGC~5157 & 13.45 & +32.03 & SBab & 11.2 & 9.5 & 10.0 & 1.6 & 20.7 & 0.026 & 106 & 129 & 0.73 & 44 \\
	NGC~5205 & 13.50 & +62.51 & SBbc & 9.9 & 2.6 & 8.3 & 0.2 & 19.9 & 0.007 & 22 & 147 & 0.67 & 49 \\
	NGC~5218 & 13.54 & +62.77 & SBab & 10.7 & 3.8 & - & - & - & 0.011 & 39 & 91 & 0.56 & 57 \\
	NGC~5267 & 13.68 & +38.79 & SBab & 11.0 & 8.1 & 9.4 & 0.6 & 19.9 & 0.022 & 84 & 44 & 0.49 & 62 \\
	NGC~5320 & 13.84 & +41.37 & SABbc & 10.3 & 6.3 & 8.4 & 0.4 & 21.3 & 0.011 & 35 & 123 & 0.55 & 57 \\
	NGC~5376 & 13.92 & +59.51 & SABb & 10.5 & 2.7 & 8.7 & 0.3 & 20.0 & 0.009 & 26 & 65 & 0.62 & 53 \\
	NGC~5378 & 13.95 & +37.80 & SBb & 10.6 & 5.6 & 7.3 & 0.6 & 24.3 & 0.012 & 40 & 60 & 0.63 & 52 \\
	NGC~5406 & 14.01 & +38.92 & SBb & 11.3 & 8.9 & 9.8 & 0.6 & 19.0 & 0.019 & 75 & 68 & 0.88 & 28 \\
	NGC~5480 & 14.11 & +50.73 & SAcd & 10.1 & 2.4 & 8.6 & 0.2 & 19.4 & 0.008 & 26 & 159 & 0.67 & 48 \\
	NGC~5519 & 14.24 & +07.52 & SBb & 10.8 & 17.8 & 9.2 & 1.8 & 22.1 & 0.027 & 107 & 89 & 0.71 & 45 \\
	NGC~5520 & 14.21 & +50.35 & SAbc & 9.9 & 2.6 & - & - & - & 0.008 & 26 & 66 & 0.57 & 56 \\
	NGC~5525 & 14.26 & +14.28 & SAa & 11.3 & 9.3 & - & - & - & 0.021 & 89 & 23 & 0.59 & 54 \\
	NGC~5533 & 14.27 & +35.34 & SAab & 11.2 & 11.8 & 10.1 & 2.3 & 20.9 & 0.015 & 53 & 30 & 0.61 & 53 \\
	NGC~5622 & 14.44 & +48.56 & SAbc & 10.2 & 5.5 & 8.4 & 0.5 & 21.7 & 0.015 & 53 & -9 & 0.50 & 61 \\
	NGC~5656 & 14.51 & +35.32 & SAb & 10.6 & 3.5 & 9.0 & 0.3 & 18.9 & 0.013 & 44 & 56 & 0.63 & 52 \\
	NGC~5657 & 14.51 & +29.18 & SBbc & 10.3 & 6.2 & 9.4 & 0.7 & 20.1 & 0.015 & 57 & 182 & 0.53 & 59 \\
	NGC~5665 & 14.54 & +08.08 & SABc & 10.2 & 3.7 & 9.0 & 0.5 & 20.5 & 0.009 & 31 & 158 & 0.84 & 34 \\
	NGC~5720 & 14.64 & +50.82 & SBbc & 10.8 & 11.6 & 9.6 & 0.6 & 19.3 & 0.028 & 111 & 129 & 0.65 & 50 \\
	NGC~5732 & 14.68 & +38.64 & SAbc & 9.9 & 5.6 & 8.4 & 0.3 & 20.9 & 0.014 & 53 & 40 & 0.58 & 55 \\
	NGC~5735 & 14.71 & +28.73 & SBbc & 10.4 & 6.8 & 8.5 & 0.3 & 20.6 & 0.015 & 53 & 87 & 0.83 & 34 \\
	NGC~5772 & 14.86 & +40.60 & SAab & 11.0 & 8.5 & 9.7 & 0.6 & 19.1 & 0.018 & 71 & 39 & 0.57 & 56 \\
	NGC~5888 & 15.22 & +41.26 & SBb & 11.2 & 10.8 & 9.7 & 1.0 & 20.1 & 0.031 & 125 & 158 & 0.54 & 58 \\
	NGC~5957 & 15.59 & +12.05 & SBb & 10.3 & 3.7 & 8.6 & 0.4 & 20.6 & 0.008 & 26 & 89 & 0.75 & 42 \\
	NGC~5971 & 15.59 & +56.46 & SABb & 10.3 & 5.6 & 9.4 & 2.8 & 22.7 & 0.016 & 48 & 128 & 0.49 & 61 \\
	NGC~6004 & 15.84 & +18.94 & SBbc & 10.7 & 6.6 & 8.7 & 0.2 & 19.1 & 0.015 & 53 & 73 & 0.94 & 20 \\
	NGC~6063 & 16.12 & +07.98 & SAbc & 10.1 & 5.2 & 7.5 & 0.3 & 22.5 & 0.011 & 39 & 152 & 0.60 & 53 \\
	NGC~6154 & 16.43 & +49.84 & SBab & 10.9 & 9.0 & 9.7 & 1.0 & 20.4 & 0.022 & 88 & 134 & 0.65 & 50 \\
	NGC~6155 & 16.44 & +48.37 & SAc & 10.1 & 2.9 & 8.4 & 0.2 & 19.5 & 0.010 & 35 & 149 & 0.68 & 47 \\
	NGC~6301 & 17.14 & +42.34 & SAbc & 11.0 & 15.1 & 9.1 & 0.7 & 21.0 & 0.029 & 120 & 109 & 0.60 & 54 \\
	NGC~6314 & 17.21 & +23.27 & SAab & 11.2 & 8.8 & 10.5 & 1.5 & 19.4 & 0.024 & 97 & 176 & 0.51 & 60 \\
	NGC~6941 & 20.61 & -04.62 & SBb & 10.9 & 11.6 & 9.6 & 0.8 & 20.0 & 0.022 & 89 & 122 & 0.73 & 44 \\
	NGC~7311 & 22.57 & +05.57 & SAa & 11.1 & 5.7 & 10.1 & 0.7 & 18.7 & 0.015 & 66 & 13 & 0.49 & 61 \\
	NGC~7321 & 22.61 & +21.62 & SBbc & 10.9 & 9.6 & 9.5 & 0.4 & 18.6 & 0.024 & 102 & 31 & 0.69 & 47 \\
	NGC~7364 & 22.74 & -00.16 & SAab & 10.9 & 6.0 & 10.5 & 6.3 & 22.2 & 0.016 & 71 & 65 & 0.65 & 50 \\
	NGC~7466 & 23.03 & +27.05 & SAbc & 10.8 & 12.3 & 9.4 & 1.8 & 21.6 & 0.025 & 107 & 201 & 0.53 & 59 \\
	NGC~7489 & 23.13 & +23.00 & SAbc & 10.5 & 10.6 & - & - & - & 0.021 & 89 & 165 & 0.55 & 58 \\
	NGC~7549 & 23.25 & +19.04 & SBbc & 10.6 & 8.8 & 10.0 & 2.8 & 21.8 & 0.016 & 66 & 140 & 0.75 & 42 \\
	NGC~7591 & 23.30 & +06.59 & SBbc & 10.8 & 8.4 & 10.0 & 1.8 & 20.9 & 0.016 & 71 & 160 & 0.59 & 55 \\
	NGC~7631 & 23.36 & +08.22 & SAb & 10.5 & 7.0 & 9.3 & 3.0 & 23.1 & 0.013 & 53 & 76 & 0.44 & 65 \\
	NGC~7653 & 23.41 & +15.28 & SAb & 10.5 & 6.3 & 9.4 & 1.5 & 21.4 & 0.014 & 62 & 163 & 0.88 & 28 \\
	NGC~7716 & 23.61 & +00.30 & SAb & 10.4 & 4.2 & 9.3 & 0.3 & 18.5 & 0.009 & 35 & 27 & 0.71 & 45 \\
	NGC~7722 & 23.64 & +15.95 & SAab & 11.2 & 9.1 & 10.1 & 1.4 & 20.1 & 0.013 & 57 & 144 & 0.87 & 29 \\
	NGC~7782 & 23.90 & +07.97 & SAb & 11.1 & 10.7 & 9.7 & 0.8 & 19.9 & 0.018 & 75 & 178 & 0.55 & 57 \\
	NGC~7787 & 23.94 & +00.55 & SABab & 10.6 & 8.8 & 9.4 & 2.3 & 22.5 & 0.022 & 97 & 3 & 0.63 & 51 \\
	NGC~7819 & 00.07 & +31.47 & SAc & 10.4 & 8.4 & 9.6 & 0.8 & 20.0 & 0.016 & 71 & 87 & 0.53 & 59 \\
	NGC~7824 & 00.09 & +06.92 & SAab & 11.2 & 11.9 & 10.2 & 1.5 & 19.9 & 0.020 & 88 & 160 & 0.75 & 42 \\
	UGC~00005 & 00.05 & -01.91 & SAbc & 10.8 & 10.0 & 8.9 & 0.5 & 20.6 & 0.024 & 106 & 235 & 0.54 & 58 \\
	UGC~00036 & 00.09 & +06.77 & SABab & 11.0 & 8.7 & 9.4 & 0.5 & 19.4 & 0.021 & 89 & 9 & 0.60 & 54 \\
	UGC~01918 & 02.46 & +25.67 & SBb & 10.7 & 8.0 & 8.9 & 0.3 & 19.5 & 0.017 & 71 & -60 & 0.54 & 58 \\
	UGC~02311 & 02.82 & -00.87 & SBbc & 10.7 & 8.2 & 9.7 & 0.6 & 19.0 & 0.023 & 102 & 43 & 0.55 & 58 \\
	UGC~02443 & 02.97 & -02.04 & SAcd & 9.6 & 3.4 & 7.3 & 0.3 & 22.6 & 0.008 & 35 & -16 & 0.55 & 58 \\
	UGC~03944 & 07.64 & +37.63 & SABbc & 10.0 & 5.6 & - & - & - & 0.014 & 53 & -56 & 0.49 & 62 \\
	UGC~03973 & 07.71 & +49.81 & SBbc & 10.8 & 7.7 & 9.5 & 0.3 & 17.0 & 0.023 & 97 & 105 & 0.54 & 58 \\
	UGC~03995 & 07.74 & +29.25 & SBb & 10.9 & 12.8 & 9.6 & 0.7 & 19.7 & 0.016 & 66 & 103 & 0.46 & 64 \\
	UGC~04145 & 07.99 & +15.39 & SABa & 11.0 & 7.6 & 10.1 & 0.5 & 17.8 & 0.016 & 88 & 133 & 0.47 & 63 \\
	UGC~04195 & 08.09 & +66.78 & SBb & 10.5 & 7.8 & 8.9 & 0.4 & 20.2 & 0.017 & 71 & 112 & 0.62 & 53 \\
	UGC~04262 & 08.32 & +83.27 & SABbc & 10.6 & 11.6 & 9.6 & 1.0 & 20.4 & 0.020 & 80 & 154 & 0.71 & 46 \\
	UGC~04308 & 08.29 & +21.69 & SBc & 10.3 & 6.0 & 8.5 & 0.3 & 19.8 & 0.013 & 48 & 111 & 0.76 & 41 \\
	UGC~05108 & 09.59 & +29.81 & SBb & 10.9 & 17.2 & - & - & - & 0.028 & 116 & 135 & 0.77 & 40 \\
	UGC~05520 & 10.25 & +65.14 & SBcd & 9.8 & 5.7 & 9.5 & 0.9 & 20.6 & 0.012 & 52 & 99 & 0.48 & 62 \\
	UGC~06312 & 11.30 & +07.84 & SAab & 11.0 & 13.2 & 10.2 & 1.7 & 20.3 & 0.023 & 93 & 48 & 0.45 & 64 \\
	UGC~07012 & 12.03 & +29.85 & SABcd & 9.4 & 4.5 & 8.6 & 0.7 & 22.0 & 0.012 & 44 & 13 & 0.54 & 58 \\
	UGC~08733 & 13.81 & +43.41 & SBdm & 9.4 & 4.5 & - & - & - & 0.010 & 31 & 187 & 0.49 & 61 \\
	UGC~08781 & 13.87 & +21.54 & SBb & 11.1 & 16.6 & 9.8 & 0.9 & 19.8 & 0.027 & 111 & 171 & 0.52 & 59 \\
	UGC~09110 & 14.24 & +15.62 & SABb & 10.2 & 9.0 & 9.2 & 0.3 & 18.7 & 0.018 & 76 & 21 & 0.45 & 64 \\
	UGC~09291 & 14.48 & +39.00 & SAcd & 10.3 & 6.1 & 7.8 & 0.4 & 22.4 & 0.012 & 39 & 107 & 0.52 & 59 \\
	UGC~09476 & 14.69 & +44.51 & SAbc & 10.2 & 4.7 & 8.4 & 0.4 & 21.1 & 0.013 & 44 & 117 & 0.63 & 51 \\
    UGC~09777 & 15.24 & +20.48 & SAbc & 10.3 & 6.7 & 9.0 & 1.5 & 22.6 & 0.018 & 66 & 145 & 0.60 & 54 \\
	UGC~09842 & 15.42 & +37.96 & SBbc & 10.7 & 13.7 & 9.6 & 2.8 & 22.8 & 0.032 & 129 & -12 & 0.50 & 61 \\
	UGC~10796 & 17.28 & +61.92 & SABcd & 9.4 & 4.8 & 9.3 & 2.1 & 22.5 & 0.012 & 44 & 114 & 0.49 & 62 \\
	UGC~11649 & 20.92 & -01.23 & SBab & 10.6 & 5.7 & 9.0 & 0.7 & 20.9 & 0.013 & 53 & 94 & 0.88 & 29 \\
	UGC~11680NED01 & 21.13 & +03.87 & SBb & 11.1 & 14.0 & - & - & - & 0.026 & 113 & 101 & 0.77 & 40 \\
	UGC~12185 & 22.79 & +31.37 & SBb & 10.7 & 8.2 & 9.3 & 0.7 & 20.1 & 0.022 & 93 & 145 & 0.47 & 63 \\
	UGC~12224 & 22.88 & +06.09 & SAc & 10.1 & 7.5 & 8.6 & 0.5 & 21.1 & 0.012 & 48 & 31 & 0.83 & 34 \\
	UGC~12633 & 23.50 & +15.76 & SABab & 10.3 & 6.8 & - & - & - & 0.014 & 57 & 146 & 0.69 & 47 \\
	UGC~12767 & 23.75 & +07.04 & SBb & 11.0 & 12.3 & - & - & - & 0.017 & 75 & 34 & 0.87 & 29 \\
	UGC~12816 & 23.86 & +03.08 & SAc & 9.8 & 8.6 & 9.6 & 2.6 & 22.6 & 0.018 & 75 & 145 & 0.62 & 52 \\
    \end{longtable}
\end{center}

\setcounter{section}{1}


\section{Profiles of abundance gradients without DIG contamination}
\label{ap02}

In this appendix, we present graphs with fits of the oxygen abundance gradients for the 32 galaxies that exhibited an inner drop in at least one of the six H{\sc ii} region selection criteria adopted in this study, after DIG decontamination. An exception is NGC~5378, excluded from the sample by the criterion of \citet{Sanchez2014} due to having fewer than 10 H{\sc ii} regions, resulting in the absence of a fitted profile for this criterion.

\begin{flushleft}
\includegraphics[width=16.25cm, height=21cm]{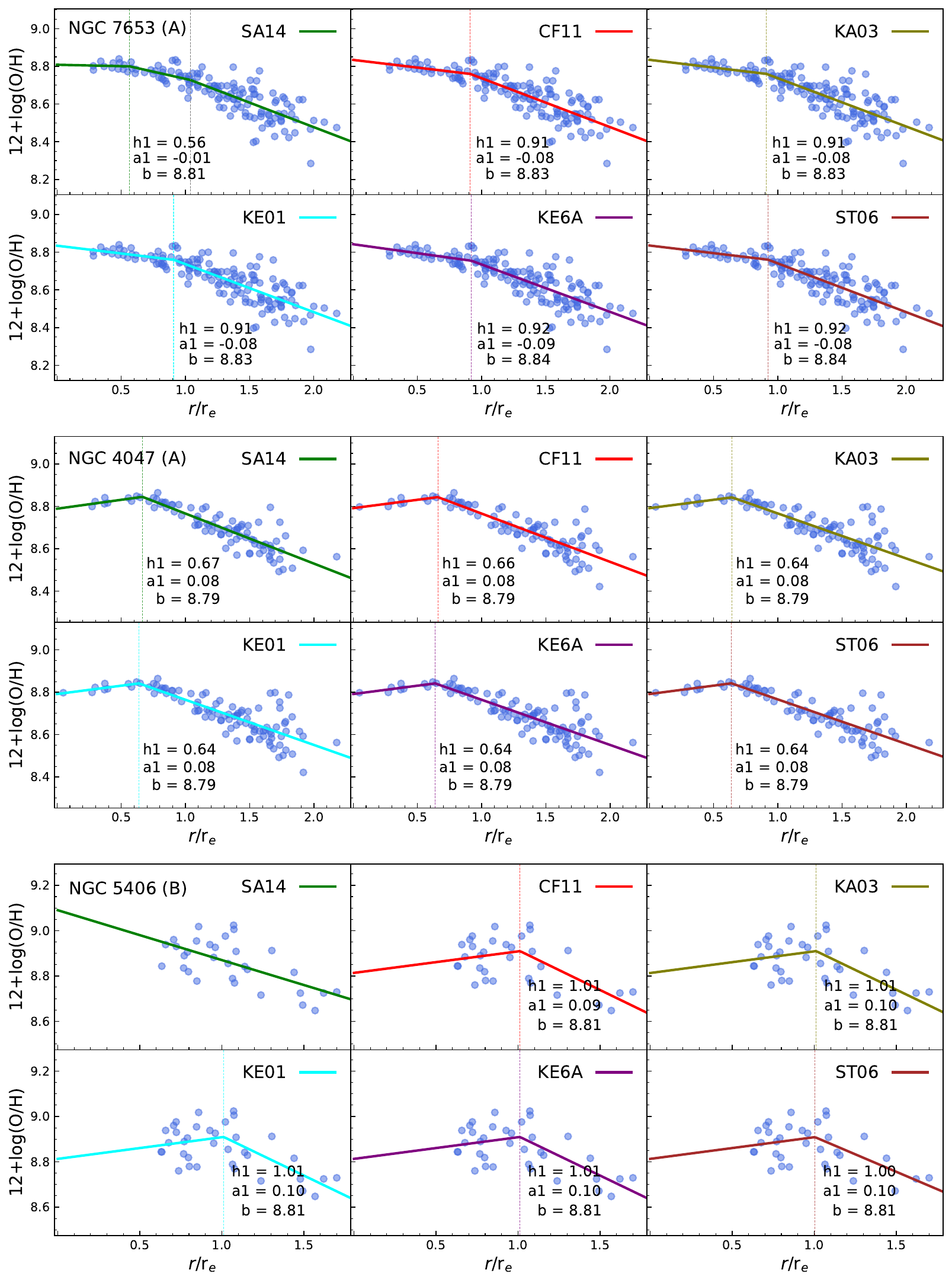}\\
\includegraphics[width=16.25cm, height=21cm]{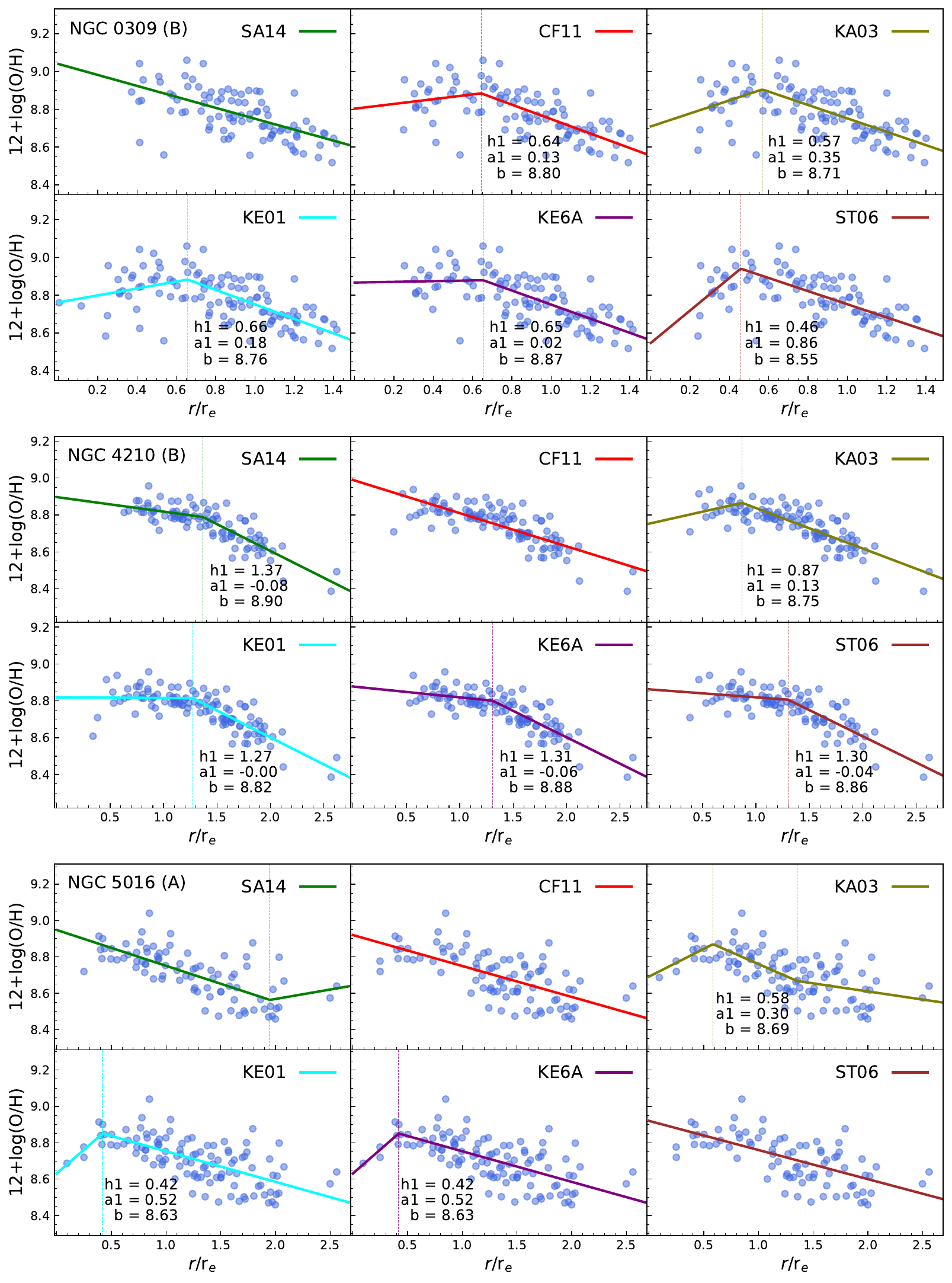}\\
\includegraphics[width=16.25cm, height=21cm]{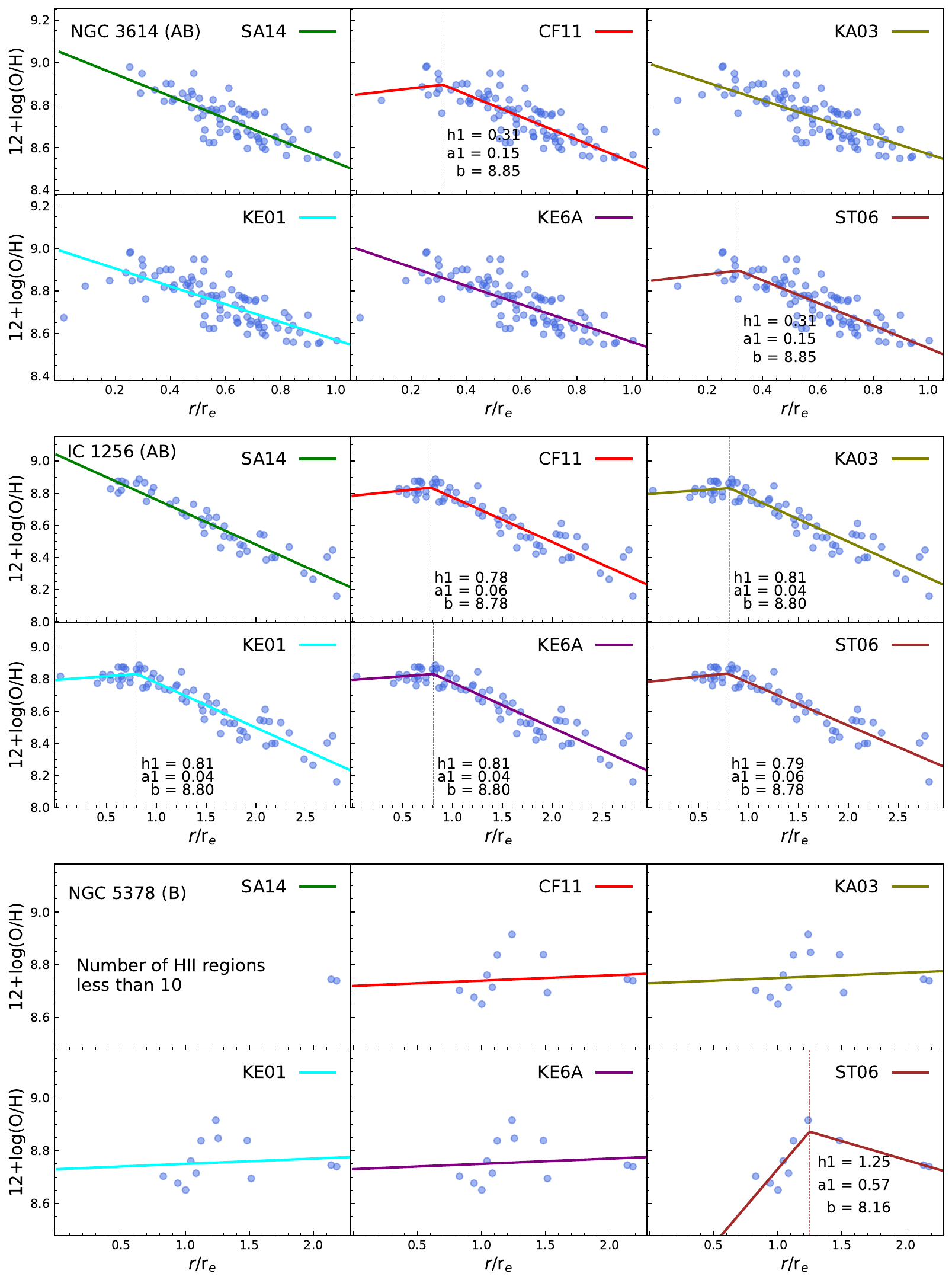}\\
\includegraphics[width=16.25cm, height=21cm]{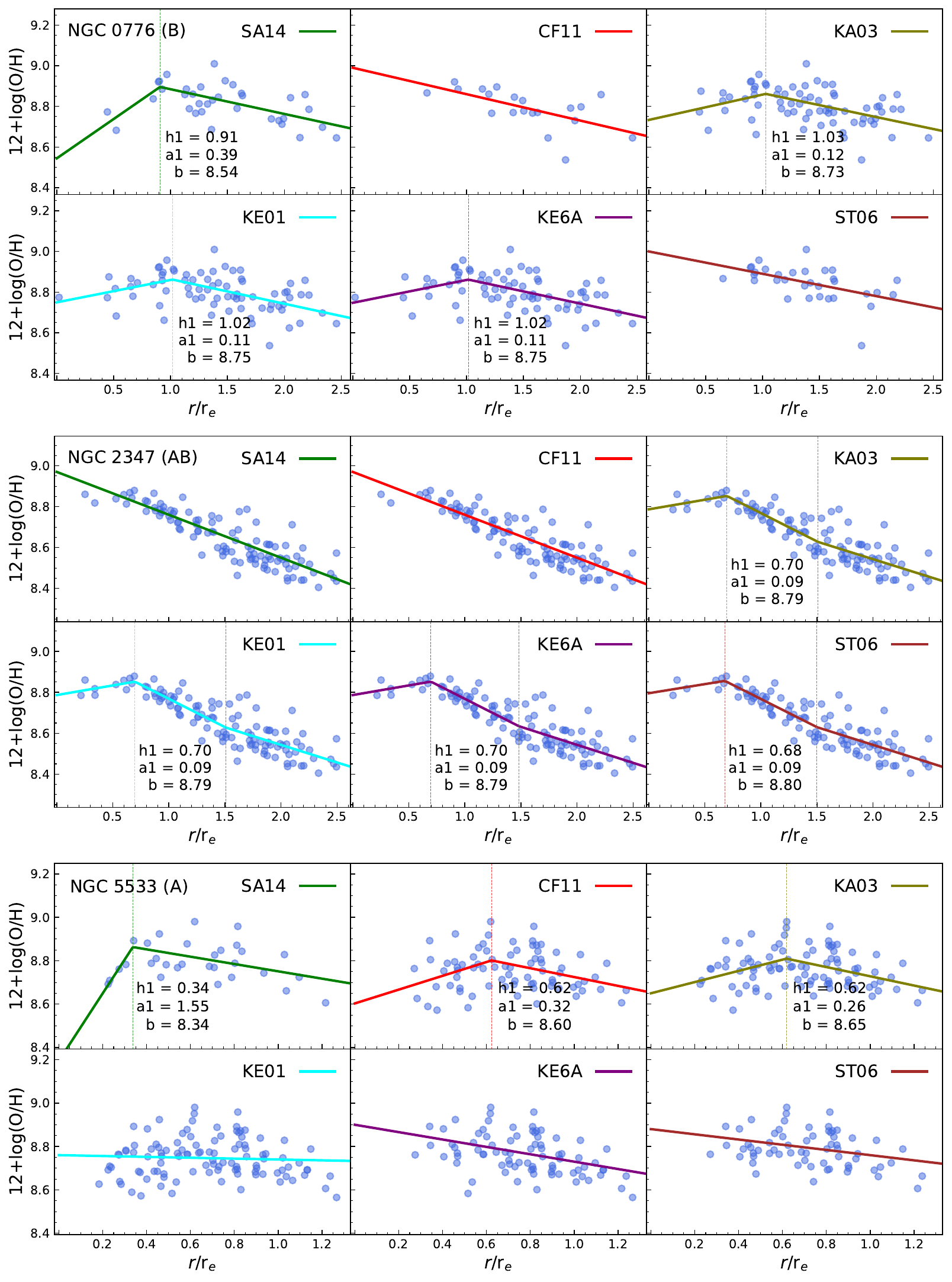}\\
\includegraphics[width=16.25cm, height=21cm]{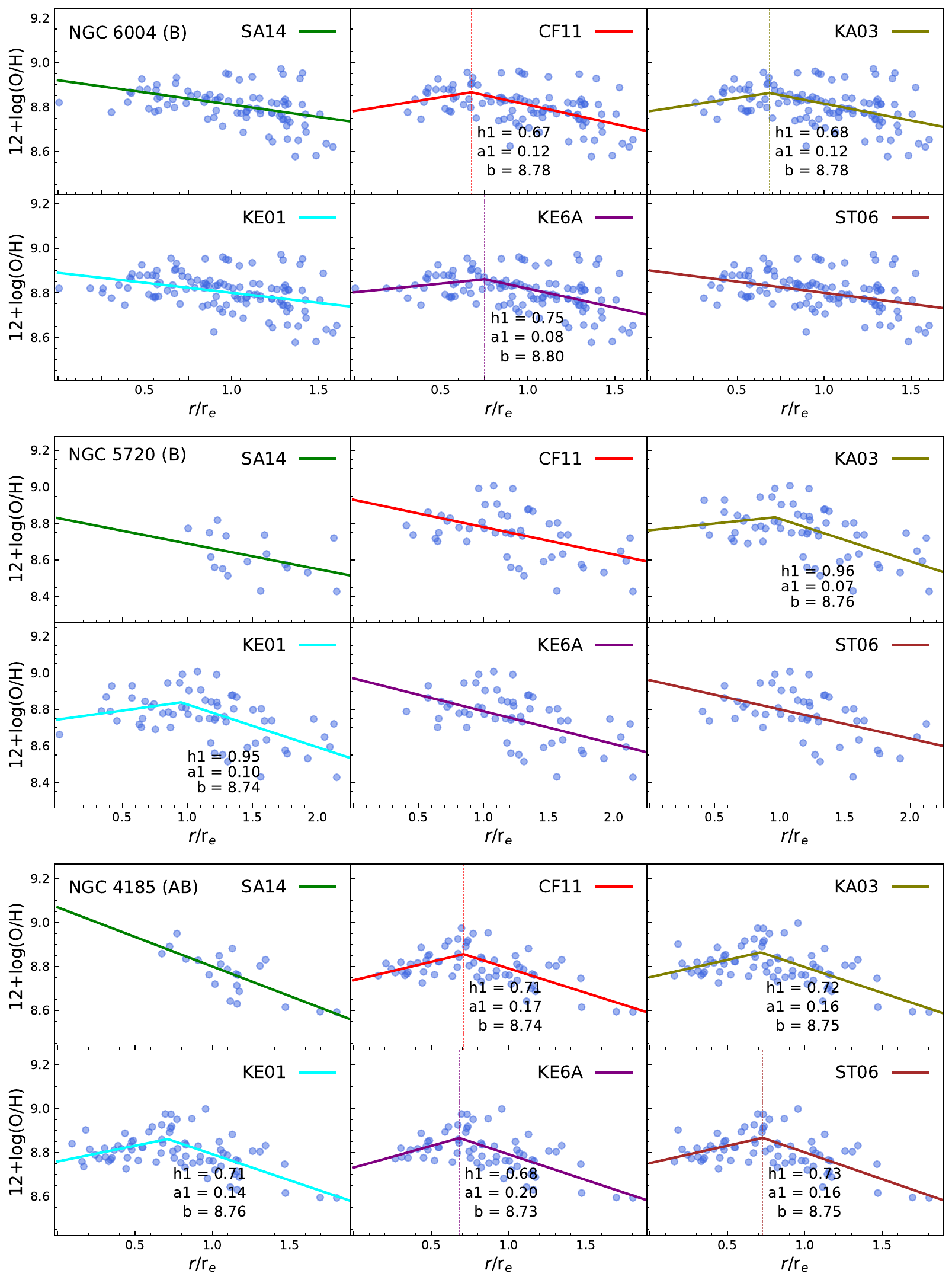}\\
\includegraphics[width=16.25cm, height=21cm]{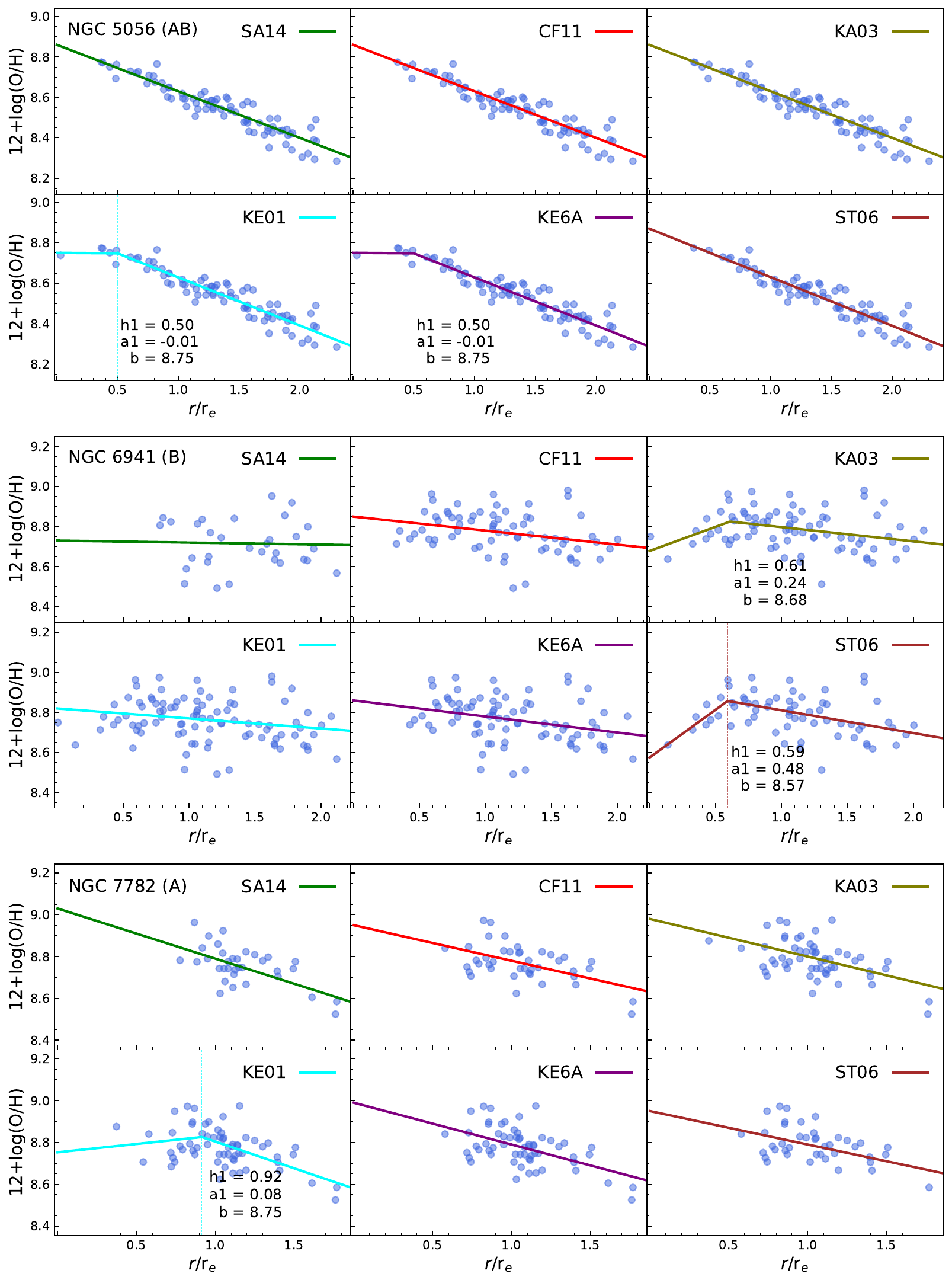}\\
\includegraphics[width=16.25cm, height=21cm]{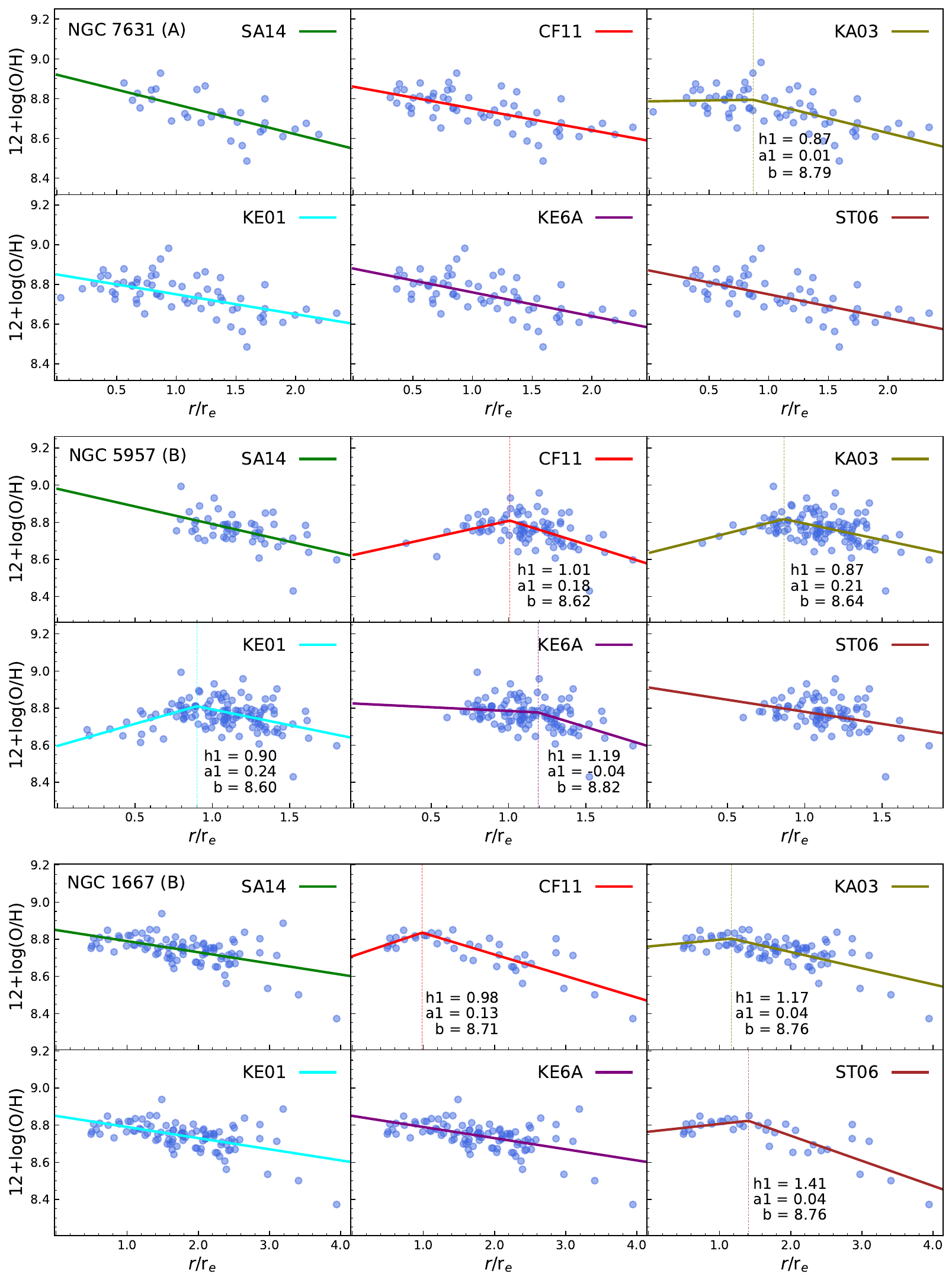}\\
\includegraphics[width=16.25cm, height=21cm]{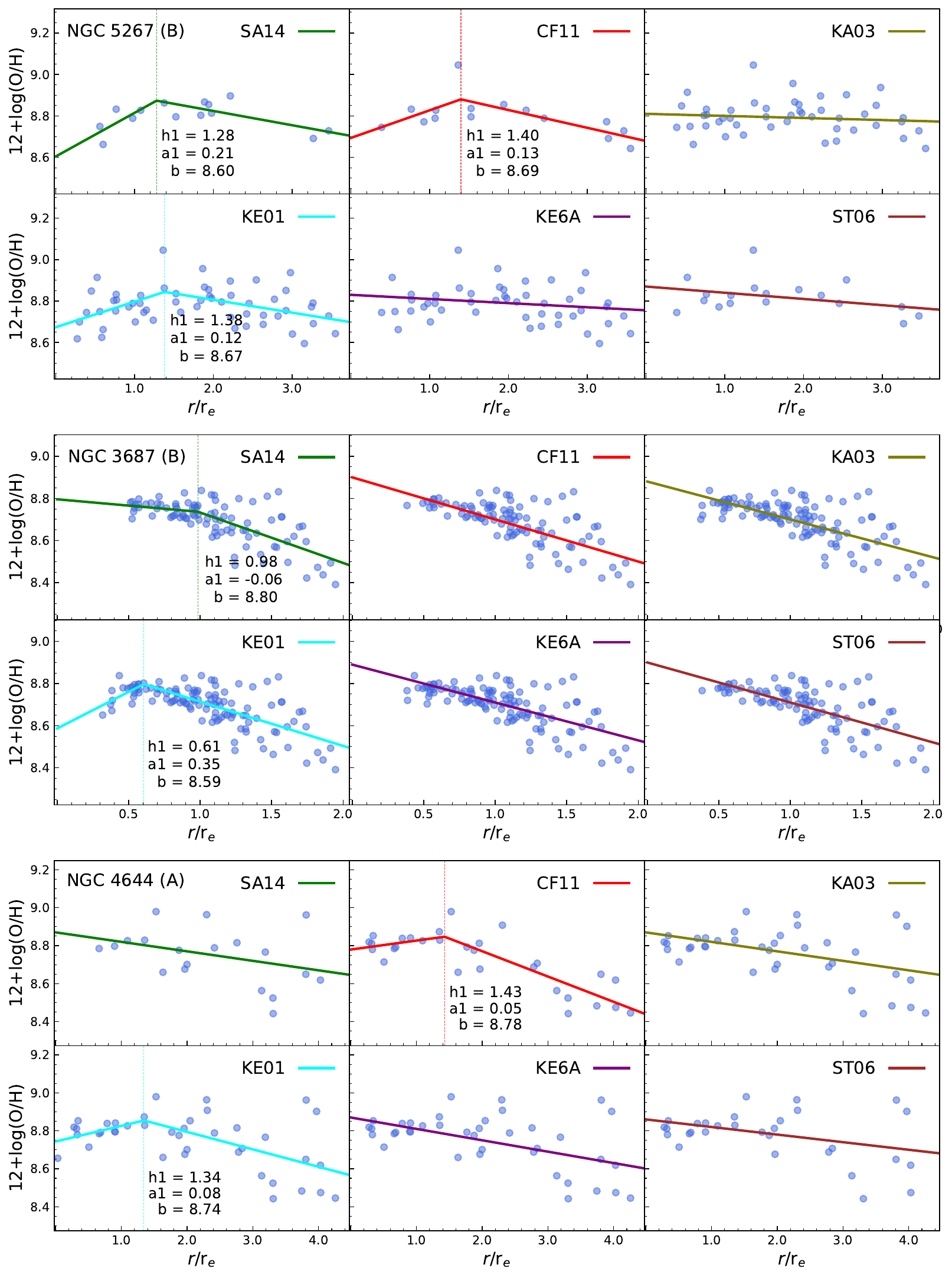}\\
\includegraphics[width=16.25cm, height=21cm]{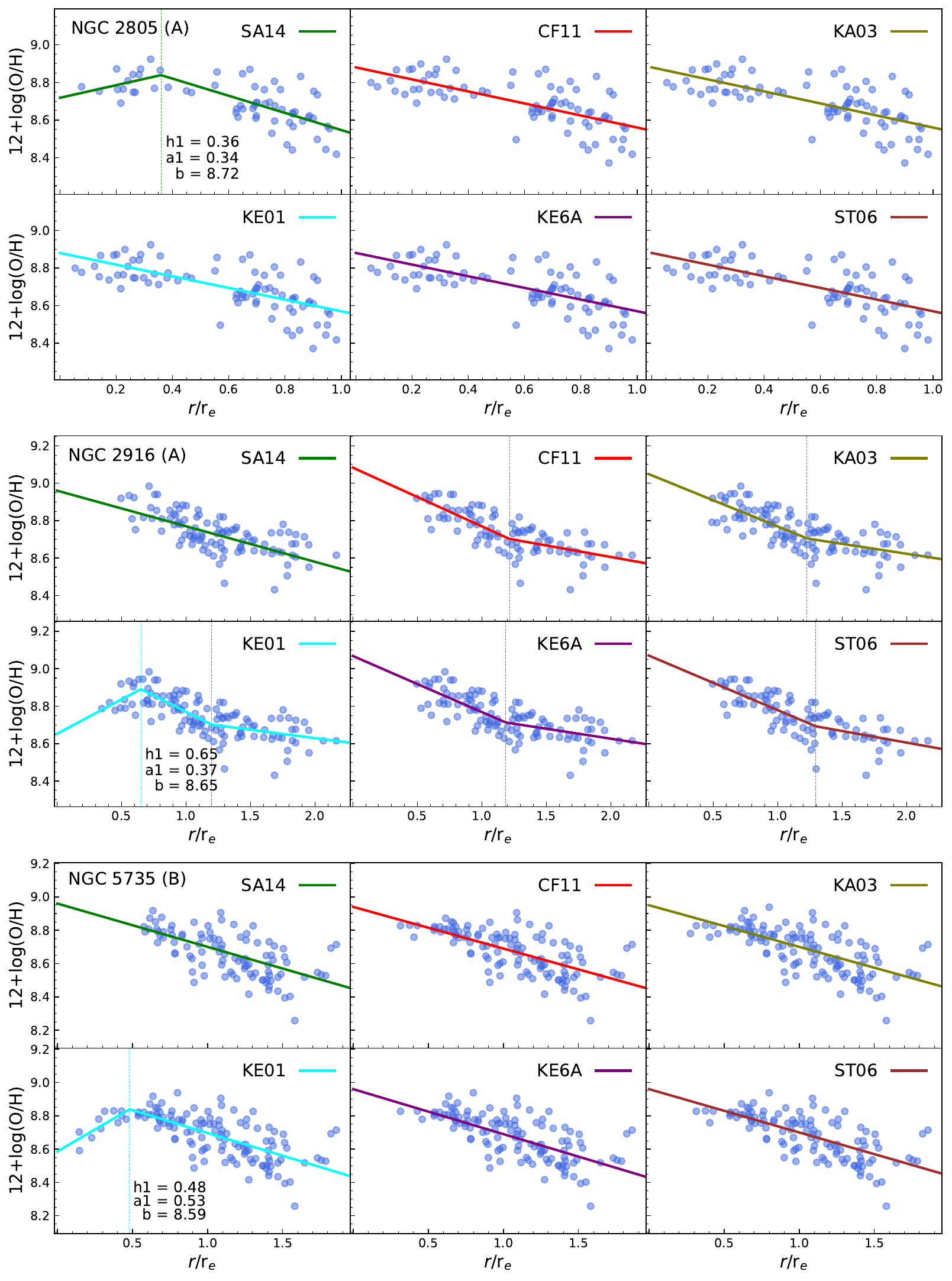}\\
\includegraphics[width=16.25cm, height=21cm]{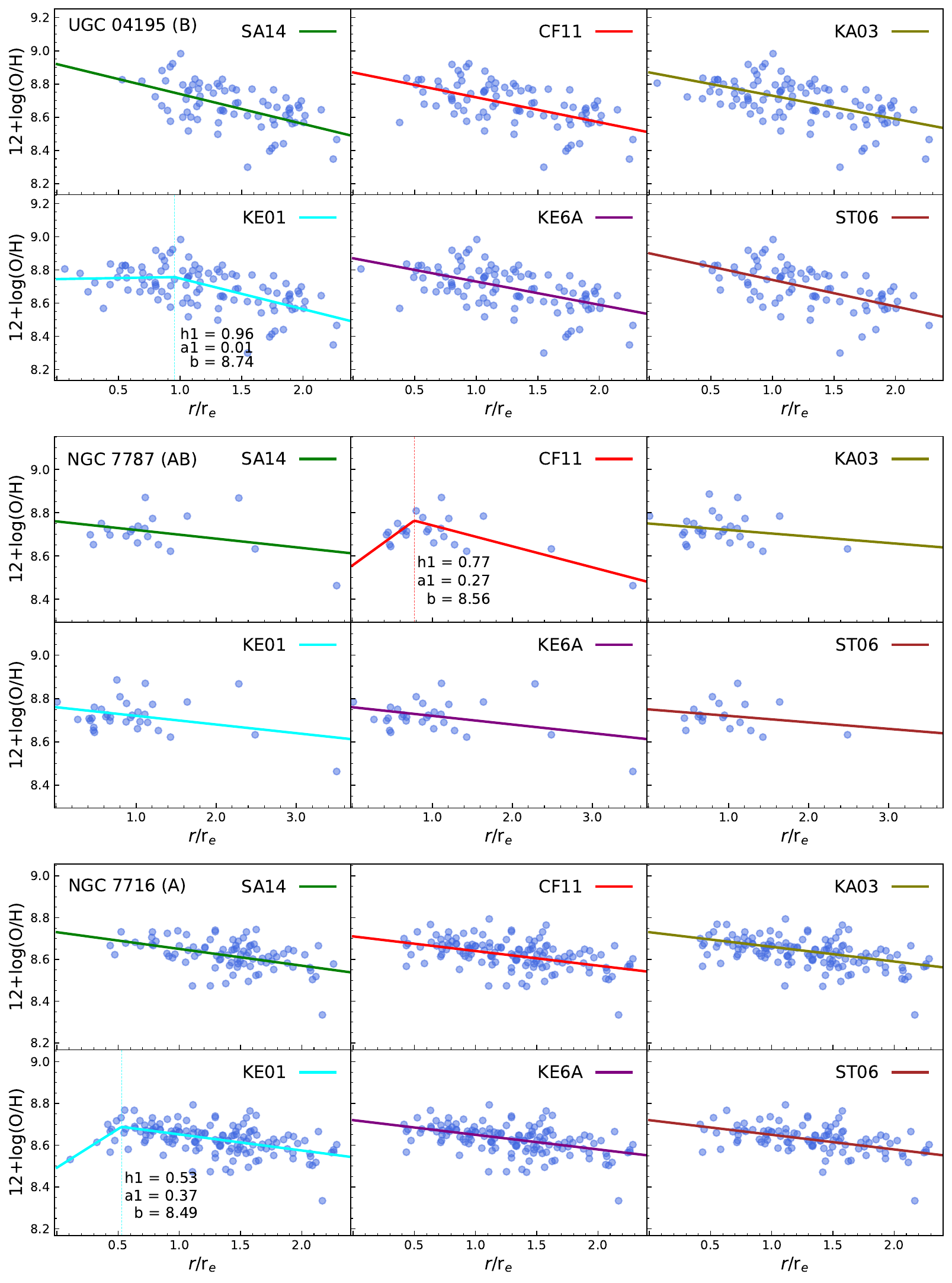}\\
\includegraphics[width=16.25cm, height=14cm]{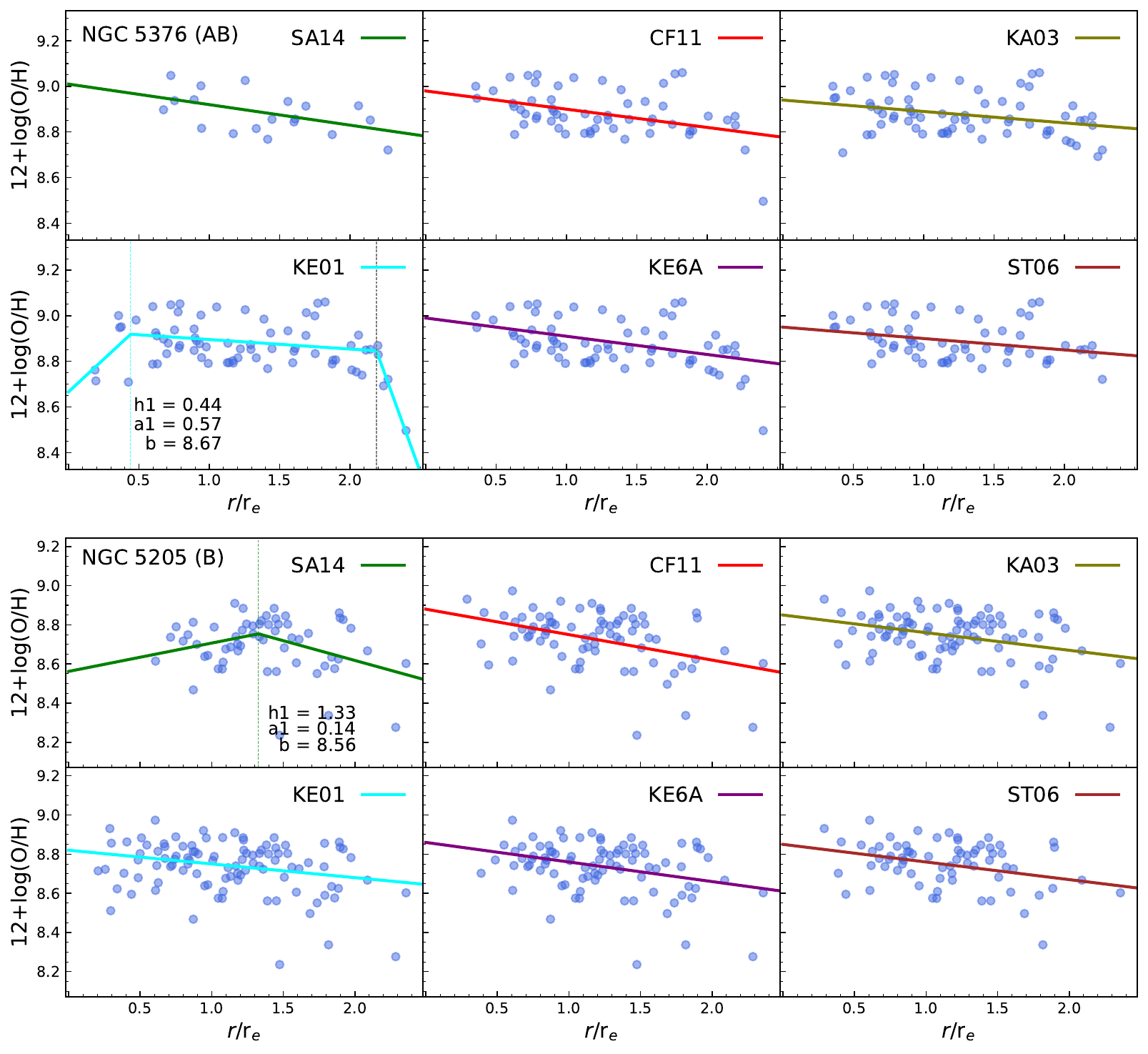}
\end{flushleft}

\bibliography{bib_ApJ}{}

\begin{thebibliography}{}
\expandafter\ifx\csname natexlab\endcsname\relax\def\natexlab#1{#1}\fi
\providecommand{\url}[1]{\href{#1}{#1}}
\providecommand{\dodoi}[1]{doi:~\href{http://doi.org/#1}{\nolinkurl{#1}}}
\providecommand{\doeprint}[1]{\href{http://ascl.net/#1}{\nolinkurl{http://ascl.net/#1}}}
\providecommand{\doarXiv}[1]{\href{https://arxiv.org/abs/#1}{\nolinkurl{https://arxiv.org/abs/#1}}}

\bibitem[{Akaike(1973)}]{akaike.test}
Akaike, H. 1973, Akad{\'e}miai Kiad{\'o}, Budapest

\bibitem[{{Aller}(1984)}]{aller_1984}
{Aller}, L.~H. 1984, {Physics of thermal gaseous nebulae}, Vol. 112 (Springer
  Dordrecht (Dordrecht, Holland: D. Reidel Publishing Company)),
  \dodoi{https://doi.org/10.1007/978-94-010-9639-3}

\bibitem[{Alloin {et~al.}(1979)Alloin, Collin-Souffrin, Joly, \&
  Vigroux}]{Alloin1979}
Alloin, D., Collin-Souffrin, S., Joly, M., \& Vigroux, L. 1979, \aap, 78, 200

\bibitem[{{Andrae}(2010)}]{andrea.bootstrap}
{Andrae}, R. 2010, arXiv e-prints, 1009.2755, \dodoi{10.48550/arXiv.1009.2755}

\bibitem[{Baldwin {et~al.}(1981)Baldwin, Phillips, \& Terlevich}]{Baldwin1981}
Baldwin, J.~A., Phillips, M.~M., \& Terlevich, R. 1981, \pasp, 93, 5,
  \dodoi{10.1086/130766}

\bibitem[{{Beifiori} {et~al.}(2012){Beifiori}, {Courteau}, {Corsini}, \&
  {Zhu}}]{beifiori.2012}
{Beifiori}, A., {Courteau}, S., {Corsini}, E.~M., \& {Zhu}, Y. 2012, \mnras,
  419, 2497, \dodoi{10.1111/j.1365-2966.2011.19903.x}

\bibitem[{{Belfiore} {et~al.}(2015){Belfiore}, {Maiolino}, {Bundy}, {Thomas},
  {Maraston}, {Wilkinson}, {S{\'a}nchez}, {Bershady}, {Blanc}, {Bothwell},
  {Cales}, {Coccato}, \& {et al.}}]{belfiore.2015}
{Belfiore}, F., {Maiolino}, R., {Bundy}, K., {et~al.} 2015, \mnras, 449, 867,
  \dodoi{10.1093/mnras/stv296}

\bibitem[{Belfiore {et~al.}(2022)Belfiore, Santoro, Groves, Schinnerer,
  Kreckel, Glover, Klessen, Emsellem, Blanc, Congiu, Barnes, Boquien, Chevance,
  Dale, \& Kruijssen}]{belfiore.2022}
Belfiore, F., Santoro, F., Groves, B., {et~al.} 2022, \aap, 659, A26,
  \dodoi{10.1051/0004-6361/202141859}

\bibitem[{{Belley} \& {Roy}(1992)}]{belley.roy.1992}
{Belley}, J., \& {Roy}, J.~R. 1992, \apjs, 78, 61, \dodoi{10.1086/191621}

\bibitem[{{Berg} {et~al.}(2020){Berg}, {Pogge}, {Skillman}, {Croxall},
  {Moustakas}, {Rogers}, \& {Sun}}]{berg2020}
{Berg}, D.~A., {Pogge}, R.~W., {Skillman}, E.~D., {et~al.} 2020, \apj, 893, 96,
  \dodoi{10.3847/1538-4357/ab7eab}

\bibitem[{{Bilitewski} \& {Sch{\"o}nrich}(2012)}]{bilitewski_2012}
{Bilitewski}, T., \& {Sch{\"o}nrich}, R. 2012, \mnras, 426, 2266,
  \dodoi{10.1111/j.1365-2966.2012.21827.x}

\bibitem[{{Binney} \& {Merrifield}(1998)}]{binney1998}
{Binney}, J., \& {Merrifield}, M. 1998, {Galactic Astronomy} (Princeton, NJ :
  Princeton University Press, 1998. (Princeton series in astrophysics))

\bibitem[{{Bird} {et~al.}(2012){Bird}, {Kazantzidis}, \&
  {Weinberg}}]{bird_2012}
{Bird}, J.~C., {Kazantzidis}, S., \& {Weinberg}, D.~H. 2012, \mnras, 420, 913,
  \dodoi{10.1111/j.1365-2966.2011.19728.x}

\bibitem[{{Bresolin} {et~al.}(2009{\natexlab{a}}){Bresolin}, {Gieren},
  {Kudritzki}, {Pietrzy{\'n}ski}, {Urbaneja}, \& {Carraro}}]{bresolin2009}
{Bresolin}, F., {Gieren}, W., {Kudritzki}, R.-P., {et~al.} 2009{\natexlab{a}},
  \apj, 700, 309, \dodoi{10.1088/0004-637X/700/1/309}

\bibitem[{{Bresolin} {et~al.}(2012){Bresolin}, {Kennicutt}, \&
  {Ryan-Weber}}]{bresolin.2012}
{Bresolin}, F., {Kennicutt}, R.~C., \& {Ryan-Weber}, E. 2012, \apj, 750, 122,
  \dodoi{10.1088/0004-637X/750/2/122}

\bibitem[{{Bresolin} {et~al.}(2009{\natexlab{b}}){Bresolin}, {Ryan-Weber},
  {Kennicutt}, \& {Goddard}}]{bresolin.2009}
{Bresolin}, F., {Ryan-Weber}, E., {Kennicutt}, R.~C., \& {Goddard}, Q.
  2009{\natexlab{b}}, \apj, 695, 580, \dodoi{10.1088/0004-637X/695/1/580}

\bibitem[{{Cavichia} {et~al.}(2010){Cavichia}, Costa, \& Maciel}]{cavichia2010}
{Cavichia}, O., Costa, R.~D.~D., \& Maciel, W.~J. 2010, \rmxaa, 46, 159,
  \dodoi{10.48550/arXiv.1003.0416}

\bibitem[{{Cavichia} {et~al.}(2023){Cavichia}, {Moll{\'a}}, \&
  {Baz{\'a}n}}]{cavichia2023}
{Cavichia}, O., {Moll{\'a}}, M., \& {Baz{\'a}n}, J.~J. 2023, \mnras, 520, 402,
  \dodoi{10.1093/mnras/stad097}

\bibitem[{Cavichia {et~al.}(2014)Cavichia, Mollá, Costa, \&
  Maciel}]{Cavichia2014}
Cavichia, O., Mollá, M., Costa, R. D.~D., \& Maciel, W.~J. 2014, \mnras, 437,
  3688, \dodoi{10.1093/mnras/stt2164}

\bibitem[{{Cid Fernandes} {et~al.}(2011){Cid Fernandes}, {Stasi{\'n}ska},
  {Mateus}, \& {Vale Asari}}]{cidfernandes2011}
{Cid Fernandes}, R., {Stasi{\'n}ska}, G., {Mateus}, A., \& {Vale Asari}, N.
  2011, \mnras, 413, 1687, \dodoi{10.1111/j.1365-2966.2011.18244.x}

\bibitem[{{Collins} \& {Rand}(2001)}]{collins2001}
{Collins}, J.~A., \& {Rand}, R.~J. 2001, \apj, 551, 57, \dodoi{10.1086/320072}

\bibitem[{{Daniel} \& {Wyse}(2015)}]{daniel_2015}
{Daniel}, K.~J., \& {Wyse}, R.~F.~G. 2015, \mnras, 447, 3576,
  \dodoi{10.1093/mnras/stu2683}

\bibitem[{{Dav{\'e}} {et~al.}(2012){Dav{\'e}}, {Finlator}, \&
  {Oppenheimer}}]{dave_2012}
{Dav{\'e}}, R., {Finlator}, K., \& {Oppenheimer}, B. 2012, \mnras, 421, 98,
  \dodoi{10.1111/j.1365-2966.2011.20148.x}

\bibitem[{{Dav{\'e}} {et~al.}(2011){Dav{\'e}}, {Finlator}, \&
  {Oppenheimer}}]{dave_2011}
{Dav{\'e}}, R., {Finlator}, K., \& {Oppenheimer}, B.~D. 2011, \mnras, 416,
  1354, \dodoi{10.1111/j.1365-2966.2011.19132.x}

\bibitem[{Davison \& Hinkley(1997)}]{davison_1997}
Davison, A.~C., \& Hinkley, D.~V. 1997, Bootstrap Methods and their
  Application, Cambridge Series in Statistical and Probabilistic Mathematics
  (Cambridge University Press)

\bibitem[{{Dopita} {et~al.}(2016){Dopita}, {Kewley}, {Sutherland}, \&
  {Nicholls}}]{dopita2016}
{Dopita}, M.~A., {Kewley}, L.~J., {Sutherland}, R.~S., \& {Nicholls}, D.~C.
  2016, \apss, 361, 61, \dodoi{10.1007/s10509-016-2657-8}

\bibitem[{Espinosa-Ponce {et~al.}(2020)Espinosa-Ponce, Sánchez, Morisset,
  Barrera-Ballesteros, Galbany, García-Benito, Lacerda, \&
  Mast}]{Espinosaponce2020}
Espinosa-Ponce, C., Sánchez, S.~F., Morisset, C., {et~al.} 2020, \mnras, 494,
  \dodoi{10.1093/mnras/staa782}

\bibitem[{{Esteban} {et~al.}(2013){Esteban}, {Carigi}, {Copetti},
  {Garc{\'\i}a-Rojas}, {Mesa-Delgado}, {Casta{\~n}eda}, \&
  {P{\'e}quignot}}]{esteban_2013}
{Esteban}, C., {Carigi}, L., {Copetti}, M.~V.~F., {et~al.} 2013, \mnras, 433,
  382, \dodoi{10.1093/mnras/stt730}

\bibitem[{{Ferguson} \& {Clarke}(2001)}]{ferguson_2001}
{Ferguson}, A.~M.~N., \& {Clarke}, C.~J. 2001, \mnras, 325, 781,
  \dodoi{10.1046/j.1365-8711.2001.04501.x}

\bibitem[{Fitzpatrick(1999)}]{Fitzpatrick1999}
Fitzpatrick, E.~L. 1999, \pasp, 111, 63, \dodoi{10.1086/316293}

\bibitem[{{Flores-Fajardo} {et~al.}(2011){Flores-Fajardo}, {Morisset},
  {Stasi{\'n}ska}, \& {Binette}}]{floresfajardo2011}
{Flores-Fajardo}, N., {Morisset}, C., {Stasi{\'n}ska}, G., \& {Binette}, L.
  2011, \mnras, 415, 2182, \dodoi{10.1111/j.1365-2966.2011.18848.x}

\bibitem[{{Fragkoudi} {et~al.}(2016){Fragkoudi}, {Athanassoula}, \&
  {Bosma}}]{fragkoudi16}
{Fragkoudi}, F., {Athanassoula}, E., \& {Bosma}, A. 2016, \mnras, 462, L41,
  \dodoi{10.1093/mnrasl/slw120}

\bibitem[{{Garnett}(1998)}]{garnett1998}
{Garnett}, D.~R. 1998, in Revista Mexicana de Astronomia y Astrofisica
  Conference Series, Vol.~7, Revista Mexicana de Astronomia y Astrofisica
  Conference Series, ed. R.~J. {Dufour} \& S.~{Torres-Peimbert}, 58

\bibitem[{Giovanelli {et~al.}(1994)Giovanelli, Haynes, Salzer, Wegner,
  da~Costa, \& Freudling}]{Giovanelli1994}
Giovanelli, R., Haynes, M.~P., Salzer, J.~J., {et~al.} 1994, \aj, 107,
  \dodoi{10.1086/117014}

\bibitem[{{Goetz} \& {Koeppen}(1992)}]{goetz_1992}
{Goetz}, M., \& {Koeppen}, J. 1992, \aap, 262, 455

\bibitem[{{Gonz{\'a}lez Delgado} {et~al.}(2016){Gonz{\'a}lez Delgado}, {Cid
  Fernandes}, {P{\'e}rez}, {Garc{\'\i}a-Benito}, {L{\'o}pez Fern{\'a}ndez},
  {Lacerda}, {Cortijo-Ferrero}, {de Amorim}, {Vale Asari}, {S{\'a}nchez},
  {Walcher}, {Wisotzki}, {Mast}, {Alves}, {Ascasibar}, {Bland-Hawthorn},
  {Galbany}, {Kennicutt}, {M{\'a}rquez}, {Masegosa}, {Moll{\'a}},
  {S{\'a}nchez-Bl{\'a}zquez}, \& {V{\'\i}lchez}}]{gonzales_delgado_2016}
{Gonz{\'a}lez Delgado}, R.~M., {Cid Fernandes}, R., {P{\'e}rez}, E., {et~al.}
  2016, \aap, 590, A44, \dodoi{10.1051/0004-6361/201628174}

\bibitem[{{González Delgado} \& {Pérez}(1997)}]{gonzalez_delgado_1997}
{González Delgado}, R.~M., \& {Pérez}, E. 1997, \aj, 108, 199,
  \dodoi{10.1086/312950}

\bibitem[{{Ho} {et~al.}(2015){Ho}, {Kudritzki}, {Kewley}, {Zahid}, {Dopita},
  {Bresolin}, \& {Rupke}}]{ho2015}
{Ho}, I.~T., {Kudritzki}, R.-P., {Kewley}, L.~J., {et~al.} 2015, \mnras, 448,
  2030, \dodoi{10.1093/mnras/stv067}

\bibitem[{{Ho} {et~al.}(1997){Ho}, {Filippenko}, \& {Sargent}}]{Ho1997}
{Ho}, L.~C., {Filippenko}, A.~V., \& {Sargent}, W.~L.~W. 1997, \apj, 487, 579,
  \dodoi{10.1086/304642}

\bibitem[{{Izotov} {et~al.}(2006){Izotov}, {Stasi{\'n}ska}, {Meynet}, {Guseva},
  \& {Thuan}}]{izotov_2006}
{Izotov}, Y.~I., {Stasi{\'n}ska}, G., {Meynet}, G., {Guseva}, N.~G., \&
  {Thuan}, T.~X. 2006, \aap, 448, 955, \dodoi{10.1051/0004-6361:20053763}

\bibitem[{Kauffmann {et~al.}(2003)Kauffmann, Heckman, Tremonti, \&
  et~al.}]{Kauffmann2003}
Kauffmann, G., Heckman, T.~M., Tremonti, C., \& et~al. 2003, \mnras, 346, 1055,
  \dodoi{10.1111/j.1365-2966.2003.07154.x}

\bibitem[{{Kennicutt} {et~al.}(1989){Kennicutt}, {Keel}, \&
  {Blaha}}]{Kennicutt1989}
{Kennicutt}, R.~C., J., {Keel}, W.~C., \& {Blaha}, C.~A. 1989, \aj, 97, 1022,
  \dodoi{10.1086/115046}

\bibitem[{Kewley {et~al.}(2001)Kewley, Dopita, Sutherland, Heisler, \&
  Trevena}]{Kewley2001}
Kewley, L.~J., Dopita, M.~A., Sutherland, R.~S., Heisler, C.~A., \& Trevena, J.
  2001, \apj, 556, 121, \dodoi{10.1086/321545}

\bibitem[{{Kewley} \& {Ellison}(2008)}]{kewley2008}
{Kewley}, L.~J., \& {Ellison}, S.~L. 2008, \apj, 681, 1183,
  \dodoi{10.1086/587500}

\bibitem[{{Kewley} {et~al.}(2006){Kewley}, {Groves}, {Kauffmann}, \&
  {Heckman}}]{kewley2006}
{Kewley}, L.~J., {Groves}, B., {Kauffmann}, G., \& {Heckman}, T. 2006, \mnras,
  372, 961, \dodoi{10.1111/j.1365-2966.2006.10859.x}

\bibitem[{{Kormendy}(1977)}]{kormendy77}
{Kormendy}, J. 1977, \apj, 218, 333, \dodoi{10.1086/155687}

\bibitem[{{Krishak} {et~al.}(2020){Krishak}, {Dantuluri}, \&
  {Desai}}]{krishak2020a}
{Krishak}, A., {Dantuluri}, A., \& {Desai}, S. 2020, \jcap, 2020, 007,
  \dodoi{10.1088/1475-7516/2020/02/007}

\bibitem[{{Kumari} {et~al.}(2019){Kumari}, {Maiolino}, {Belfiore}, \&
  {Curti}}]{kumari.2019}
{Kumari}, N., {Maiolino}, R., {Belfiore}, F., \& {Curti}, M. 2019, \mnras, 485,
  367, \dodoi{10.1093/mnras/stz366}

\bibitem[{{Lacerda} {et~al.}(2018){Lacerda}, {Cid Fernandes}, {Couto},
  {Stasi{\'n}ska}, {Garc{\'\i}a-Benito}, {Vale Asari}, {P{\'e}rez},
  {Gonz{\'a}lez Delgado}, {S{\'a}nchez}, \& {de Amorim}}]{lacerda2018}
{Lacerda}, E.~A.~D., {Cid Fernandes}, R., {Couto}, G.~S., {et~al.} 2018,
  \mnras, 474, 3727, \dodoi{10.1093/mnras/stx3022}

\bibitem[{{Lacey} \& {Fall}(1985)}]{laceyfall1985}
{Lacey}, C.~G., \& {Fall}, S.~M. 1985, \apj, 290, 154, \dodoi{10.1086/162970}

\bibitem[{{Lopez} {et~al.}(2011){Lopez}, {Krumholz}, {Bolatto}, {Prochaska}, \&
  {Ramirez-Ruiz}}]{lopez_2011}
{Lopez}, L.~A., {Krumholz}, M.~R., {Bolatto}, A.~D., {Prochaska}, J.~X., \&
  {Ramirez-Ruiz}, E. 2011, \apj, 731, 91, \dodoi{10.1088/0004-637X/731/2/91}

\bibitem[{{L{\'o}pez-Cob{\'a}} {et~al.}(2022){L{\'o}pez-Cob{\'a}},
  {S{\'a}nchez}, {Lin}, {Anderson}, {Lin}, {Cruz-Gonz{\'a}lez}, {Galbany}, \&
  {Barrera-Ballesteros}}]{lopez-coba22}
{L{\'o}pez-Cob{\'a}}, C., {S{\'a}nchez}, S.~F., {Lin}, L., {et~al.} 2022, \apj,
  939, 40, \dodoi{10.3847/1538-4357/ac937b}

\bibitem[{{Maiolino} \& {Mannucci}(2019)}]{maiolino19}
{Maiolino}, R., \& {Mannucci}, F. 2019, \aapr, 27, 3,
  \dodoi{10.1007/s00159-018-0112-2}

\bibitem[{Marino {et~al.}(2013)Marino, Rosales-Ortega, Sánchez, \&
  et~al.}]{Marino2013}
Marino, R.~A., Rosales-Ortega, F.~F., Sánchez, S.~F., \& et~al. 2013, \aap,
  559, \dodoi{10.1051/0004-6361/201321956}

\bibitem[{{Martin} \& {Roy}(1994)}]{martin1994}
{Martin}, P., \& {Roy}, J.~R. 1994, \apj, 424, 599, \dodoi{10.1086/173917}

\bibitem[{Martin \& {Roy}(1995)}]{martin.1995}
Martin, P., \& {Roy}, J.~R. 1995, \apj, 445, 161, \dodoi{10.1086/175682}

\bibitem[{{Matteucci} \& {Francois}(1989)}]{matteucci1989}
{Matteucci}, F., \& {Francois}, P. 1989, \mnras, 239, 885,
  \dodoi{10.1093/mnras/239.3.885}

\bibitem[{McCall {et~al.}(1985)McCall, Rybski, \& Shields}]{McCall1985}
McCall, M.~L., Rybski, P.~M., \& Shields, G.~A. 1985, \apj, 57, 1,
  \dodoi{10.1086/190994}

\bibitem[{{M{\'e}ndez-Abreu} {et~al.}(2017){M{\'e}ndez-Abreu}, {Ruiz-Lara},
  {S{\'a}nchez-Menguiano}, {de Lorenzo-C{\'a}ceres}, {Costantin},
  {Catal{\'a}n-Torrecilla}, {Florido}, \& {et al.}}]{mendez.abreu.2017}
{M{\'e}ndez-Abreu}, J., {Ruiz-Lara}, T., {S{\'a}nchez-Menguiano}, L., {et~al.}
  2017, \aap, 598, A32, \dodoi{10.1051/0004-6361/201629525}

\bibitem[{{M{\'e}ndez-Delgado} {et~al.}(2022){M{\'e}ndez-Delgado}, {Amayo},
  {Arellano-C{\'o}rdova}, {Esteban}, {Garc{\'\i}a-Rojas}, {Carigi}, \&
  {Delgado-Inglada}}]{mendez-delgado22}
{M{\'e}ndez-Delgado}, J.~E., {Amayo}, A., {Arellano-C{\'o}rdova}, K.~Z.,
  {et~al.} 2022, \mnras, 510, 4436, \dodoi{10.1093/mnras/stab3782}

\bibitem[{{Minchev} \& {Famaey}(2010)}]{minchev_2010}
{Minchev}, I., \& {Famaey}, B. 2010, \apj, 722, 112,
  \dodoi{10.1088/0004-637X/722/1/112}

\bibitem[{{Mollá} {et~al.}(2019){Mollá}, {Díaz}, {Cavichia}, {Gibson},
  {Maciel}, {Costa}, {Ascasibar}, \& {Few}}]{molla2019}
{Mollá}, M., {Díaz}, {\'A}.~I., {Cavichia}, O., {et~al.} 2019, \mnras, 482,
  3071, \dodoi{10.1093/mnras/sty2877}

\bibitem[{{Muggeo}(2003)}]{muggeo2003}
{Muggeo}, V.~M. 2003, Statist. Med., 22, 3055, \dodoi{10.1002/sim.1545}

\bibitem[{{Narisetty}(2020)}]{akaike}
{Narisetty}, N.~N. 2020, in Handbook of Statistics, Vol.~43, Principles and
  Methods for Data Science, ed. A.~S.~R. {Srinivasa Rao} \& C.~R. {Rao}
  (Elsevier), 207--248, \dodoi{https://doi.org/10.1016/bs.host.2019.08.001}

\bibitem[{Oey {et~al.}(2003)Oey, Parker, Mikles, \& Zhang}]{Oey_2003}
Oey, M.~S., Parker, J.~S., Mikles, V.~J., \& Zhang, X. 2003, \aj, 126, 2317,
  \dodoi{10.1086/378163}

\bibitem[{{Oppenheimer} \& {Dav{\'e}}(2008)}]{oppenheimer_2008}
{Oppenheimer}, B.~D., \& {Dav{\'e}}, R. 2008, \mnras, 387, 577,
  \dodoi{10.1111/j.1365-2966.2008.13280.x}

\bibitem[{{Oppenheimer} {et~al.}(2010){Oppenheimer}, {Dav{\'e}}, {Kere{\v{s}}},
  {Fardal}, {Katz}, {Kollmeier}, \& {Weinberg}}]{oppenheimer_2010}
{Oppenheimer}, B.~D., {Dav{\'e}}, R., {Kere{\v{s}}}, D., {et~al.} 2010, \mnras,
  406, 2325, \dodoi{10.1111/j.1365-2966.2010.16872.x}

\bibitem[{Osterbrock(1989)}]{Osterbrock1989}
Osterbrock, D.~E. 1989, Annals of the New York Academy of Sciences, 571, 99,
  \dodoi{10.1111/j.1749-6632.1989.tb50500.x}

\bibitem[{{Osterbrock} \& {Ferland}(2006)}]{osterbroack2006}
{Osterbrock}, D.~E., \& {Ferland}, G.~J. 2006, Astrophysics of gaseous nebulae
  and active galactic nuclei (Sausalito, California, USA: University Science
  Books), 461 p.

\bibitem[{{Pagel} {et~al.}(1979){Pagel}, {Edmunds}, {Blackwell}, {Chun}, \&
  {Smith}}]{pagel1979}
{Pagel}, B.~E.~J., {Edmunds}, M.~G., {Blackwell}, D.~E., {Chun}, M.~S., \&
  {Smith}, G. 1979, \mnras, 189, 95, \dodoi{10.1093/mnras/189.1.95}

\bibitem[{{Pagel} {et~al.}(1992){Pagel}, {Simonson}, {Terlevich}, \&
  {Edmunds}}]{pagel_1992}
{Pagel}, B.~E.~J., {Simonson}, E.~A., {Terlevich}, R.~J., \& {Edmunds}, M.~G.
  1992, \mnras, 255, 325, \dodoi{10.1093/mnras/255.2.325}

\bibitem[{{Pearson}(1895)}]{pearson.coef}
{Pearson}, K. 1895, Proceedings of the Royal Society of London Series I, 58,
  240

\bibitem[{{Peimbert} \& {Costero}(1969)}]{peimbert_1969}
{Peimbert}, M., \& {Costero}, R. 1969, Boletin de los Observatorios
  Tonantzintla y Tacubaya, 5, 3

\bibitem[{{P{\'e}rez-Montero} \& {Contini}(2009)}]{perezmontero2009}
{P{\'e}rez-Montero}, E., \& {Contini}, T. 2009, \mnras, 398, 949–960,
  \dodoi{10.1111/j.1365-2966.2009.15145.x}

\bibitem[{Pettini \& Pagel(2004)}]{Pettini2004}
Pettini, M., \& Pagel, B. E.~J. 2004, \mnras, 348, L59–L63,
  \dodoi{10.1111/j.1365-2966.2004.07591.x}

\bibitem[{Pilgrim(2021)}]{Pilgrim2021}
Pilgrim, C. 2021, Journal of Open Source Software, 6, 3859,
  \dodoi{10.21105/joss.03859}

\bibitem[{{Pilyugin} {et~al.}(2018){Pilyugin}, {Grebel}, {Zinchenko},
  {Nefedyev}, {Shulga}, {Wei}, \& {Berczik}}]{pilyugin.2018}
{Pilyugin}, L.~S., {Grebel}, E.~K., {Zinchenko}, I.~A., {et~al.} 2018, \aap,
  613, A1, \dodoi{10.1051/0004-6361/201732185}

\bibitem[{{Prugniel} \& {Simien}(1997)}]{prugniel.1997}
{Prugniel}, P., \& {Simien}, F. 1997, \aap, 321, 111

\bibitem[{{Qu} {et~al.}(2011){Qu}, {Di Matteo}, {Lehnert}, {van Driel}, \&
  {Jog}}]{qu_2011}
{Qu}, Y., {Di Matteo}, P., {Lehnert}, M.~D., {van Driel}, W., \& {Jog}, C.~J.
  2011, \aap, 535, A5, \dodoi{10.1051/0004-6361/201116502}

\bibitem[{{Quillen} {et~al.}(2009){Quillen}, {Minchev}, {Bland-Hawthorn}, \&
  {Haywood}}]{quillen_2009}
{Quillen}, A.~C., {Minchev}, I., {Bland-Hawthorn}, J., \& {Haywood}, M. 2009,
  \mnras, 397, 1599, \dodoi{10.1111/j.1365-2966.2009.15054.x}

\bibitem[{{Reynolds} {et~al.}(2001){Reynolds}, {Sterling}, {Haffner}, \&
  {Tufte}}]{reynolds2001}
{Reynolds}, R.~J., {Sterling}, N.~C., {Haffner}, L.~M., \& {Tufte}, S.~L. 2001,
  \apjl, 548, L221, \dodoi{10.1086/319119}

\bibitem[{{Rosales-Ortega} {et~al.}(2011){Rosales-Ortega}, {D{\'\i}az},
  {Kennicutt}, \& {S{\'a}nchez}}]{rosales.ortega.2011}
{Rosales-Ortega}, F.~F., {D{\'\i}az}, A.~I., {Kennicutt}, R.~C., \&
  {S{\'a}nchez}, S.~F. 2011, \mnras, 415, 2439,
  \dodoi{10.1111/j.1365-2966.2011.18870.x}

\bibitem[{{Ro{\v{s}}kar} {et~al.}(2012){Ro{\v{s}}kar}, {Debattista}, {Quinn},
  \& {Wadsley}}]{roskar_2012}
{Ro{\v{s}}kar}, R., {Debattista}, V.~P., {Quinn}, T.~R., \& {Wadsley}, J. 2012,
  \mnras, 426, 2089, \dodoi{10.1111/j.1365-2966.2012.21860.x}

\bibitem[{{Roy}(1996)}]{roy1996}
{Roy}, J.~R. 1996, in Astronomical Society of the Pacific Conference Series,
  Vol.~91, IAU Colloq. 157: Barred Galaxies, ed. R.~{Buta}, D.~A. {Crocker}, \&
  B.~G. {Elmegreen}, 63

\bibitem[{S{\'{a}}nchez(2006)}]{sanchez2006}
S{\'{a}}nchez, S.~F. 2006, Astronomische Nachrichten, 327, 850,
  \dodoi{10.1002/asna.200610643}

\bibitem[{{S{\'a}nchez} {et~al.}(2012){S{\'a}nchez}, {Rosales-Ortega},
  {Marino}, \& {et al.}}]{Sanchez2012b}
{S{\'a}nchez}, S.~F., {Rosales-Ortega}, F.~F., {Marino}, R.~A., \& {et al.}
  2012, \aap, 546, A2, \dodoi{10.1051/0004-6361/201219578}

\bibitem[{{S{\'a}nchez} {et~al.}(2015){S{\'a}nchez}, {P{\'e}rez},
  {Rosales-Ortega}, {Miralles-Caballero}, {L{\'o}pez-S{\'a}nchez},
  {Iglesias-P{\'a}ramo}, {Marino}, {S{\'a}nchez-Menguiano},
  {Garc{\'\i}a-Benito}, \& {et al.}}]{sanchez.2015.bpt}
{S{\'a}nchez}, S.~F., {P{\'e}rez}, E., {Rosales-Ortega}, F.~F., {et~al.} 2015,
  \aap, 574, A47, \dodoi{10.1051/0004-6361/201424873}

\bibitem[{{Sani} {et~al.}(2011){Sani}, {Marconi}, {Hunt}, \&
  {Risaliti}}]{sani.2011}
{Sani}, E., {Marconi}, A., {Hunt}, L.~K., \& {Risaliti}, G. 2011, \mnras, 413,
  1479, \dodoi{10.1111/j.1365-2966.2011.18229.x}

\bibitem[{Scarano {et~al.}(2008)Scarano, Madsen, Roy, \& Lépine}]{Scarano2008}
Scarano, S., Madsen, F. R.~H., Roy, N., \& Lépine, J. R.~D. 2008, \mnras, 386,
  963–972, \dodoi{10.1111/j.1365-2966.2008.13079.x}

\bibitem[{{Searle}(1971)}]{searle.1971}
{Searle}, L. 1971, \apj, 168, 327, \dodoi{10.1086/151090}

\bibitem[{{Sellwood} \& {Binney}(2002)}]{sellwood_2002}
{Sellwood}, J.~A., \& {Binney}, J.~J. 2002, \mnras, 336, 785,
  \dodoi{10.1046/j.1365-8711.2002.05806.x}

\bibitem[{{Singh} {et~al.}(2013){Singh}, {van de Ven}, {Jahnke}, {Lyubenova},
  {Falc{\'o}n-Barroso}, {Alves}, {Cid Fernandes}, {Galbany},
  {Garc{\'\i}a-Benito}, {Husemann}, {Kennicutt}, \& {et al.}}]{singh.2013}
{Singh}, R., {van de Ven}, G., {Jahnke}, K., {et~al.} 2013, \aap, 558, A43,
  \dodoi{10.1051/0004-6361/201322062}

\bibitem[{{Stasi{\'n}ska} {et~al.}(2006){Stasi{\'n}ska}, {Cid Fernandes},
  {Mateus}, {Sodré}, \& {Asari}}]{stasinska2006}
{Stasi{\'n}ska}, G., {Cid Fernandes}, R., {Mateus}, A., {Sodré}, L., \&
  {Asari}, N.~V. 2006, \mnras, 371, 972,
  \dodoi{10.1111/j.1365-2966.2006.10732.x}

\bibitem[{{Stasi{\'n}ska} {et~al.}(2008){Stasi{\'n}ska}, {Vale Asari}, {Cid
  Fernandes}, {Gomes}, {Schlickmann}, {Mateus}, {Schoenell}, {Sodr{\'e}}, \&
  {Seagal Collaboration}}]{stasinska2008}
{Stasi{\'n}ska}, G., {Vale Asari}, N., {Cid Fernandes}, R., {et~al.} 2008,
  \mnras, 391, L29, \dodoi{10.1111/j.1745-3933.2008.00550.x}

\bibitem[{{Storchi-Bergmann} {et~al.}(1994){Storchi-Bergmann}, {Calzetti}, \&
  {Kinney}}]{storchi1994}
{Storchi-Bergmann}, T., {Calzetti}, D., \& {Kinney}, A.~L. 1994, \apj, 429,
  572, \dodoi{10.1086/174345}

\bibitem[{{Sánchez} {et~al.}(2016){Sánchez}, {García-Benito}, {Zibetti}, \&
  {et al.}}]{Sanchez2016}
{Sánchez}, S.~F., {García-Benito}, R., {Zibetti}, S., \& {et al.} 2016, \aap,
  594, A36, \dodoi{10.1051/0004-6361/201628661}

\bibitem[{Sánchez {et~al.}(2012)Sánchez, Kennicutt, de~Paz, \&
  et~al.}]{Sanchez2012a}
Sánchez, S.~F., Kennicutt, R.~C., de~Paz, A.~G., \& et~al. 2012, \aap, 538,
  A8, \dodoi{10.1051/0004-6361/201117353}

\bibitem[{Sánchez {et~al.}(2014)Sánchez, Rosales-Ortega, Iglesias-Páramo, \&
  et~al.}]{Sanchez2014}
Sánchez, S.~F., Rosales-Ortega, F.~F., Iglesias-Páramo, J., \& et~al. 2014,
  \aap, 563, A49, \dodoi{10.1051/0004-6361/201322343}

\bibitem[{Sánchez-Menguiano {et~al.}(2016)Sánchez-Menguiano, Sánchez,
  Pérez, \& et~al.}]{Sanchez-Menguiano2016}
Sánchez-Menguiano, L., Sánchez, S.~F., Pérez, I., \& et~al. 2016, \aap, 587,
  A70, \dodoi{10.1051/0004-6361/201527450}

\bibitem[{Sánchez-Menguiano {et~al.}(2018)Sánchez-Menguiano, Sánchez,
  Pérez, Ruiz-Lara, Galbany, Anderson, Krühler, Kuncarayakti, \&
  Lyman}]{Sanchez-Menguiano2018}
Sánchez-Menguiano, L., Sánchez, S.~F., Pérez, I., {et~al.} 2018, \aap, 609,
  A119, \dodoi{10.1051/0004-6361/201731486}

\bibitem[{{Vale Asari} {et~al.}(2019){Vale Asari}, {Couto}, {Cid Fernandes},
  {Stasi{\'n}ska}, {de Amorim}, {Ruschel-Dutra}, {Werle}, \&
  {Florido}}]{vale.asari.2019}
{Vale Asari}, N., {Couto}, G.~S., {Cid Fernandes}, R., {et~al.} 2019, \mnras,
  489, 4721, \dodoi{10.1093/mnras/stz2470}

\bibitem[{{Vila-Costas} \& {Edmunds}(1992)}]{vilacostas1992}
{Vila-Costas}, M.~B., \& {Edmunds}, M.~G. 1992, \mnras, 259, 121,
  \dodoi{10.1093/mnras/259.1.121}

\bibitem[{{Vilchez} \& {Esteban}(1996{\natexlab{a}})}]{vilchez_1996}
{Vilchez}, J.~M., \& {Esteban}, C. 1996{\natexlab{a}}, \mnras, 280, 720,
  \dodoi{10.1093/mnras/280.3.720}

\bibitem[{{Vilchez} \& {Esteban}(1996{\natexlab{b}})}]{vilchez1996}
---. 1996{\natexlab{b}}, \mnras, 280, 720, \dodoi{10.1093/mnras/280.3.720}

\bibitem[{{Vogt} {et~al.}(2014){Vogt}, {Dopita}, {Kewley}, {Sutherland},
  {Scharw{\"a}chter}, {Basurah}, {Ali}, \& {Amer}}]{vogt.2014}
{Vogt}, F.~P.~A., {Dopita}, M.~A., {Kewley}, L.~J., {et~al.} 2014, \apj, 793,
  127, \dodoi{10.1088/0004-637X/793/2/127}

\bibitem[{{Vogt} {et~al.}(2017){Vogt}, {P{\'e}rez}, {Dopita},
  {Verdes-Montenegro}, \& {Borthakur}}]{vogt2017}
{Vogt}, F.~P.~A., {P{\'e}rez}, E., {Dopita}, M.~A., {Verdes-Montenegro}, L., \&
  {Borthakur}, S. 2017, \aap, 601, A61, \dodoi{10.1051/0004-6361/201629853}

\bibitem[{{Walcher} {et~al.}(2008){Walcher}, {Lamareille}, {Vergani}, \& {et
  al.}}]{walcher2008}
{Walcher}, C.~J., {Lamareille}, F., {Vergani}, D., \& {et al.} 2008, \aap, 491,
  713, \dodoi{10.1051/0004-6361:200810704}

\bibitem[{Walcher {et~al.}(2014)Walcher, Wisotzki, Bekeraité, \&
  et~al.}]{Walcher2014}
Walcher, C.~J., Wisotzki, L., Bekeraité, S., \& et~al. 2014, \aap, 569, A1,
  \dodoi{10.1051/0004-6361/201424198}

\bibitem[{Zaritsky {et~al.}(1994)Zaritsky, Kennicutt, \& Huchra}]{Zaritsky1994}
Zaritsky, D., Kennicutt, R. C.~J., \& Huchra, J.~P. 1994, \apj, 420, 87,
  \dodoi{10.1086/173544}

\bibitem[{{Zinchenko} {et~al.}(2019){Zinchenko}, {Just}, {Pilyugin}, \&
  {Lara-Lopez}}]{zinchenko.2019}
{Zinchenko}, I.~A., {Just}, A., {Pilyugin}, L.~S., \& {Lara-Lopez}, M.~A. 2019,
  \aap, 623, A7, \dodoi{10.1051/0004-6361/201834364}

\bibitem[{{Zinchenko} {et~al.}(2016){Zinchenko}, {Pilyugin}, {Grebel},
  {S{\'a}nchez}, \& {V{\'\i}lchez}}]{Zinchenko.2016}
{Zinchenko}, I.~A., {Pilyugin}, L.~S., {Grebel}, E.~K., {S{\'a}nchez}, S.~F.,
  \& {V{\'\i}lchez}, J.~M. 2016, \mnras, 462, 2715,
  \dodoi{10.1093/mnras/stw1857}

\bibitem[{{Zurita} {et~al.}(2021){Zurita}, {Florido}, {Bresolin},
  {P{\'e}rez-Montero}, \& {P{\'e}rez}}]{zurita2021}
{Zurita}, A., {Florido}, E., {Bresolin}, F., {P{\'e}rez-Montero}, E., \&
  {P{\'e}rez}, I. 2021, \mnras, 500, 2359, \dodoi{10.1093/mnras/staa2246}

\bibitem[{{Zurita} {et~al.}(2000){Zurita}, {Rozas}, \& {Beckman}}]{zurita2000}
{Zurita}, A., {Rozas}, M., \& {Beckman}, J.~E. 2000, \aap, 363, 9

\end{thebibliography}
\bibliographystyle{aasjournal}



\end{document}